\documentclass[prd,aps,preprint,tightenlines,superscriptaddress,nofootinbib,showpacs,showkeys]{revtex4}
\usepackage{float}
\usepackage{mathrsfs}
\usepackage{amsfonts}
\usepackage{array}
\usepackage{epsfig}
\usepackage{amsmath}    
\usepackage{amssymb}   
\usepackage{graphicx}   
\usepackage{verbatim}   
\usepackage{color}      
\usepackage{subfigure}  
\usepackage{hyperref}   
\usepackage{url}

\usepackage{array,multirow}

\def\gev{{\ifmmode{\mathrm{\:GeV}}\else${\mathrm{\:GeV}}$\fi}}
\def\tev{{\ifmmode{\mathrm{\:TeV}}\else${\mathrm{\:TeV}}$\fi}}

\begin{document}

\title{
CT14 Intrinsic Charm Parton Distribution Functions \\
from CTEQ-TEA Global Analysis}

\author{Tie-Jiun Hou}
\email{tjhou@msu.edu}
\affiliation{
Department of Physics, Southern Methodist University,\\
 Dallas, TX 75275-0181, U.S.A. }

\author{Sayipjamal Dulat}
\email{sdulat@msu.edu}
\affiliation{
School of Physics Science and Technology, Xinjiang University,\\
 Urumqi, Xinjiang 830046 China }
 \affiliation{
Center for Theoretical Physics, Xinjiang University,\\
 Urumqi, Xinjiang 830046 China }
\affiliation{
Department of Physics and Astronomy, Michigan State University,\\
 East Lansing, MI 48824 U.S.A. }

\author{Jun Gao}
\affiliation{School of Physics and Astronomy, INPAC,\\
Shanghai Key Laboratory for Particle Physics and Cosmology,\\
Shanghai Jiao-Tong University, Shanghai 200240, China}
\email{jung49@sjtu.edu.cn}

\author{ Marco Guzzi}
\email{mguzzi@kennesaw.edu}
\affiliation{School of Physics and Astronomy, University of Manchester, Manchester M13 9PL, United Kingdom}
\affiliation{Department of Physics, Kennesaw State University, Kennesaw, GA 30144, USA}

\author{Joey Huston}
\email{huston@pa.msu.edu}
\affiliation{
Department of Physics and Astronomy, Michigan State University,\\
East Lansing, MI 48824 U.S.A. }

\author{Pavel Nadolsky}
\email{nadolsky@physics.smu.edu}
\affiliation{
Department of Physics, Southern Methodist University,\\
 Dallas, TX 75275-0181, U.S.A. }

\author{Carl Schmidt}
\email{schmidt@pa.msu.edu}
\affiliation{
Department of Physics and Astronomy, Michigan State University,\\
 East Lansing, MI 48824 U.S.A. }

\author{Jan Winter}
\email{jwinter@pa.msu.edu}
\affiliation{
Department of Physics and Astronomy, Michigan State University,\\
 East Lansing, MI 48824 U.S.A. }

\author{Keping Xie}
\email{kepingx@mail.smu.edu}
\affiliation{
Department of Physics, Southern Methodist University,\\
 Dallas, TX 75275-0181, U.S.A. }

\author{ C.--P. Yuan}
\email{yuan@pa.msu.edu}
\affiliation{
Department of Physics and Astronomy, Michigan State University,\\
 East Lansing, MI 48824 U.S.A. }

\begin{abstract}

  We investigate the possibility of a (sizable) nonperturbative
  contribution to the charm parton distribution function (PDF)
  in a nucleon, theoretical issues arising in its interpretation, and its
  potential impact on LHC scattering processes.
  The ``fitted charm'' PDF obtained in various QCD analyses
  contains a process-dependent component that is partly
  traced to power-suppressed radiative contributions in DIS
  and is generally different at  the LHC.
  We discuss separation of the universal component of the nonperturbative charm
  from the rest of the radiative contributions and estimate
  its magnitude in the CT14 global QCD analysis at the
  next-to-next-to leading order in the QCD coupling strength,
  including the latest experimental data from HERA and the Large Hadron
  Collider.  Models for the nonperturbative charm PDF are examined as
  a function of the charm quark mass and other parameters. The
  prospects for testing these models in
  the associated production of a Z boson and a charm jet at the LHC
  are studied under realistic assumptions, including effects of the
  final-state parton showering.

\end{abstract}



\maketitle

\section{Introduction: CTEQ distributions with intrinsic charm
\label{sec:INTRODUCTION}}
The principle of the global analysis is to use QCD theory to analyze a broad
range of experimental data, including precision data from
HERA, the Tevatron, and the Large Hadron Collider (LHC).
In particular, theoretical predictions
for short-distance scattering processes allow the measurement, within
some approximations, of universal parton distribution functions (PDFs)
for the proton. These functions can then be used to predict hadronic
cross sections in the QCD and electroweak theories, and in beyond-the-standard-model
theories. With the new high-precision data becoming available from the LHC,
the ultimate goal for the global QCD analysis is to be able to make predictions that are accurate to about one percent. This, in turn, requires improvements
in theoretical predictions to allow for an accurate extraction of the parton content of the proton in global fits.

A recently published CTEQ-TEA (CT) analysis of QCD data
~\cite{Dulat:2015mca} produced the CT14NNLO PDFs,
referred to as the CT14 PDFs in this paper. The analysis is
based on the next-to-next-to-leading order (NNLO)
approximation for perturbative QCD.
That is, NNLO expressions are used for the running coupling
$\alpha_{\rm S}(Q)$, for the Dokshitzer-Gribov-Lipatov-Altarelli-Parisi (DGLAP) evolution equations~\cite{Gribov:1972ri, Gribov:1972rt,Lipatov:1974qm,Dokshitzer:1977sg,Altarelli:1977zs},
and for those hard matrix elements for which the
NNLO approximation is available, such as the
deep-inelastic scattering (DIS) neutral-current data from HERA and
fixed-target experiments, and the Drell-Yan data from the Tevatron,
fixed-target experiments,
and the LHC~\cite{Moch:2004pa,Vogt:2004mw,SanchezGuillen:1990iq,vanNeerven:1991nn,
Zijlstra:1991qc,Zijlstra:1992qd,Laenen:1992zk,Riemersma:1994hv,Buza:1995ie,Anastasiou:2003yy,Anastasiou:2003ds}.
Next-to-leading order (NLO) is used only for inclusive jet data from
the Tevatron and the LHC
and for deep-inelastic scattering (DIS) charged-current data from HERA
and fixed-target experiments. The NNLO predictions for these processes
~\cite{Berger:2016inr, Currie:2016bfm,Currie:2017ctp} were not available
or incomplete at the time of the CT14 study, and we have argued \cite{Gao:2013xoa, Dulat:2015mca} that the effect of missing NNLO terms in jet production on the PDFs is small relatively to the experimental uncertainties in the CT14 data sets. Similarly, the NNLO contribution for charged-current DIS, including massive charm scattering contributions, is modest compared to the experimental uncertainties.

In the global analysis, all QCD parameters, such as $\alpha_s$ and the quark
masses, are correlated with the PDFs.
The determination of the PDFs
depends not only on the data sample included in the fits, but also on
the specific theory assumptions and underlying physics models.
As one such choice made in the standard CT PDF sets, the charm quark and
antiquark PDFs are taken to be zero below a low energy scale $Q_c=Q_0$ of
order of the charm mass. In the CT14 analysis, the charm quark and
antiquark PDFs were turned on at the scale $Q_c=Q_0=m_c=1.3$ GeV,
with an initial ${\cal O}(\alpha_s^2)$ distribution consistent with
NNLO matching~\cite{Buza:1995ie,Buza:1996wv} to the three-flavor
result. At higher $Q$, most of the charm PDF is generated from
the DGLAP evolution that proceeds through perturbative splittings of gluons and light-flavor quarks.
Hence, the charm PDF from a standard global analysis is called
``perturbative'', for it was obtained by perturbative relations from light-parton PDFs
at scale $Q_c$ and perturbatively evolved to the experimental data scale $Q$.

In addition to the perturbative charm production mechanism, it is
believed that ``intrinsic charm quarks'' may emerge
from the nonperturbative structure of the hadronic bound
state. The plausibility of the intrinsic charm (IC) component,
its dynamical origin, and its actual magnitude have been a subject of
a long-standing debate. Indeed, QCD theory rigorously predicts
existence of power-suppressed (higher-twist)
channels for charm quark production that are independent of
the leading-power (twist-2, or perturbative) production of
charm quarks. The intrinsic
charm (IC) quarks have been associated with the higher
$|uudc\overline{c}\rangle$ Fock state of the proton wave
function~\cite{Brodsky:1980pb,Brodsky:1981se,Pumplin:2005yf,Chang:2011vx,Blumlein:2015qcn,Brodsky:2015fna}
and predicted by meson-baryon models~\cite{Navarra:1995rq,Paiva:1996dd,Steffens:1999hx,Hobbs:2013bia}.
On the other hand, Refs.~\cite{Jimenez-Delgado:2014zga,Jimenez-Delgado:2015tma}
concluded that the momentum fraction carried by intrinsic charm quarks
is at most 0.5\% at the 4$\sigma$ level,
though this conclusion has been challenged in Ref.~\cite{Brodsky:2015uwa}.
This is to be compared to the earlier CT10IC study~\cite{Dulat:2013hea}, which
concluded that the existing data may tolerate a much larger momentum fraction
carried by intrinsic charm quarks. For a valence-like model, it was found to be
less than about 2.5\%, at the 90\% confidence level (C.L.).
Recently, several analyses by the NNPDF group
\cite{Ball:2015tna,Ball:2015dpa,Ball:2016neh,Ball:2017nwa} established a smaller
fitted charm momentum fraction.
NNPDF determined a fitted charm momentum fraction equal to $(0.26 \pm 0.42$)\%
at 68\% C.L. just above the charm mass threshold, with the charm quark
pole mass taken to be 1.51 GeV~\cite{Ball:2017nwa}, and equal to
$(0.34 \pm 0.14$)\% when the EMC data \cite{Aubert:1982tt}
 on SIDIS charm production were included.

The current paper revisits the issue in the context of the CT14
analysis~\cite{Dulat:2015mca}, also including more recent advances that
were made in the follow-up CT14HERA2 study~\cite{Hou:2016nqm}.
It updates the previous work~\cite{Dulat:2013hea} on
fitting the charm PDFs based on the CT10 NNLO
framework~\cite{Gao:2013xoa}, as well as the CTEQ6.6 IC study
\cite{Pumplin:2007wg} done at NLO.
In addition to implementing the combined
HERA I+II data on DIS, the new LHC data, and improved parametrizations for
light-parton distributions, we shall address some fundamental
questions: What dynamics produces the nonperturbative $c$ and $\overline{c}$
components of the proton? Is there a universal description of this type of charm
component that is supported by the QCD factorization theorem, such that the
same charm PDF can be used in both lepton-hadron
and hadron-hadron scattering processes?

These core questions must be raised to appraise the range
of validity of the PDF models with nonperturbative charm in our work
and in the other recent
studies~\cite{Jimenez-Delgado:2014zga,Jimenez-Delgado:2015tma,Lyonnet:2015dca,Ball:2016neh,Ball:2017nwa}.
We address them by starting from the fundamental QCD result,
the factorization theorem for DIS cross sections with massive
fermions. We start by discussing the definition to the ``intrinsic
charm'', the term that has been used inconsistently in the literature. In the
theoretical section, we advance a viewpoint that the ``intrinsic
charm'' can refer to related, but non-equivalent concepts
of  either the ``fitted charm'' PDF parametrization, on one hand,  or
the genuine nonperturbative charm contribution
defined by the means of power counting of radiative contributions to
DIS. This means that the generic notion of the ``intrinsic charm'' may cover
several kinds of unalike radiative contributions.
After we draw this consequential distinction, and assuming
that the nonperturbative charm scattering cross section can be approximated by a
factorized form, our global analysis examines agreement of various
models for the nonperturbative charm with the modern QCD
experimental data.

The nonperturbative charm content is normally assumed to
be suppressed by powers of $(\Lambda^2/m^2_c)$, where $\Lambda$
is a nonperturbative QCD scale. But, since this ratio
is not very small, it may be relevant in some processes such as precise DIS.
The allowed magnitude of the nonperturbative charm is
influenced by other theoretical assumptions that a global fit makes,
especially by the heavy-quark factorization scheme~\cite{Aivazis:1993kh,Aivazis:1993pi,Buza:1996wv,Thorne:1997ga,Martin:2010db,Forte:2010ta},
the $\alpha_s$ order of the calculation, the assumed
charm mass $m_c$, and the parametrization forms for the PDFs of all flavors.
We study such effects in turn and find that, among the listed factors,
the IC component is strongly correlated with the
assumed charm mass.

Dependence on $m_c$ in the absence of the nonperturbative charm has
been addressed at NNLO in the CT10 NNLO framework~\cite{Gao:2013wwa}
and in other references~\cite{Alekhin:2012vu,Harland-Lang:2015qea,Ball:2016neh,Bertone:2016ywq,Alekhin:2017kpj,Gizhko:2017fiu}.
In the context of the CT10 analysis~\cite{Gao:2013wwa},
the general dependence on the charm quark mass was studied, and a
preferred value of $m_c(m_c) = 1.15^{+0.18}_{-0.12}$ GeV
was obtained at 68\% C.L.,
where the error is a sum in quadrature of PDF and theoretical uncertainties.
Here, $m_c(m_c)$ denotes the running mass of the charm quark, defined in the
modified minimal-subtraction ($\overline {\rm MS}$) scheme and
evaluated at the scale of $m_c$. This value, constrained primarily by
a combination of inclusive and charm production measurements in HERA
deep-inelastic scattering, translates  into the pole mass $m_c^{\rm
  pole}=1.31^{+0.19}_{-0.13}$ GeV and $1.54^{+0.18}_{-0.12}$ GeV when
using the conversion formula in Eq.~(17) of Ref.~\cite{Chetyrkin:2000yt}
at the one-loop and two-loop order, respectively.
As the pole mass of 1.3-1.8 GeV borders the nonperturbative region, accuracy of
its determination is limited by significant radiative contributions associated with renormalons \cite{Bigi:1994em,Beneke:1994sw,Beneke:1998ui}.
In this light both converted values are compatible with the value of
$m_c^{\rm pole} = 1.3$ GeV, which was assumed by CT10 and CT14 and provides the best fit to HERAI+II data at NNLO with the chosen PDF parametric form.
We shall use it as our standard charm quark pole mass value
in this paper, unless specified otherwise.

To establish robustness of our conclusions, in our fits we varied
the selection of data and the analysis setup. Constraints
on the IC from both CT14~\cite{Dulat:2015mca}
and CT14HERA2 sets~\cite{Hou:2016nqm} of experimental data were
compared. As the CT14HERA2 fit prefers a smaller strangeness PDF than
CT14, comparison of the CT14 and CT14HERA2 allowed us to estimate
the sensitivity of the IC to the strangeness content. [The sensitivity
  to the treatment of bottom quarks is expected to be  marginal.]

Finally, we consider the impact of the possible nonperturbative
charm on predictions for the present and future experimental data.
The momentum sum rule, one of the key QCD constraints, implies that
introduction of a fitted charm PDF
modifies the gluon and sea (anti)quark PDFs, particularly,
for $\bar u$ and $\bar d$. Hence, accurate predictions of the $c$ and
$\overline{c}$ parton distributions will be relevant to various important
LHC measurements, such as production of $W^{\pm}$, $Z^{0}$, and
Higgs boson, or associated production of a charm jet and a $Z^{0}$.

The remainder of this paper is organized as follows.
In Sec.~\ref{sec:QCDFactorization} we review
the theoretical foundations of the CTEQ global PDF analysis with
contributions of massive quarks.
In particular, we discuss issues related to the factorization of the
charm PDF in the proton, after clarifying the meaning of the PDFs for
the leading-power (perturbative) charm, power-suppressed
charm, and the fitted charm.
Several theoretical models of the intrinsic charm PDF
at the $Q_0$ scale will be presented in Sec.~\ref{sec:ModelsForFittedCharm}.
The results of our global fits, called the CT14IC PDFs, are discussed in
Sec.~\ref{sec:RESULTS}, where the quality of the data description is
documented, and a detailed comparison of the CT14IC PDFs with the CT14
PDFs and other PDF sets is provided.  The dependence of the CT14IC PDF
fits on the charm-quark mass is detailed in Sec.~\ref{sec:McDependence}.
In Sec.~\ref{sec:EMCdata}, we discuss the impact of including the EMC
data in the global fits for the fitted charm PDFs, as predicted by
those theoretical models introduced in Sec.~\ref{sec:ModelsForFittedCharm}.
We examine the impact of the CT14IC PDFs on the production of the
electroweak $W^\pm$, $Z$ and Higgs bosons at the LHC in
Sec.~\ref{sec:key-obs}, and on a charm jet production associated with
a $Z$ boson at the LHC in Sec.~\ref{sec:PREDICTIONS}. Finally, our
conclusions are presented in Sec.~\ref{sec:Conclusions}.

\section{QCD factorization with power-suppressed charm contributions \label{sec:QCDFactorization}}

Particle interactions with energies of hundreds of GeV,
at modern colliders such as the LHC or the Tevatron,
are not directly sensitive to the masses of most Standard Model (SM) fermions.
At such high energy, one may safely neglect the mass of any quark in a
short-distance scattering cross section, except for the top quark.
Protons, the initial-state nucleons at the LHC, behave as bound states composed
of strongly interacting constituents
lighter than the top, including light quarks
($u,$ $d,$ $s$), heavy quarks ($c$ and $b$), and gluons
$g$.\footnote{Without loss of generality, we focus on a situation when
  neither top quarks nor photons are classified as nucleon's partonic
  constituents. } A parton $a$ knocked out of an initial-state proton by a hard collision
moves essentially as a massless particle; however, the probability for knocking
the parton out, quantified by the parton distribution function
$f_{a/p}(\xi,\mu)$, or $a(\xi, \mu)$ for short, depends on the parton's flavor
and, ultimately, the parton's mass.

A charm quark with mass $m_{c}\sim1.3-1.6$ GeV is heavier than a
proton at rest, with mass 0.938 GeV. If we introduce a parton distribution
for the charm, what is the physical origin of this PDF?

The answer is not as clear-cut as for the lighter quarks, whose PDFs
are dominated by nonperturbative QCD contributions arising from energies
smaller than the proton mass. The light-quark PDFs are essentially
nonperturbative; we parametrize each light-quark PDF by a phenomenological
function $f_{a/p}(x,Q_{0})$ at an initial energy scale $Q_{0}$
of order 1 GeV and evolve the PDFs to higher energies using the DGLAP
equations~\cite{Gribov:1972ri,Gribov:1972rt,Lipatov:1974qm,Dokshitzer:1977sg,Altarelli:1977zs}.
For the charm and anticharm contributions, on the other hand,
the respective PDFs at such low $Q_{0}$ are not mandatory. Only some QCD
factorization schemes introduce them, with the goal to improve perturbative
convergence at scales $Q$ much larger than $Q_{0}$. The perturbative
component of the charm PDF dominates in conventional treatments, such
as those implemented in the general-purpose QCD analyses by CTEQ-TEA
and other groups. However, a nonperturbative component in the charm PDF cannot
be excluded either -- we will explore it in this paper. What are
the theoretical motivation and experimental constraints for the nonperturbative
component? Can it be relevant for the LHC applications?

We can systematically approach these questions by reviewing QCD
factorization, and the associated factorization theorem, for a perturbative
QCD calculation of a radiative contribution with heavy quarks. Let us
focus on predictions for neutral-current
DIS structure functions $F(x,Q)$ with 3 and 4 active flavors, at
a relatively low momentum transfer $Q$ that is comparable to the
mass $m_{c}$ of the charm quark. Our considerations can be extended
readily to situations with more than four active flavors, and
to higher $Q$ values. Moreover, among the experimental processes included in the
global QCD analysis, the neutral-current DIS is the most sensitive to charm
scattering dynamics
\cite{Gao:2013wwa,Harland-Lang:2015qea,Ball:2016neh,Bertone:2016ywq,Alekhin:2017kpj,Gizhko:2017fiu}
with the rest of the processes providing weaker constraints.
Therefore, it is natural to focus on DIS as the starting point.

\subsection{Exact and approximate factorization formulas \label{sec:ExactApproxFactorization}}

We first write down a \emph{phenomenological} form for the DIS structure
function that is implemented in the CTEQ-TEA PDF analysis:
\begin{eqnarray}
F(x,Q) & =&\sum_{a=0}^{N_{f}}\int_{x}^{1}\frac{d\xi}{\xi}\,{\cal
  C}_{a}^{(N_{ord})}\left(\frac{x}{\xi},\frac{Q}{\mu},\frac{m_{c}}{\mu};\alpha_s(\mu)\right)\,f_{a/p}^{(N_{ord})}(\xi,\mu)
\nonumber \\
& \equiv & \sum_{a=0}^{N_{f}} \left[{\cal
  C}_{a}^{(N_{ord})} \otimes f_{a/p}^{(N_{ord})}\right](x,Q).\label{Fpheno}
\end{eqnarray}
This is a standard convolution formula, consisting of the coefficient
function ${\cal C}_{a}^{(N_{ord})}(x/\xi,Q/\mu,m_{c}/\mu;\alpha_{s}(\mu))$
and the PDFs $f_{a/p}^{(N_{ord})}(\xi,\mu)$ dependent on the light-cone
partonic momentum fraction $\xi$ and factorization scale $\mu$ of
order $Q$ (set to coincide with the renormalization scale to simplify
the notation). The index $a$ denotes the initial-state parton's flavor, running
from $a=0$, corresponding to the gluon, to the number $N_{f}$ of
active quark flavors assumed in the QCD coupling
strength $\alpha_{s}(\mu)$ and the PDFs $f_{a/p}(\xi,\mu).$ Implicitly,
summation over quarks and antiquarks is assumed. We reserve the index
``$h"$ for a heavy-quark flavor, $h=c$ in DIS charm production.\footnote{
  Beyond the NNLO accuracy considered in this paper, DIS includes
  contributions with both $c$ and $b$ quarks. Treatment of such
  contributions in the ACOT formalism is explained in Refs.~\cite{Guzzi:2011ew, WangBowenThesis}.} The superscripts $(N_{ord})$ in both $C^{(N_{ord})}$ and $f_{a/p}^{(N_{ord})}$
emphasize that their perturbative coefficients are computed up to a fixed order
$N_{ord}$ of $\alpha_{s}$.

Let us highlight several aspects of this formula. First, $N_{f}$, the number
of active flavors, is not measurable, it is a theoretical
parameter of the renormalization and factorization schemes chosen
for the perturbative calculation. $N_{f}$ should be distinguished
from $N_{f}^{fs}$ \cite{Tung:2006tb,Guzzi:2011ew}, the number of
(anti-)quark species that can be physically produced in the final
state in DIS at given collision energy. The optimal value of
$N_{f}$ is \emph{chosen } as a part of the QCD factorization scheme
to optimize perturbative convergence. $N_{f}^{fs}$ can be determined
from an experimental observable, such as the final-state hadronic mass
in the neutral-current DIS process.

\begin{figure}[h]
\begin{centering}
\includegraphics[height=4in]{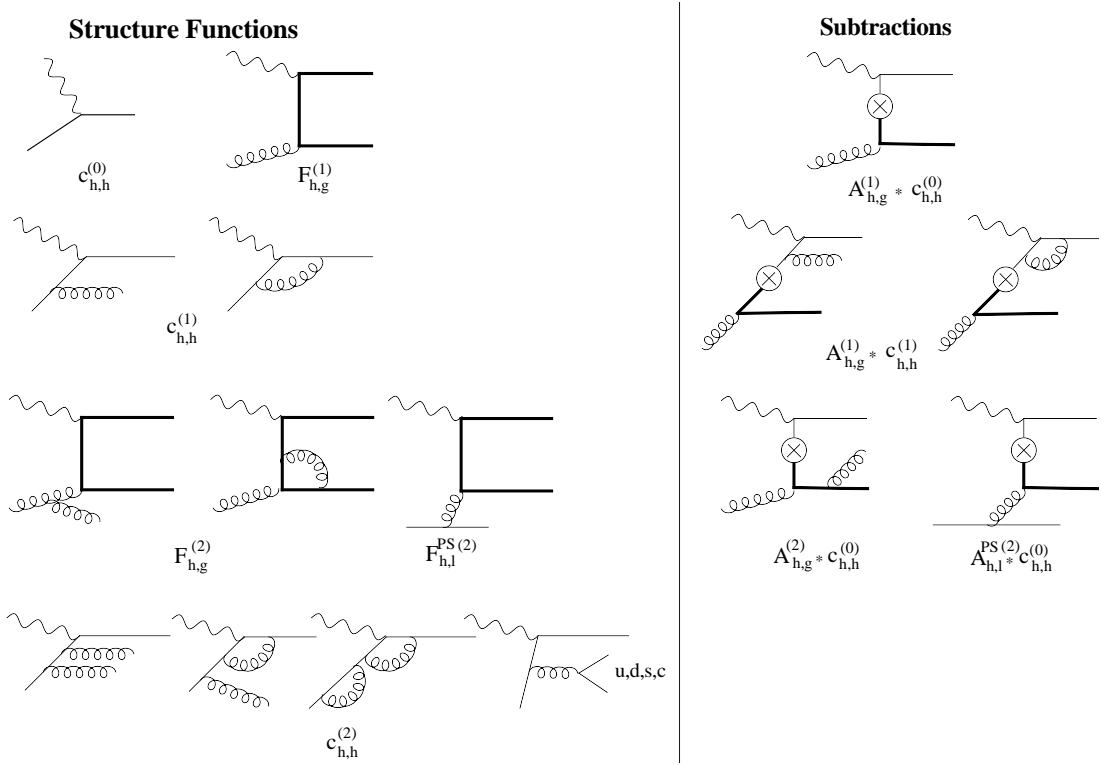}
\par\end{centering}

\caption{\label{fig:PertCharm} Leading-power (perturbative) radiative contributions
for neutral-current DIS charm production and scattering, included
up to ${\cal O}(\alpha_{s}^{2})$ in the S-ACOT-$\chi$ scheme. The
figure is reproduced from Ref.~\cite{Guzzi:2011ew}. }
\end{figure}

Second, the CTEQ-TEA group computes the
perturbative coefficients of ${\cal C}_{a}^{(N_{ord})}$ in the S-ACOT-$\chi$
scheme~\cite{Aivazis:1993pi,Collins:1998rz,Kramer:2000hn,Tung:2001mv},
a general-purpose factorization scheme for lepton-hadron and
hadron-hadron scattering processes. For neutral-current DIS,
${\cal C}_{a}^{(N_{ord})}$ were derived
in this scheme up to ${\cal O}(\alpha_{s}^{2})$,
or NNLO \cite{Guzzi:2011ew}. Figure~\ref{fig:PertCharm} is reproduced here
from Ref.~\cite{Guzzi:2011ew} and shows the Feynman diagrams and
notations for the perturbative coefficients of the ``charm
production'' structure function $F_c(x,Q)$
up to NNLO in the S-ACOT-$\chi$ approach.
Our discussion will turn to these diagrams for an illustration.
The remaining NNLO charm scattering
contributions in NC DIS,
arising in the light-quark structure function $F_l(x,Q)$
and not as important numerically, can also be
found in Ref.~\cite{Guzzi:2011ew}.

Third, in a general-purpose analysis such as CT14 NNLO, we start with
non-zero PDF parametrizations for the gluon and 3 light (anti-)quark
flavors at the initial scale slightly below the charm mass,
$Q_{0}=m_{c}-\epsilon.$ The input charm mass can be either the
$\overline{MS}$ mass $m_{c}(m_{c})$, or the pole mass
$m_{c}^{pole}$: the two are related by NNLO perturbative
relations~\cite{Chetyrkin:1997un,Chetyrkin:2000yt}, both are
implemented in CT14 PDFs.\footnote{
The past CTEQ-TEA analyses traditionally used $m_c^{pole}$ as an input, but $m_c(m_c)$ may be preferable in future precise calculations.
The pole mass cannot be used to arbitrarily high accuracy because
of nonperturbative infrared effects in QCD, related to
the fact that the full quark propagator has no pole
because of the quark confinement~\cite{Olive:2016xmw}.
}
As $f_{a/p}^{(N_{ord})}(\xi,Q)$
are evolved upward from the initial scale $Q_0$, they
are converted from $N_{f}=3$ to $4$, and from $4$ to $5,$ at the
corresponding switching points $Q_i$. The perturbative
coefficients of ${\cal C}_{a}^{(N_{ord})}$ are converted
concurrently to preserve the factorization
scheme invariance at each order of $\alpha_s$. The CT14 analysis
switches from $N_f$ to $N_{f+1}$ exactly at the heavy quark mass;
so for the charm quark the switching takes place at the energy scale $Q_c=m_c$.

In this conventional setup, we assume a zero charm PDF,
$f_{c/p}(\xi,Q_0)=0$, for $N_{f}=3$ at the initial scale $Q_0$
slightly below $Q_{c}=m_c$, and obtain a \emph{small}
non-zero $f_{c/p}(\xi,Q_c)$  for $N_{f}=4$ at scale $Q_c$
via perturbative matching.
Of course, $Q_c$ is arbitrary, we could equally choose a $Q_c$ value
below $Q_0$ and then expect a {\it non-zero} charm PDF also at $Q_0$.
This alternative suggests the possibility of including a non-zero
initial charm PDF parametrization, or the ``fitted charm'' parametrization, at the
initial scale $Q_0$ that would now correspond to $N_f=4$.  However, if the charm quarks
are produced exclusively from perturbative gluon splittings,
the dependence on the fitted $f_{c/p}(\xi,Q_0)$
cancels up to the higher $\alpha_{s}$ order {\it in the cross section},
not the PDF alone. It only
makes a difference, compared to the higher-order uncertainty,
if another mechanism adds up to perturbative charm-quark production.

To demonstrate this, compare the above \emph{approximate} fixed-order
formula (\ref{Fpheno}), which either includes the fitted charm
PDF, or not, to the all-order expression for $F(x,Q)$ with massive
quarks that follows from the QCD factorization theorem~\cite{Collins:1998rz,Collins:2011zzd}:
\begin{align}
F(x,Q) & =\sum_{a=0}^{N_{f}}\int_{x}^{1}\frac{d\xi}{\xi}\,{\cal C}_{a}\left(\frac{x}{\xi},\frac{Q}{\mu},\frac{m_{c}}{\mu};\alpha_s(\mu)\right)\,f_{a/p}(\xi,\mu)+\mathcal{O}(\Lambda^{2}/m_{c}^{2},\Lambda^{2}/Q^{2}).\label{F}
\end{align}
Eq.~(\ref{F}) underlies all modern computations for the inclusive
DIS observables, in the factorization schemes with fixed or varied
$N_{f}$ values. The convolution of ${\cal C}_{a}$ with $f_{a/p}(\xi,\mu)$
in Eq.~(\ref{F})
includes all ``leading-power'' radiative contributions that do not
vanish when the physical scales $\sqrt{s}$, $Q,$ $m_{c}$ are much
larger than the nonperturbative hadronic scale $\Lambda$ of order
less than 1 GeV. In Eq.~(\ref{Fpheno}), as implemented in the fits, this
leading-power ${\cal C}_{a}\otimes f_{a/p}$ is approximated just
up to order $N_{ord}.$

This means that, in the all-order factorization theorem (\ref{F}),
$[\mathcal{C}_{a}\otimes f_{a/p}](x,Q)$, the first term on the
right-hand side, captures all contributions associated with
the leading-power, perturbative, charm production.
On the other hand, when a non-zero initial condition for $f_{c/p}^{(N_{ord})}(\xi,Q_{0})$
is introduced in the fitted formula (\ref{Fpheno}),
it plays the role of a placeholder for several kinds of missing
contributions that appear in the full factorization formula
(\ref{F}), but not in the approximate formula (\ref{Fpheno}).
For example, it substitutes in part for the leading-power perturbative
contributions beyond the order $N_{ord}$.
The ${\cal O}(\alpha_{s}^{2})$, or NNLO, radiative contribution
to neutral-current DIS heavy-quark production is large numerically.
If a global fit is done at NLO, as in
Refs.~\cite{Pumplin:2007wg,Jimenez-Delgado:2014zga,Ball:2016neh},
it prefers an augmented
fitted charm $f_{c/p}^{(NLO)}(\xi,Q_{0})$ of a certain shape in part
to compensate for the missing NNLO DIS Wilson coefficients.

The fitted charm may also absorb part of the last,
power-suppressed, term on the right-hand
side of Eq.~(\ref{F}). The ``power counting'' analysis of Feynman integrals
shows that the ordinary power-suppressed contribution in
unpolarized inclusive DIS
is proportional to $(\Lambda/Q)^n$ with integer $n\geq2$ (``twist-4'',
see, e.g., \cite{Jaffe:1983hp,Jaffe:1982pm}).
In the DIS scattering of charm
quarks, the lowest power-suppressed contribution also includes terms
of order
$\Lambda^2/m_c^2$~\cite{Brodsky:1980pb,Brodsky:1981se,Collins:1998rz}. The
latter term clearly does not vanish with increasing $Q$ and,
furthermore, at very high $Q$ it is enhanced
logarithmically and behaves as
$(\Lambda^{2}/m_{c}^{2})\ln^{d}(Q^{2}/m_{c}^{2})$ with $d\geq 0$ due to contributions
from collinear scattering. The power-suppressed charm contribution, once
introduced at low scale $Q \sim m_c$, will survive to the much
higher scales relevant to the LHC.

\begin{figure}[tb]
\begin{centering}
\includegraphics[height=4in]{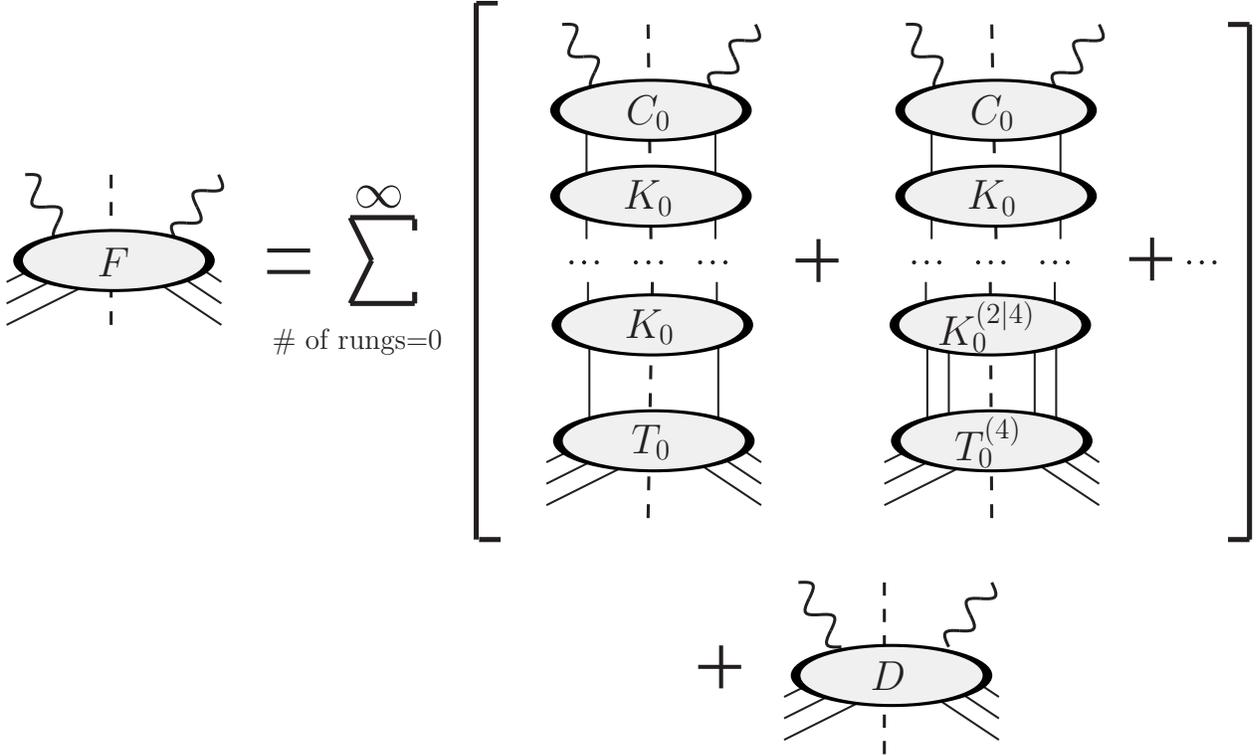}
\par\end{centering}

\caption{\label{fig:ICKT} Dominant squared leading-power amplitudes
  in DIS charm
production in the $Q^{2}\gg m_{c}^{2}\gg\Lambda^{2}$ limit. Here
$F$ is the DIS structure function, $C_{0}$, $K_{0}$ and $D$ are
two-particle irreducible (2PI) subgraphs,
$T_{0}$ and $T_{0}^{(4)}$ are the twist-$2$
and $4$ target hadron subgraphs, and $K_{0}^{(2|4)}$ is the heavy-quark
``mixed-twist'' 2PI subgraph.}
\end{figure}

\subsection{Charm contributions in 3-flavor and 4-flavor schemes \label{sec:Charm3and4Flavors}}

While the complete analysis of the twist-4 contribution
is far too extensive, we present a heuristic
explanation of its logarithmic growth by following an analogy with
the leading-power, or twist-2, terms \cite{Guzzi:2011ew}.
It is useful
to compare the relevant Feynman graphs in the $N_f=3$ factorization
scheme, the most appropriate scheme to use in the threshold kinematical region,
where $Q$ is comparable to $m_c$, and in the $N_f=4$ scheme, which
is most appropriate at $Q^2 \gg m_c^2$, where
the charm density has the most physical interpretation.

First, recall that in the $N_f=3$ scheme all subgraphs
containing heavy-quark propagators are assigned to the Wilson
coefficients ${\cal C}_{a}$ and
not to the PDFs $f_{a/p}$. Among the leading-power hard-scattering amplitudes
in Fig.~\ref{fig:PertCharm}, the only contributions arising in the
3-flavor scheme are those attached to the external gluons and light quarks,
denoted by $F_{h,a}^{(k)}$. The explanation for this is that the
$N_f=3$ scheme applies zero-momentum subtraction
to UV singularities with heavy-quark propagators and strongly suppresses
highly off-shell charm quark propagators as a consequence
of their manifest decoupling.
Therefore, the non-negligible Feynman integrals
in this scheme contain the charm propagators only in the
hard-scattering subgraphs, where the virtualities of all particle momenta
are comparable to $Q^{2}$ and $m_{c}^{2}$. The nonperturbative subgraphs
with virtualities much less than $Q^2$ contain only light-parton
propagators, as those are renormalized in the $\overline{MS}$ scheme.

A twist-4 $N_f=3$ hard-scattering matrix element for $F_{h,a}^{(k)}$
can be thought of as a twist-2 $N_f=3$ hard-scattering matrix element
connected to the parent hadron by an additional light-parton propagator
at any point in the hard subgraph.
Both  twist-2 and twist-4 terms with charm
take the factorized form illustrated in Fig.~\ref{fig:ICKT}, while Fig.~\ref{fig:ICdiagrams} shows representative twist-4 {\it squared}
matrix elements obtained after attaching the second initial-state gluon to some
of the twist-2 matrix elements in Fig.~\ref{fig:PertCharm}.
In the hadronic cross section,
every twist-4 hard scattering cross section shown in Fig.~\ref{fig:ICdiagrams}
is multiplied by a twist-4
(double-parton) nonperturbative function, such as $f_{gg/p}(\xi_1,\xi_2, \mu)$.
Insertion of two QCD vertices suppresses the twist-4 cross section
by a power of $\alpha_{s}$
compared to the counterpart twist-2 cross section, while the insertion of
two propagators and multiplication by a twist-4 function further suppresses
it by a power of $\Lambda^{2}/p^{2}$ with $p^{2}$ of order $Q^{2}\sim m_{c}^{2}$.

At twist-4, we encounter several new nonperturbative functions that
are not constrained by the data and obey their own evolution equations
at the scale $Q$ \cite{Braun:2009vc,Ji:2014eta}. The complete analysis of twist-4 is lengthy -- we will
refer to the vast literature on the subject, including Refs.~\cite{Gross:1971wn,Anikin:1978tj,Jaffe:1983hp,Jaffe:1982pm,Ellis:1982cd,Ellis:1982wd,Balitsky:1989jb,Qiu:1990xxa,Qiu:1990xy,Jaffe:1991ra,Jaffe:1996zw,Mueller:1998fv,Blumlein:1999sc,Geyer:1999uq,Geyer:2000ma}.

We further note that, in
the limit $Q^{2}\gg m_{c}^{2}\gg\Lambda^{2},$
the twist-4 charm scattering cross sections contain
ladder subgraphs of essentially twist-2 topology.
They can be seen in Fig.~\ref{fig:ICKT},
illustrating a decomposition of the structure function $F$ containing
the ladder contributions. $D$ denotes a two-particle irreducible (2PI)
part (in the vertical channel) of the
structure function $F$. The first graph on the right-hand side is
a generic twist-2 ladder contribution recognized from the
calculation of NLO splitting functions in the massless case by Curci, Furmanski,
and Petronzio \cite{Curci:1980uw}. It is composed of 2PI subgraphs
$C_{0},$ $K_{0},$ and $T_{0}$ (without an upper index),
where $C_{0}$ and $T_{0}$ are coupled to the virtual photon and target
hadron, respectively.
The decomposition in terms of $D$ and $C*K_0...*T_0$ for twist-2 also appears
in the Collins' proof of QCD factorization for DIS
with massive quarks~\cite{Collins:1998rz}.

The ladder graphs are different in the $N_f=3$
and $N_f=4$ schemes. The $N_f=4$ scheme introduces additional terms
with heavy quarks that approximate the leading contribution in the
$Q^2 \gg m_c^2$ limit. In Fig.~\ref{fig:PertCharm}, these ladders
correspond to the contributions proportional to the ``flavor-excitation'' Wilson
coefficient functions $c_{h,h}^{(k)}$. Such terms are absent in the $N_{f}=3$ scheme,
and their purpose is to resum collinear logs $\ln^{d}(Q^{2}/m_{c}^{2})$
from higher orders with the help of DGLAP
equations. In this case both the light- and
heavy-parton subgraphs are renormalized in
the $\overline{MS}$ scheme. Importantly, apart from a finite renormalization of
$\alpha_s$, the perturbative expansions
of the structure functions in the $N_{f}=3$
and $N_{f}=4$ schemes are equal up to the first
unknown order in $\alpha_{s}$ -- the condition that we expect to
hold both for the twist-2 and twist-4 heavy-quark contributions.

Next to the twist-2 term in Fig.~\ref{fig:ICKT}
we show a ladder attached to a twist-4 target subgraph $T_{0}^{(4)}$
[with an upper index ``(4)''], connected to the twist-2 kernels $K_0$
in the upper part via a ``mixed-twist'' kernel $K_{0}^{(2|4)}$ containing
a real heavy-quark emission. As $K_{0}^{(2|4)}$ is connected to $T_0$ by four propagators, at $Q^{2}\gg m_{c}^{2} \gg \Lambda^2$
it scales as $\Lambda^2/p^2$. Since it includes
loop integrals with  {\it massive} quark propagators $1/(k\!\!\!/ - m_c)$,
the momentum scale $p$ can be either $Q$ or $m_c$; but the
$\Lambda^2/m_c^2$ term is less suppressed than $\Lambda^2/Q^2$.
[It is crucial that two large QCD scales, $m_c$ and $Q$, are present, in contrast to the massless-quark case.]
On the other hand, apart
from the replacement of $T_0\cdot K_0$ by $T^{(4)}_0\cdot K^{(2|4)}_0$,
the second ladder has the structure of the first one.

\subsubsection{Factorization for twist-2 contributions \label{sec:FactorizationTwist2}}

We assume that the Feynman diagrams in Fig.~\ref{fig:ICKT} are \emph{unrenormalized}
and indicate this by a subscript ``0''. Ref.~\cite{Collins:1998rz}
shows how to recast the full sum of
twist-2 diagrams into a factorized convolution
\begin{equation}
F(x,Q)=\sum_{a}[{\cal C}_{a}\otimes f_{a/p}](x,Q)+r\label{Ftwist2}
\end{equation}
by recursively applying
a factorization operator $Z$ and renormalizing the UV singularities.
$Z$ is a projection operator that is inserted recursively between
the rungs of the ladder diagram, e.g., at the location indicated by
the circle markers. The action of the $Z$ operator is to replace the exact
ladder graph by a simpler, factorized expression which provides a
good approximation to the full graph in the $Q^{2}\gg m_{c}^{2}$
limit, and which is valid up to a power-suppressed remainder $r$. In
particular, $Z$ replaces the off-shell intermediate parton propagator
at the insertion point by an on-shell external state with zero
transverse momentum in the Breit
frame. By considering recursive
insertions of the $Z$ operators to all orders, one demonstrates
factorization for $F(x,Q)$ in either the $N_f=3$ scheme or the $N_{f}=4$
scheme of the Aivasis-Collins-Olness-Tung (ACOT) class \cite{Aivazis:1993pi}.
By its construction, the remainder $r$ is of order
\begin{equation}
 \left(\frac{\textrm{highest virtuality in}~ T_0}{\textrm{lowest virtuality in}~ C_0}\right)^2=\left(\Lambda/p\right)^2,
\end{equation}
with $p=Q$ or $m_c$.

While the $Z$ operation in the $\overline{MS}$
scheme is uniquely defined for intermediate light states, for a heavy
quark, it encounters an additional ambiguity. The projection operator
acting upon an intermediate heavy quark, denoted by $Z_{h}$, may include
additional powers of $(m_{c}^{2}/Q^{2})$ that vary among the conventions
\cite{Collins:1998rz,Guzzi:2011ew}. The ambiguity in $Z_{h}$ gives
rise to several versions of the ACOT-like schemes, all equivalent
up to a higher order in $\alpha_{s}$. The form of $Z_{h}$ may
be even made dependent on the type and $\alpha_{s}$ order of the
scattering contribution: some choices for $Z_{h}$, such as the one
made in the SACOT-$\chi$ scheme \cite{Kramer:2000hn,Tung:2001mv,Guzzi:2011ew},
simplify perturbative coefficients and enable fast perturbative convergence.

In a practical calculation of a twist-2 cross section illustrated
by Fig.~\ref{fig:PertCharm}, the $Z$ operation defines the prescription
for constructing the perturbative coefficients $\mathcal{C}_{i,b}^{(k)}$
of Wilson coefficient functions from the structure functions $F_{i,b}^{(k)}$
computed in DIS $e+b\rightarrow e+X$ on a partonic target
$b$. Here $i$ denotes an (anti)quark struck by the virtual photon.\footnote{Up to NNLO, we use a simplified decomposition of the neutral-current
DIS structure function over the quark flavors probed by the virtual
photons: $F(e+b\rightarrow e+X)\equiv\sum_{i=1}^{N_{f}^{fs}}e_{i}^{2}F_{i,b}$,
where $e_{i}$ is the (anti)quark's electric charge \cite{Guzzi:2011ew}. The $SU(N_f)$ decomposition of the ACOT structure functions for higher orders was derived in Ref.~\cite{WangBowenThesis}.}
The parton-scattering structure functions, coefficient functions,
and PDFs are expanded as a series in $a_{s}\equiv\alpha_{s}(\mu,N_{f})/(4\pi)$:
\begin{eqnarray}
F_{i,b} & = & F_{i,b}^{(0)}+a_{s}\,F_{i,b}^{(1)}+a_{s}^{2}\,F_{i,b}^{(2)}+\dots,\nonumber \\
{\cal C}_{i,a} & = & {\cal C}_{i,a}^{(0)}+a_{s}\,{\cal C}_{i,a}^{(1)}+a_{s}^{2}{\cal C}_{i,a}^{(2)}+\dots,\nonumber \\
f_{a/b}(x) & = & \delta_{ab}\delta(1-x)+a_{s}\,A_{a,b}^{(1)}+a_{s}^{2}A_{a,b}^{(2)}+\dots,\label{Fcf}
\end{eqnarray}
where $A_{a,b}^{(k)}$ $(k=0,1,2,\dots)$ are perturbative coefficients
\cite{Buza:1995ie} of operator matrix elements for finding a parton $a$ in a parton
$b.$ A perturbative coefficient ${\cal C}_{i,a}^{(k)}$ of the Wilson
coefficient function at $a_{s}$ order $k$ can be found by comparing
the perturbative coefficients on the left and right sides of
\begin{equation}
F_{i,b}=\sum_{a}{\cal C}_{i,a}\otimes f_{a/b}.\label{FibCf}
\end{equation}
The comparison does not specify the form of the perturbative coefficients
$c_{i,h}^{(k)}$ with an initial-state heavy quark; those are specified
by $Z_{h}$ at each $a_{s}$ order $k$ and re-used in
exactly the same form in all occurrences of $c_{i,h}^{(k)}$ in the
contributions of orders $k+1$ and higher.
The freedom in selecting $Z_h$ affects $c_{i,h}^{(k)}$ and not the
partonic PDF coefficients $A^{(k)}_{a,b}$ that remain defined in the
$\overline{MS}$ scheme.
With such self-consistent definition,
the dependence on $Z_{h}$ cancels up to the first unknown order in
$a_{s}$, as it was verified numerically up to ${\cal O}(\alpha_{s}^{2})$
in Ref.~\cite{Guzzi:2011ew}.

\subsubsection{Factorization of twist-4 contributions: a
  sketch \label{sec:FactorizationTwist4}}

Going back to Fig.~\ref{fig:ICKT}, we recall that, while in the twist-2 factorization formula (\ref{Ftwist2}) the $K_{0}^{(2|4)}\cdot T_{0}^{(4)}$ subgraph is counted as a part of the remainder $r\sim \Lambda^2/m_c^2$, diagrammatically, it
 is attached to the
upper ladder subgraphs in exactly the same way as the twist-2
$K_{0}\cdot T_{0}$. We can treat
the sum $K_{0}\cdot T_{0}+K_{0}^{(2|4)}\cdot T_{0}^{(4)}$ as a modified target
contribution of twist-2, which now includes some power-suppressed correction.
The derivation of the factorization for $Q^{2}\gg m_{c}^{2}$ can be repeated
for $N_{f}=4$ as in the previous subsection. The factorized cross section
reproduces the structure function up to the terms of order $\Lambda^{2}/Q^{2}$
or $(\Lambda^{4}/m_{c}^{4})$.

At the level of individual contributions, the $K_{0}^{(2|4)}\cdot
T_{0}^{(4)}$ target subgraph introduces a
non-zero term in the charm PDF at the switching scale from 3 to 4
flavors. We can continue to use the DGLAP equations and
the same coefficient functions as in the pure twist-2 case, and the latter
are again dependent on the definition of operator $Z_{h}$ (the heavy-quark
mass scheme). In particular, if the flavor-excitation coefficient
function $c_{h,h}^{(k)}$ is modified by a term of order $(m_{c}^{2}/Q^{2}),$
the twist-4 component of the structure function $[c_{h,h}^{(k)}\otimes
  f_{c}]_{\mbox{twist-4}}$ is modified by a term of order
$(m_{c}^{2}/Q^{2})\cdot(\Lambda^{2}/m_{c}^{2})=\Lambda^{2}/Q^{2}$.
The net change does not exceed the total error $\Lambda^{2}/Q^{2}$
of the factorized approximation.

This implies that the twist-4 component of the charm PDF is compatible
with any available version of the ACOT scheme, the
differences between the structure functions in these schemes are of
order $\Lambda^{2}/Q^{2}$ for twist-4 and even weaker for higher
twists. Furthermore, by the structure of the ACOT schemes, the scheme
differences cancel order-by-order in $\alpha_s$. Therefore, the claim in
Refs.~\cite{Ball:2015tna,Ball:2015dpa,Ball:2016neh} that the nonperturbative
charm is only consistent with the ``full'' version of the
ACOT scheme or its analog schemes, such as the
fully massive FONLL scheme, is not correct. In our analysis, it
suffices to use the S-ACOT-$\chi$ scheme, with or without the
power-suppressed component. Since open charm is produced in $c\bar c$
pairs in neutral-current DIS, and not as lone $c$ (anti)quarks, the
$\chi$ rescaling in the S-ACOT-$\chi$ scheme~\cite{Tung:2001mv},
requiring production in pairs only, approximates
energy-momentum conservation better than its full ACOT counterpart that also
tolerates production of single $c$ or $\bar c$ quarks.

\begin{figure}[tb]
\begin{centering}
\includegraphics[width=1\textwidth]{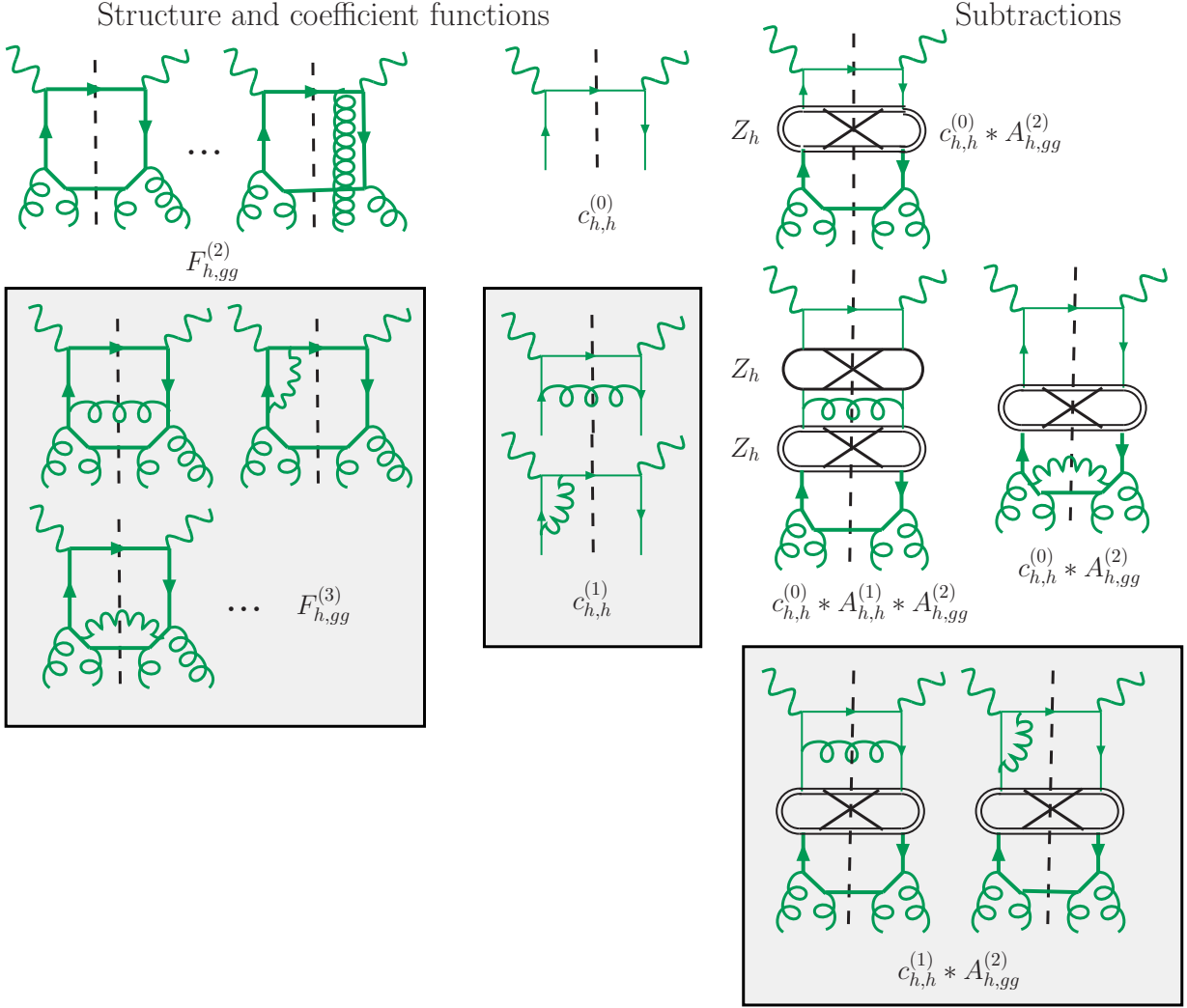}
\par\end{centering}

\caption{\label{fig:ICdiagrams} Examples of subleading-power contributions
to charm production originating from double-gluon initial states.}
\end{figure}

Let us illustrate the calculation
of the simplest twist-4 charm contributions on an example of select
twist-4 squared amplitudes in Fig.~\ref{fig:ICdiagrams}.
Again, we follow a close analogy
to the twist-2 S-ACOT-$\chi$ calculation in Sec.~\ref{sec:FactorizationTwist2},
see also \cite{Aivazis:1993pi} and \cite{Guzzi:2011ew}.

The first line in Fig.~\ref{fig:ICdiagrams} shows the lowest-order
twist-4 contributions of order $\mathcal{O}(\alpha_{s}^{2}),$ the
remaining lines show some radiative contributions of order $\mathcal{O}(\alpha_{s}^{3}).$
As before, a superscript in the parentheses indicates the order $k$ of
the perturbative coefficient.

In either the 3- or 4-flavor
scheme, we start by computing ``flavor production'' structure functions
$F_{h,ab}^{(k)}$, such as $F_{h,gg}^{(2)}$ or $F_{h,gg}^{(3)}$ shown
in Fig.~\ref{fig:ICdiagrams}, with $b$ standing for a $gg$ or another
double-parton initial state. Many more diagrams besides the ones
shown arise at each order
depending on the locations of the extra gluon attachments in the
hard subgraph. The coefficient functions associated
with twist-4 are derived by matching the perturbative coefficients
order-by-order as in Eqs.~(\ref{Fcf}) and (\ref{FibCf}).

For instance, at order $\alpha_s^2$, the double-convolution
integral $F_{h,gg}^{(2)}\otimes\otimes f_{gg/p}$ over the gluon-pair
light-cone momentum fractions $\xi_{1}$ and $\xi_{2}$ scales as
$\alpha_{s}^{2}(Q)\Lambda^{2}/p^{2},$ where $p^2$ is at least
as large as $Q^2$ or $m_{c}^2$. In the limit $\Lambda^{2}\ll m_{c}^{2}\ll Q^{2}$,
the ${\cal O}(\alpha_{s}^{2})$ contribution with a smaller hard scale
$p^2=m_{c}^2$ still survives. A part of it is resummed in the flavor-excitation
term $c_{h,h}^{(0)}\otimes f_{c/p}$, added across all
$Q \geq Q_c$ in order to obtain a smooth prediction for
$F(x,Q)$.\footnote{The discontinuity of $F(x,Q)$ at the switching point
  $Q_c$ is very mild at NNLO and reduced with including higher $\alpha_s$
  orders. Smoothness of $F(x,Q)$ is desirable for the convergence of
  PDF fits.}

The twist-4 $\mathcal{O}(\alpha_{s}^{2})$ remainder of $F_{h,gg}^{(2)}\otimes\otimes f_{gg/p}$
that is not absorbed in $c_{h,h}^{(0)}\otimes f_{c/p}$ may be of the same order
as $c_{h,h}^{(0)}\otimes f_{c/p}$ at relatively low $Q$. The remainder
is given by $\mathcal{C}_{h,gg}^{(2)}\otimes \otimes f_{gg/p},$ where $\mathcal{C}_{h,gg}^{(2)}$
is found from the comparison of the ${\cal {\cal O}}(\alpha_{s}^{2})$
coefficients in Eq.~(\ref{FibCf}):
\begin{equation}
\mathcal{C}_{h,gg}^{(2)}=F_{h,gg}^{(2)}-c_{h,h}^{(0)}\otimes A_{h,gg}^{(2)}.\label{Chgg2-2}
\end{equation}
The Feynman diagram for the ``subtraction term'' $c_{h,h}^{(0)}\otimes A_{h,gg}^{(2)}$
is shown in the first row of Fig.~\ref{fig:ICdiagrams}. It is obtained
by inserting $Z_{h}$ into the Feynman graph for $F_{h,gg}^{(2)}$
in order to constrain the momentum of the cut charm
propagator to be collinear to that of the target hadron, and
to replace the part of the graph for $F_{h,gg}^{(2)}$ above the insertion by
a simpler subgraph given by $c_{h,h}^{(0)}$. Clearly, the remainder is process-dependent.

The next-order contribution $F_{h,gg}^{(3)}$ with an added gluon line
develops a logarithmic enhancement at $m_{c}^{2}\ll Q^{2}$,
\begin{equation}
F_{h,gg}^{(3)}\otimes\otimes f_{gg/p}\sim\alpha_{s}^{3}(Q)\left(\Lambda^{2}/m_{c}^{2}\right)\ln(Q^{2}/m_{c}^{2}),
\end{equation}
which is resummed as a part of $c_{h,h}^{(0)}\otimes f_{c/p}$ and
$c_{h,h}^{(1)}\otimes f_{c/p}$. Again, the $\mathcal{O}(\alpha_{s}^{3})$
remainder that is not resummed must still be included in the full
result, it takes the form of $\mathcal{C}_{h,gg}^{(3)}\otimes f_{gg/p},$
where
\begin{equation}
{\cal C}_{h,gg}^{(3)}=\widehat{F}_{h,gg}^{(3)}-c_{h,h}^{(0)}\otimes\left[A_{h,h}^{(1)}\otimes A_{h,gg}^{(2)}+A_{h,gg}^{(3)}\right]-c_{\,h,h}^{(1)}\otimes A_{\,h,gg}^{(2)}.\label{Chgg3}
\end{equation}
$\widehat{F}_{h,gg}^{(3)}$ stands for the infrared-safe part (with respect to light partons)
 of $F_{h,gg}^{(3)}$ in the $\overline{MS}$ scheme \cite{Guzzi:2011ew}.

The rest of the coefficient functions can be computed along the same lines.

\section{Models for the fitted charm\label{sec:ModelsForFittedCharm}}
\subsection{Overview \label{sec:Overview}}

To recap the previous sections, a non-zero initial condition at $Q_c$ for the
``intrinsic charm PDF'', interpreted in the sense of the
``fitted charm'', may be used to test for the power-suppressed
charm scattering contribution of order ${\cal O}(\alpha_s^2)$,
of the kind shown in Fig.~\ref{fig:ICdiagrams}.  To be
sensitive to these contributions, the twist-2 cross sections must be
evaluated at least to NNLO to reduce contamination
by the higher-order twist-2 terms.
The complete set of power-suppressed massive contributions
can be organized according to the method of the ACOT scheme. It is comprised
of numerous matrix elements $F_{h,b}^{(k)},$
$A_{h,b}^{(k)}$ for double-parton initial-states $b$,
as well as of twist-4 nonperturbative
functions such as $f_{gg/p}(\xi_{1},\xi_{2},\mu)$.

Various model estimates suggest
a power-suppressed charm cross section of a modest size:
of order of a fraction of the $\alpha_{s}^{2}$ component in DIS charm
production, carrying less than about a percent of the proton's momentum.
To estimate sensitivity of the QCD data before resorting to the full
twist-4 calculation, we utilize an update of the phenomenological method
of the CTEQ6.6 IC NLO and CT10 IC NNLO
analyses \cite{Pumplin:2007wg,Dulat:2013hea}.
In contrast to the previous analyses, we examine a more extensive
list of nonperturbative models, fit the most complete set of DIS data
from HERA as well as the data from the LHC and (optionally) the EMC,
and utilize a PDF parametrization that results in a more
physical behavior.

Four models for the charm-quark PDF $c(x,Q_{0})\equiv \widehat{c}(x)$
at the initial scale $Q_0$ will be considered. [$Q_0$ is
  set to be less than $Q_c=m_c^{pole}$ in all cases.]
Besides the conventional
CT14 model that sets $\widehat{c}(x)=0$,
the other three models allow for $\widehat{c}(x)$ of an arbitrary
magnitude. In all models, the
charm PDF is convoluted with the S-ACOT-$\chi$ coefficient functions $c_{h,h}^{(k)},$
with $k\leq2$. It remains constant below the switching scale $Q_{c}$
and is combined with the perturbative charm component at $Q_{c}$ and evolved
to $Q>Q_{c}$ by the 4- and 5-flavor DGLAP equations.

Neither the present fit, nor the contemporary fits by the other groups
include the twist-4 remainders of DIS cross sections discussed in Sec.~\ref{sec:FactorizationTwist4}:
$\mathcal{C}_{h,gg}^{(2)}\otimes f_{gg/p},$
$\mathcal{C}_{h,gg}^{(3)}\otimes f_{gg/p},$ etc. The remainders
are process-dependent and comparable to the $c_{h,h}^{(k)}\otimes f_{c/p}$
convolutions at energies close to $m_c$.
Without including these process-dependent terms explicitly, the
fitted charm PDF found in a fit to DIS is not a truly universal nonperturbative
function; it
absorbs the above process-dependent
remainders. Furthermore, in DIS at very low $Q$ or $W$, separation of the
$\Lambda^2/Q^2$ and $\Lambda^2/m_c^2$ terms presents an additional
challenge. The experimental data in the CT14(HERA2) fits is selected
with the cuts $Q^2 > 4\mbox{ GeV}^2$, $W^2 > 12.5\mbox{ GeV}^2$
so as to minimize sensitivity to the $\Lambda^2/Q^2$
terms. This is usually sufficient to minimize the
$\Lambda^2/Q^2$ contributions below the PDF uncertainty from other sources.
We examine the possibility of the impact of the $\Lambda^2/Q^2$ terms
on the best-fit $c(x,Q_0)$ in Sec.~\ref{sec:DataImpact}.

\subsection{Valence-like and sea-like parametrizations \label{sec:ValenceSea}}
Given that several mechanisms may give rise to
the fitted charm, we will parametrize it by two generic shapes,
a ``\emph{valence-like}'' and a ``\emph{'sea-like}'' shape.
The two shapes arise in a variety of dynamical models.

A valence-like shape has a local maximum at $x$ above
0.1 and satisfies $f_{q/p}(x,Q_{c})\sim x^{-a_1}$ with $a_1\lesssim 1/2$
for $x\rightarrow0$ and $f_{q/p}(x,Q_{c})\sim (1-x)^{a_2}$ with $a_2 \gtrsim 3$
for $x\rightarrow 1$. The distributions for valence $u$ and $d$ quarks
fall into this broad category, as well as the ``intrinsic'' sea-quark
distributions that can be naturally generated in several ways
\cite{Pumplin:2005yf}, e.g.,
for all flavors, nonperturbatively from a $|uudQ\overline{Q}\rangle$
Fock state in light-cone
\cite{Brodsky:1980pb,Brodsky:1981se,Chang:2011vx,Blumlein:2015qcn,Brodsky:2015fna}
and meson-baryon models
\cite{Navarra:1995rq,Paiva:1996dd,Steffens:1999hx,Hobbs:2013bia}; for $\bar{u}$ and $\bar{d}$, from connected diagrams in lattice
QCD \cite{Liu:2012ch}.\footnote{In contrast to the light flavors,
in lattice QCD a charm PDF arises exclusively from
disconnected diagrams \cite{KFLiuPrivate}. This suggests
that $c$ and $\bar c$ contributions in DIS are connected
to the hadron target by gluon insertions, in accord
with the physical picture of the
QCD factorization in Sec.~\ref{sec:ExactApproxFactorization}.}

A sea-like component is usually monotonic in $x$
and satisfies $f_{q/p}(x,Q_{c})\sim x^{-a_1}$ for $x\rightarrow0$
and $f_{q/p}(x,Q_{c})\sim (1-x)^{a_2}$ for $x\rightarrow 1$,
with $a_1$ slightly above
1, and $a_2 \gtrsim 5$. This behavior is typical for the
leading-power, or ``extrinsic'' production.
For example, an (anti)quark PDF with this
behavior originates from $g\rightarrow q\bar{q}$ splittings in perturbative
QCD, or from disconnected diagrams in lattice QCD (see Ref.~\cite{Liu:2012ch}
for details). Even a missing next-to-next-to-next-to-leading order (N3LO)
leading-power correction may
produce a sea-like contribution at $x\ll 0.1$, where the valence-like
components are suppressed.

One may wonder why the charm quark PDF cannot be fitted to a more general
parametrization, in the same manner as the light-quark PDFs. We find
that the primary problem is that there are not enough precision data
available to provide meaningful constraints on the power-suppressed
IC content in the $\{x,Q\}$ regions where it can be important
(see the discussion of the EMC charm data in Sec.~\ref{sec:EMCdata}).
There is also a danger that the charm quark distribution, being relatively
unconstrained, may behave unphysically, for example, when the fit
allows a valence-like $c(x,Q_{0})$ to be almost the same in size
as $\bar{u}(x,Q_{0})$ or $\bar{d}(x,Q_{0})$ at $Q_0 \sim m_c$ and $x\rightarrow1$,
where the experimental constraints are weak. We must also demand conceivable cross sections to be non-negative, even though the PDFs themselves
can generally have a negative sign.
Adopting a too flexible fitted charm PDF parametrization may mask unrelated higher-order radiative contributions
to the data, hence lead to misinterpreted fits.
Thus, we restrict the
freedom of the charm quark somewhat by constraining it to be non-negative
and have either a valence-like or sea-like form,
with only one free multiplicative parameter.
The positivity of the BHPS form enables positive charm-scattering cross sections at large $x$, while a negative-valued SEA form is not statistically distinguishable in the fit from a positive SEA form at a larger $m_c$ value. [The dependence of SEA fits on $m_c$ is reviewed in the next Section.]
We have verified that a mixed charm parametrization that interpolates
between the valence-like and sea-like parametrizations only slightly
increases the range of the allowed charm momentum fraction, without
impacting the main outcomes.

\subsection{The charm distribution models in detail \label{sec:CharmModels}}

\begin{figure}[tb]
\begin{center}
\includegraphics[width=0.48\textwidth]{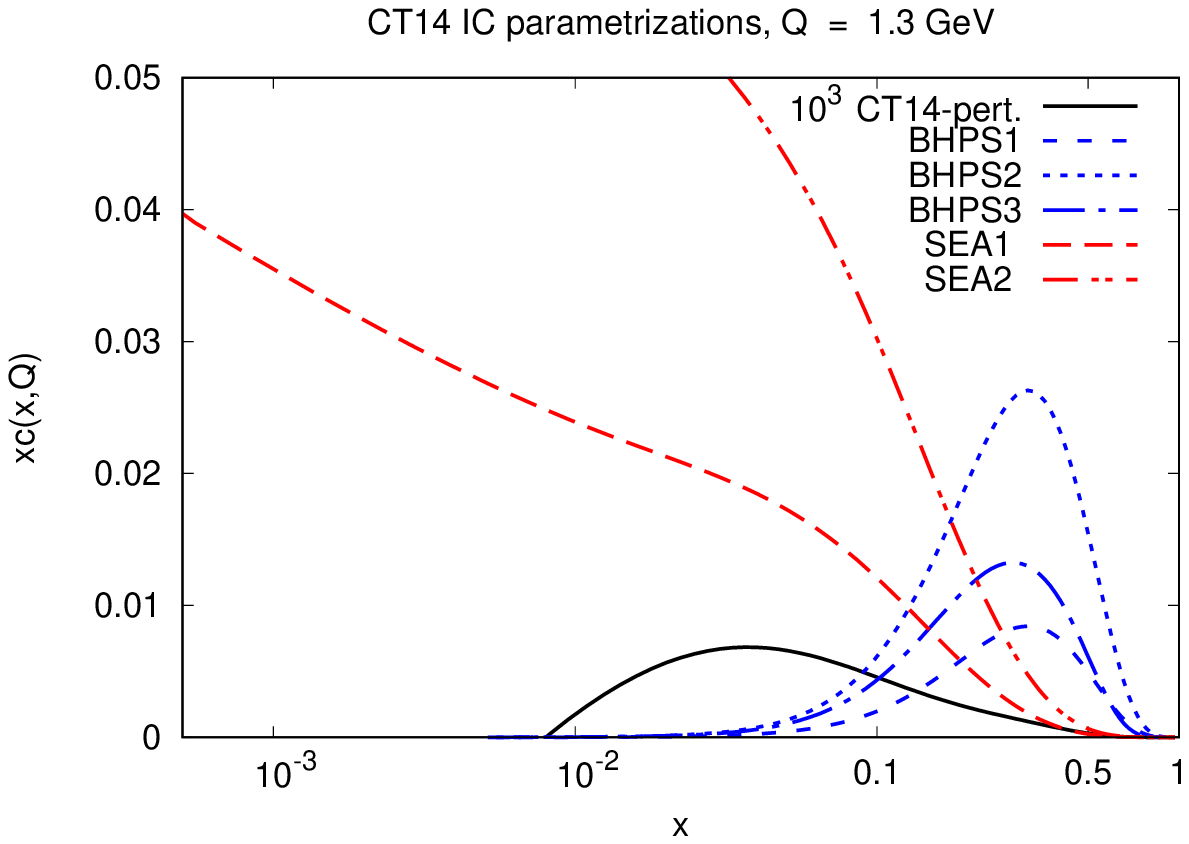}
\includegraphics[width=0.48\textwidth]{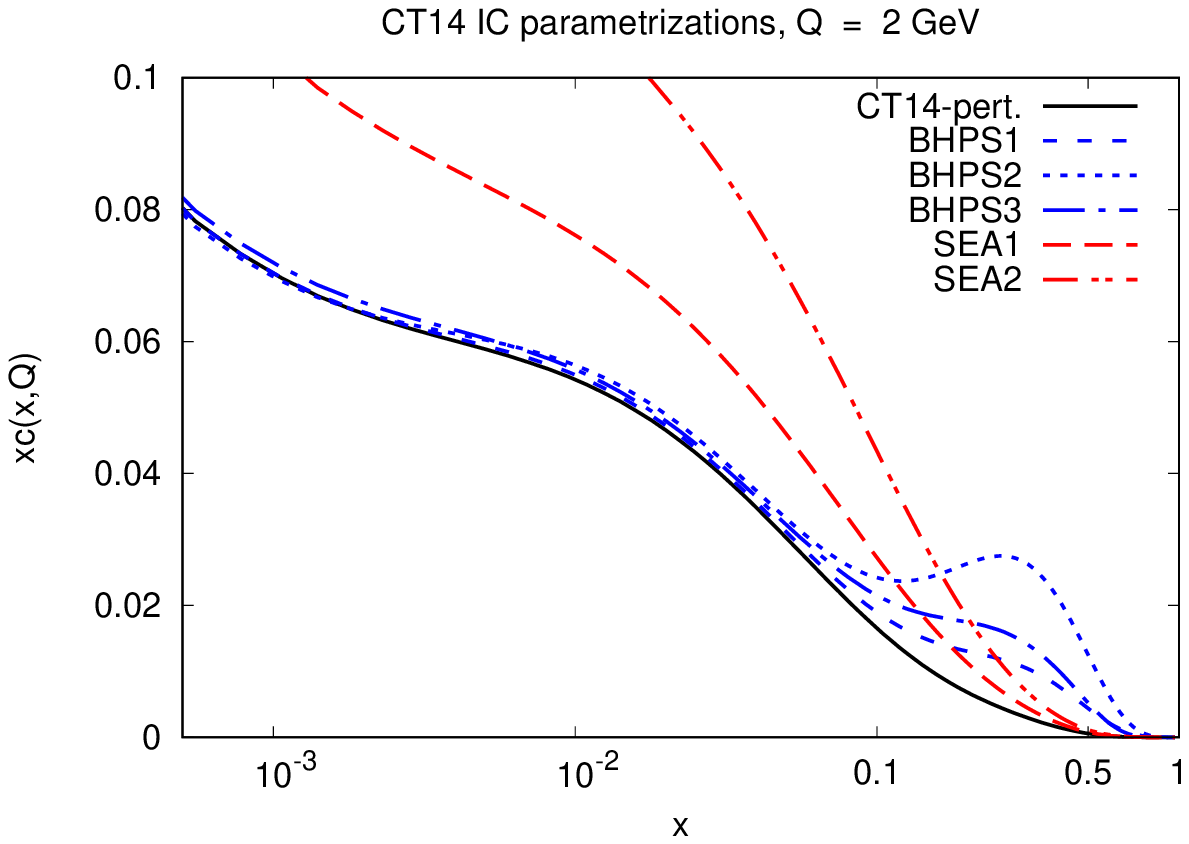}
\end{center}
\caption{$x c(x,Q)$ distributions for various models, evaluated
at $Q=1.3$ GeV and $Q=2$ GeV, respectively.
\label{fig:fourmodels}}
\end{figure}

We will now review these four models, whose $x c(x,Q)$ distributions
at $Q=1.3$ GeV and $Q=2$ GeV are depicted in Fig.~\ref{fig:fourmodels}
for later reference. These models are implemented in five fits,
BHPS1,2,3 and SEA1,2, summarized in the next section.

\textit{i)} \textbf{Perturbative charm.} The first model is the one
used in the standard CT14 (and CT14HERA2) PDF fits, in which a non-zero
charm PDF is produced entirely perturbatively by NNLO switching from
the 3-flavor to the 4-flavor scheme at the scale $Q_{c}$. The size
of the preferred charm distribution at a given $Q$ significantly
depends on the values of the physical
charm quark mass $m_c$ and QCD coupling strength $\alpha_{s}(m_{Z})$.
On the other hand, its dependence on the auxiliary theoretical scales of order $m_{c}$,
including the switching scale $Q_{c}$ and the scale in the rescaling
variable $\chi$, cancels up to N3LO and thus is
relatively weak; see a practical illustration in Fig.~1 of \cite{Gao:2013wwa}.
The net momentum fraction of the proton carried
by charm quark starts off close to zero at $Q\approx Q_{c}$ and effectively
saturates at high $Q$ values at a level of approximately 2.5\%, see
Fig.~\ref{fig:xQ}.

\textit{ii}) \textbf{The approximate Brodsky-Hoyer-Peterson-Sakai
(BHPS) model} \cite{Brodsky:1980pb,Brodsky:1981se} parametrizes
the charm PDF at $Q_{0}$ by a ``valence-like'' nonperturbative
function
\begin{equation}
\widehat{c}(x)=\frac{1}{2}A~x^{2}\left[\frac{1}{3}(1-x)(1+10x+x^{2})-2x(1+x)\ln{\left(1/x\right)}\right].\label{eq:modelB}
\end{equation}
This function is obtained from a light-cone momentum distribution by
taking the charm mass to be much heavier than the masses of the proton
and light quarks:
$m_{c}\gg M_{p},m_{u},m_{d}$. Here and in the following,
$A$ is the normalization factor that is to be determined from the fit.
This parametrization choice is employed in two global fits named BHPS1 and BHPS2, corresponding to two values of $A$ in Eq.~\ref{eq:modelB}.
The parametrizations for
$\bar{u}(x,Q_{0})$ and $\bar{d}(x,Q_{0})$ in this case are taken to be the same
as in the CT14/CT14HERA2 fits, i.e., they do not have a ``valence-like''
component and monotonically decrease at $x\rightarrow1$. The parametrizations
of this kind tend to have enhanced $\bar{c}/\bar{u}$ and $\bar{c}/\bar{d}$
ratios at $x\rightarrow1$, see Fig.~\ref{fig:other_ratios}.

\textit{iii)} \textbf{The exact solution of the BHPS model} is realized
in the BHPS3 fit. Instead of approximating the probability integral
as in model \textit{ii}),
the $\widehat{c}(x)$ is obtained by solving the BHPS model
for the $|uudc\bar{c}\rangle$ Fock state numerically and keeping
the exact dependence on $M_{p},m_{u}$, and $m_{d}$.
This fit also includes small BHPS contributions to the
$\bar{u}$ and $\bar{d}$ antiquarks generated
from the $|uudu\bar{u}\rangle$ and $|uudd\bar{d}\rangle$ Fock states
according to the same method. In the BHPS model, the quark
distributions are determined by starting from a $|uudq\bar{q}\rangle$
proton Fock state, where the probability differential for a quark
$i$ to carry a momentum fraction $x_{i}$ is given by
\begin{equation}
d{\cal P}(x_{1},\dots,x_{5})=A\ dx_{1}\dots dx_{5}\
\delta(1-\sum_{i=1}^{5}x_{i})\frac{1}{\left[M_{p}^{2}-\sum_{i=1}^{5}\frac{m_{i}^{2}}{x_{i}^{2}}\right]^{2}}\,.\label{bhps-prob}
\end{equation}
The standard BHPS result, used in \textit{ii}), is given by
letting $q=c$ and taking the limit $m_{c}\gg M_{p},m_{u},m_{d}$
to produce Eq.~(\ref{eq:modelB}). However, Ref.~\cite{Chang:2011vx}
has shown that the solution that keeps the masses finite, including those of the
light quarks, modifies the shape of $\widehat{c}(x)$,
slightly shifting the peak to smaller $x$.
A similar conclusion was reached in Ref.~\cite{Blumlein:2015qcn},
where a kinematic condition on the intrinsic charm was determined
analytically by neglecting the masses of the three light valence quarks
and retaining the ratio $M_{p}^{2}/m_{c}^{2}$.

The change in the BHPS charm quark PDF from including the full mass dependence,
although visible, is small compared to the uncertainties in
the global analysis. However, by using this generalized BHPS
model (BHPS3) in the context of the CT14HERA2 fit, and also including
the BHPS $\bar{u}$ and $\bar{d}$ components,
we obtain physically consistent ratios of the
charm-quark and light-antiquark PDFs at large $x$, cf. Fig.~\ref{fig:other_ratios}. We do not, however,
include the BHPS contribution to the $s$ quark PDF, because
it is overwhelmed by the very large strange PDF uncertainty.
The presence
of a BHPS component for the strange quark does not affect our
conclusions about the nonperturbative charm, so we leave this
topic for a separate CTEQ study of the strange content of the proton.

\textit{iv}) \textbf{In the SEA model}, the charm PDF is parametrized
by a ``sea-like'' nonperturbative function that is proportional
to the light quark distributions:
\begin{equation}
\widehat{c}(x)=A~\left(\overline{d}(x,Q_{0})+\overline{u}(x,Q_{0})\right)\,.\label{eq:modelS}
\end{equation}
This model is assumed with the SEA1 and SEA2 PDF sets from the two global fits
distinguished by the value of normalization $A$ in Eq.~\ref{eq:modelS}.

Finally, the normalization coefficient $A$ in models \textit{ii})\textit{-iv})
can be derived from the charm momentum
fraction (first moment) at scale $Q$:
\begin{equation}
\langle{x}\rangle_{{\rm IC}}=\int_{0}^{1}x\left[c(x,Q_{0})+\bar{c}(x,Q_{0})\right]dx.\label{xICdef}
\end{equation}

By its definition, $\langle x\rangle_{\rm IC}$ is evaluated
at the initial scale $Q_{0}$. It is to be distinguished from the full
charm momentum fraction $\langle x\rangle_{c+\bar{c}}(Q)$ at $Q > Q_c$,
which rapidly increases with $Q$ because of the admixture of the
twist-2 charm component.

\section{Features of the CT14 intrinsic charm \label{sec:RESULTS}}
\subsection{Settings of the fits \label{sec:Settings}}
The BHPS1, BHPS2, SEA1, and SEA2 parametrizations
are obtained by following the setup of the CT14 analysis
\cite{Dulat:2015mca}. BHPS3 is obtained with
the CT14HERA2 setup \cite{Hou:2016nqm}. The CT14HERA2 NNLO fit is very similar to the CT14 fit except
that the HERA Run I and II combined cross sections were used in place of
the Run I cross sections. One of the poorly fit NMC data
sets~\cite{Arneodo:1996qe}
was dropped in CT14HERA2, and the low-$x$ behavior of the strange
(anti)quarks was no longer tied to that of the $\bar{u}$ and $\bar{d}$
antiquarks. This extra flexibility in $s(x,Q_0)$ of CT14HERA2
resulted in a reduction of $s(x,Q_0)$ over the entire $x$
range relatively to CT14. This feature has
potential implications for the models of $\widehat{c}(x)$ with a sea-like
behavior. In some exploratory fits, we include the EMC data \cite{Aubert:1982tt}
on semiinclusive DIS charm production, while in the other fits we
examine sensitivity on the input pole charm
mass.\footnote{CTEQ-TEA fits can also take a $\overline{MS}$ charm mass, rather than
the pole mass as the input \cite{Gao:2013wwa}, with similar
conclusions. }

The PDFs for light partons are parametrized at an initial
scale slightly below $Q_{0}=m_c^{pole}=1.3$ GeV,
with the exception of the study of the $m_{c}^{pole}$ dependence, in
which it was more convenient to start at a
lower initial scale $Q_0=1.0$ GeV. For all models, the QCD coupling
constant is set to $\alpha_{s}(M_{Z})=0.118$,
compatible with the world average value~\cite{Olive:2016xmw}
$\alpha_{s}(M_{Z})=0.1184\pm0.0007$, as in the standard CT PDF
fits. The PDFs are evolved at NNLO with the {\sc HOPPET}
code~\cite{Salam:2008qg}. NLO {\sc ApplGrid} \cite{Carli:2010rw}
and {\sc FastNLO} \cite{Wobisch:2011ij} interpolation interfaces,
combined with NNLO/NLO factor look-up tables, were utilized for fast
estimation of some NNLO cross sections.

\subsection{Dependence on the charm momentum fraction \label{sec:xICDependence}}
\begin{figure}[tb]
\begin{center}
\includegraphics[width=0.45\textwidth]{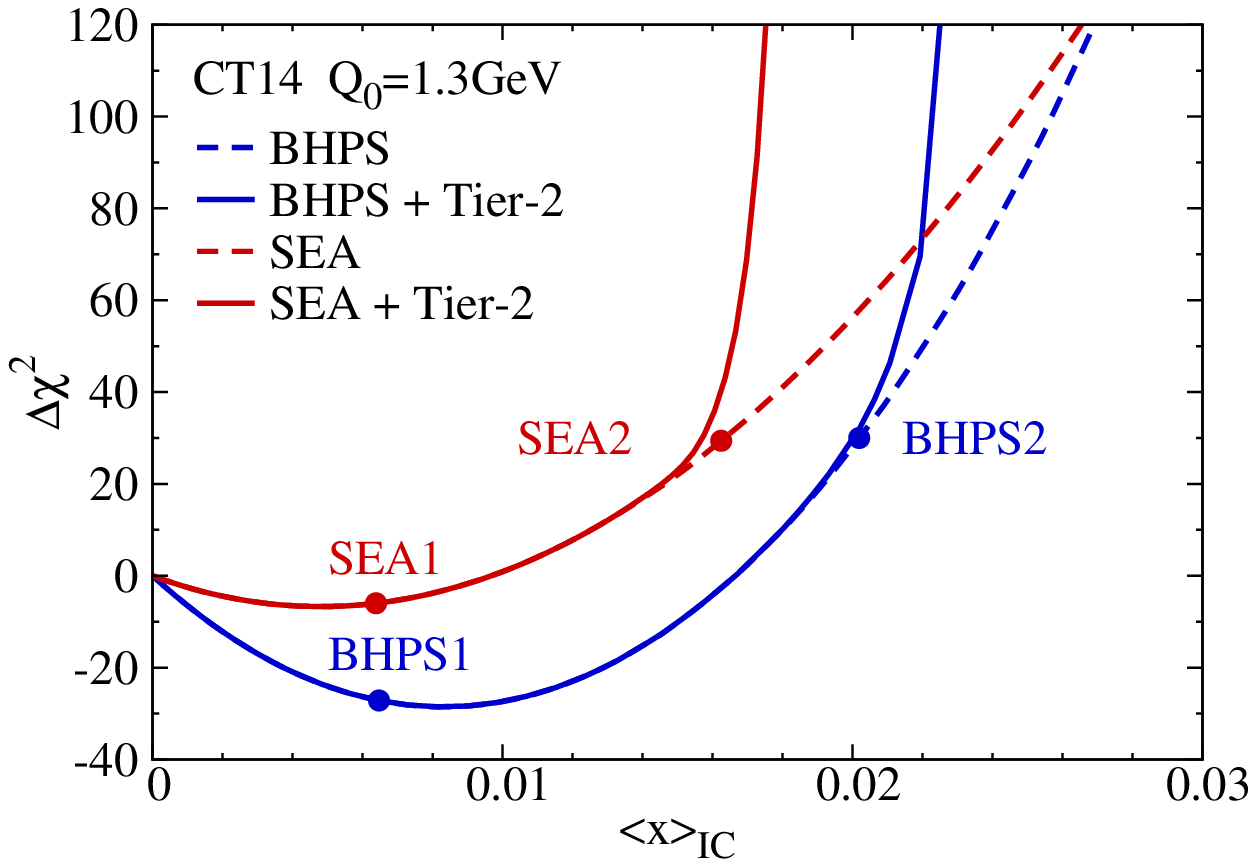}
\includegraphics[width=0.45\textwidth]{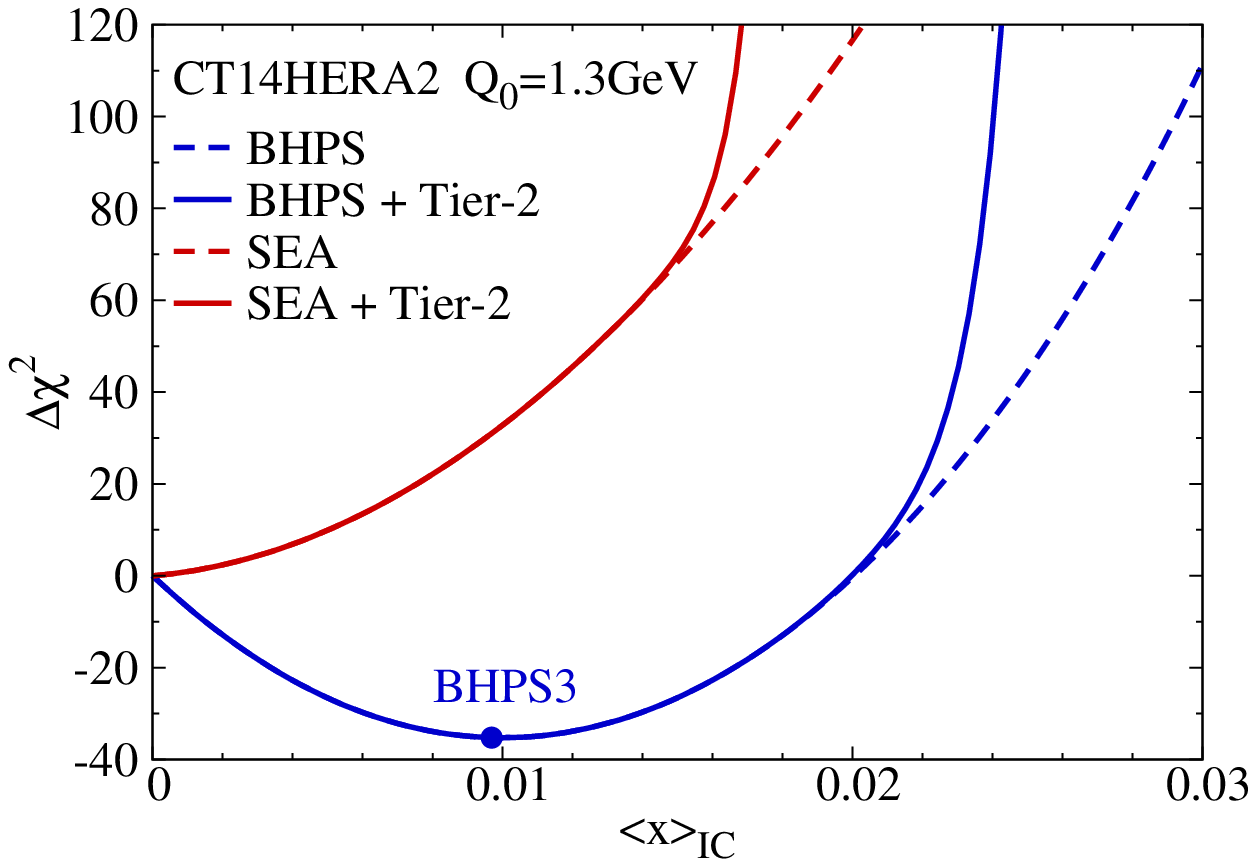}
\end{center}
\caption{The change $\Delta\chi^{2}$ in the goodness of fit to the
  CT14 (left) and CT14HERA2 (right) data sets
  as a function of the charm momentum fraction
  $\langle{x}\rangle_{\rm IC}$ for the BHPS (blue) and SEA (red) models.
Solid (dashed) lines represent the total $\chi^2$ and the partial
$\chi^2_{global}$, as defined in Eq.~(\ref{chi2chi2P}).
\label{fig:delta_chisqVxic}}
\end{figure}

In the models in Sec.~\ref{sec:CharmModels}, the magnitude
of $\widehat{c}(x)$ is controlled by its normalization $A$, correlated
uniquely with the net momentum fraction $\langle x \rangle_{\rm IC}$
of $c(x,Q_0)+\bar{c}(x,Q_0)$ defined in Eq.~(\ref{xICdef}).
The choice of the $\langle
x\rangle_{\rm IC}$ affects theoretical predictions in a number of ways, either
directly by modifying the charm scattering contributions, or
indirectly via the proton sum rule that changes the
momentum fractions available to other parton flavors.

To gauge the preference of the global QCD data to a specific $\langle
x\rangle_{\rm IC}$, we examine the goodness-of-fit function
\begin{equation}
  \chi^2 \equiv \chi^2_{global}+ P \label{chi2chi2P},
\end{equation}
constructed in the CT14 method from the global
$\chi^2_{global}$ and a ``tier-2'' statistical penalty $P$
\cite{Dulat:2015mca}. It is convenient to compare each fit with an
$\langle x\rangle_{\rm IC}\neq 0$ to the ``null-hypothesis'' fit
obtained assuming $\langle x\rangle_{\rm IC}=0$. Thus, we start by computing
\begin{equation}
  \Delta \chi^2 \equiv \chi^2 -
  \chi^2_0, \label{Deltachi2}
\end{equation}
where $\chi^2$ and $\chi^2_0$ are given for $\langle x\rangle_{\rm
  IC}\neq 0$ and $\langle x\rangle_{\rm
  IC}=0$, respectively, at 50 values of $\langle x\rangle_{\rm IC}$ and default
$Q_0=m_c^{pole}=1.3$ GeV. We plot the resulting $\Delta \chi^2$ behavior in
Fig.~\ref{fig:delta_chisqVxic}. The CT14
(CT14HERA2) data sets are compared against the
approximate (exact) solution of the BHPS model, respectively.
The SEA charm parametrizations are constructed as in
Eq.~(\ref{eq:modelS}) in terms of the respective CT14 or CT14HERA2
light-antiquark parametrizations.

For each series of fits, we show curves for two types of estimators: a
dashed curve for
$\Delta \chi^2_{global}$ without the tier-2 penalty $P$, and a solid
one for $\chi^2$ that includes $P$ according to Eq.~(\ref{chi2chi2P}). The
$\chi_{global}^2$ function estimates the global quality of fit and
is equal to the sum of $\chi^2$ contributions from all experiments
and theoretical constraints. A non-negative ``Tier-2'' penalty $P$ is added to
$\chi_{global}^2$ to quantify agreement with each individual
experiment~\cite{Dulat:2013hea,Gao:2013xoa}. Being negligible in good
fits, $P$ grows very rapidly when some experiment turns out to be
inconsistent with theory. The net effect of $P$ is to quickly increase
the full $\chi^2$ if an inconsistency with some experiment occurs, even
when $\chi^2_{global}$ remains within the tolerable limits.

We see from Fig.~\ref{fig:delta_chisqVxic} that large amounts
of intrinsic charm are disfavored for all models under scrutiny.
A mild reduction in $\chi^{2}$, however, is observed for the BHPS
fits, roughly at $\langle{x}\rangle_{\rm IC} = 1\%$,
both in the CT14 and CT14HERA2 frameworks.

The significance of this reduction and the upper
limit on $\langle x\rangle_{\rm IC}$
depends on the assumed criterion. In CTEQ
practice, a set of PDFs with $\Delta\chi^2$ smaller (larger) than 100 units
is deemed to be accepted (disfavored) at about 90\% C.L.
Thus, a reduction of $\chi^2$ by less than forty units for the BHPS curves has
significance roughly of order one standard deviation. We also obtain
the new upper limits on $\langle{x}\rangle_{\rm IC}$ in the CT14 and
CT14HERA2 analyses at the 90\% C.L.:
\begin{eqnarray}
\langle{x}\rangle_{\rm IC} &\lesssim & 0.021 \mbox{ for CT14 BHPS}, \nonumber \\
\langle{x}\rangle_{\rm IC} &\lesssim & 0.024~\mbox{ for CT14HERA2 BHPS}, \nonumber \\
\langle{x}\rangle_{\rm IC} &\lesssim & 0.016~\mbox{for CT14 and
  CT14HERA2 SEA}. \label{xICconstraints}
\end{eqnarray}

In keeping with the previous analysis of Ref.~\cite{Dulat:2013hea}, we
define specific fits with particular choices of $\langle x\rangle_{\rm IC}$
for both examined models. The fits BHPS1 and SEA1 correspond
to $\langle{x}\rangle_{\rm IC} = 0.6 \%$, while BHPS2 has
$\langle{x}\rangle_{\rm IC} = 2.1 \%$ and SEA2 has
$\langle{x}\rangle_{\rm IC} = 1.6 \%$.
Both the BHPS2 and SEA2 charm parametrizations lie
near the edge of disagreement with some experiments in the global
analysis data according to the CTEQ-TEA tolerance criterion,
cf. Fig.~\ref{fig:delta_chisqVxic}.
In the CT14HERA2 fit,
the BHPS3 point corresponds to $\langle{x}\rangle_{\rm IC} = 1\%$, which
represents the best-fit momentum fraction in the CT14HERA2 analysis.
We remind the reader that, in addition to fitting more recent
experimental data from the LHC and other experiments,
the BHPS3 analysis also employs a general numerical
solution to the BHPS probability distributions and small
valence-like contributions for both the $\bar{u}$ and $\bar{d}$ quarks.

The results in Fig.~\ref{fig:delta_chisqVxic} are compatible with
the findings of the previous CT10 NNLO IC analysis~\cite{Dulat:2013hea}.
In particular, comparing to CT14 in the left frame of
Fig.~\ref{fig:delta_chisqVxic} and to Fig. 2 in
Ref.~\cite{Dulat:2013hea},\footnote{In
  Ref.~\cite{Dulat:2013hea}, $\chi^2_{global}$, $P$, and $\chi^2$ are
  denoted by $\chi^2_F$, $T_2$, and $\chi^2_F+T_2$.} we see
that the minimum in $\Delta\chi^2$ in the right frame of
Fig.~\ref{fig:delta_chisqVxic} deepened  by approximately 10 units for
BHPS3/CT14HERA2 -- a minor reduction caused mostly by the change to
the CT14HERA2 setup, either for the exactly or approximately solved
BHPS model.

Also, for the CT14HERA2 analysis in Fig.~\ref{fig:delta_chisqVxic}
(right), we note that $\Delta\chi^2$ of the SEA model rises more
rapidly with increasing $\langle{x}\rangle_{\rm IC}$ than it does in
the comparable CT14 fit. This is due to the greater flexibility in the
low-$x$ behavior of the strange-quark distribution in the CT14HERA2
framework discussed previously. More freedom reduces $s(x,Q)$ at low
$x$ and thus increases
$\bar{u}(x,Q)$ and $\bar{d}(x,Q)$ at the same $x$. In the
CT14 fit with the SEA charm component, the $\Delta \chi^2$
minimum is at $\langle x
\rangle_\textrm{IC} \approx 0.004$, and it is largely washed out in
the CT14HERA2 case. The $\Delta \chi^2$ for SEA grows faster
for CT14HERA2 compared to CT14: at $\langle x\rangle_{\textrm{IC}} =
1.6\%$ it is higher by about 40 units
in Fig.~\ref{fig:delta_chisqVxic}(right) relatively
to Fig.~\ref{fig:delta_chisqVxic}(left).

The reduction in $\chi^2$ for the NNLO BHPS fits at $\langle{x}\rangle_{\rm
  IC}$ = 0.01, relatively to the fit with $\langle x\rangle_{\rm IC}=0$,
thus remains a persistent feature of the CT10, CT14, and CT14HERA2
analyses. While the $\Delta \chi^2$ reduction
is not statistically significant, it raises one's curiosity:
is it a sign of a genuine charm
component or of the other circumstantial factors identified in
Sec.~\ref{sec:Overview}? It will be discussed in Sec.~\ref{sec:DataImpact}
that $\chi^2$ is reduced primarily in  a few fixed-target
experiments (the $F_2$ measurements from BCDMS and the E605 Drell-Yan
data) that are not overtly sensitive to charm production.
Conversely, the description of the other experiments that might
be expected to be most sensitive to intrinsic charm is not improved.

\subsection{Dependence on the charm-quark mass and energy scale\label{sec:McDependence}}

We have checked that these conclusions are not strongly dependent on the
PDF parametrizations of the light partons. However, the SEA
parametrization at the initial $Q_0$ is very sensitive to the assumed charm
mass.

Distinct from the auxiliary QCD mass parameters -- $Q_0$, $Q_c$, and
the mass in the $\chi$ rescaling variable -- the physical charm-quark
mass of the QCD Lagrangian enters the DIS hard matrix elements
through the ``flavor-creation'' coefficient functions, such as the ones
for the photon-gluon fusion. The NNLO fit to DIS
is mostly sensitive to the primordial QCD mass parameter $m_c$, not to the
auxiliary parameters of order $m_c$ \cite{Gao:2013wwa}. The
$m_{c}^{pole}$ dependence remains mild, the $m_{c}^{pole}$ values in the range
$1.1-1.5$ GeV are broadly consistent with the CT14 data.

\begin{figure}[tb]
\begin{center}
\includegraphics[width=0.45\textwidth]{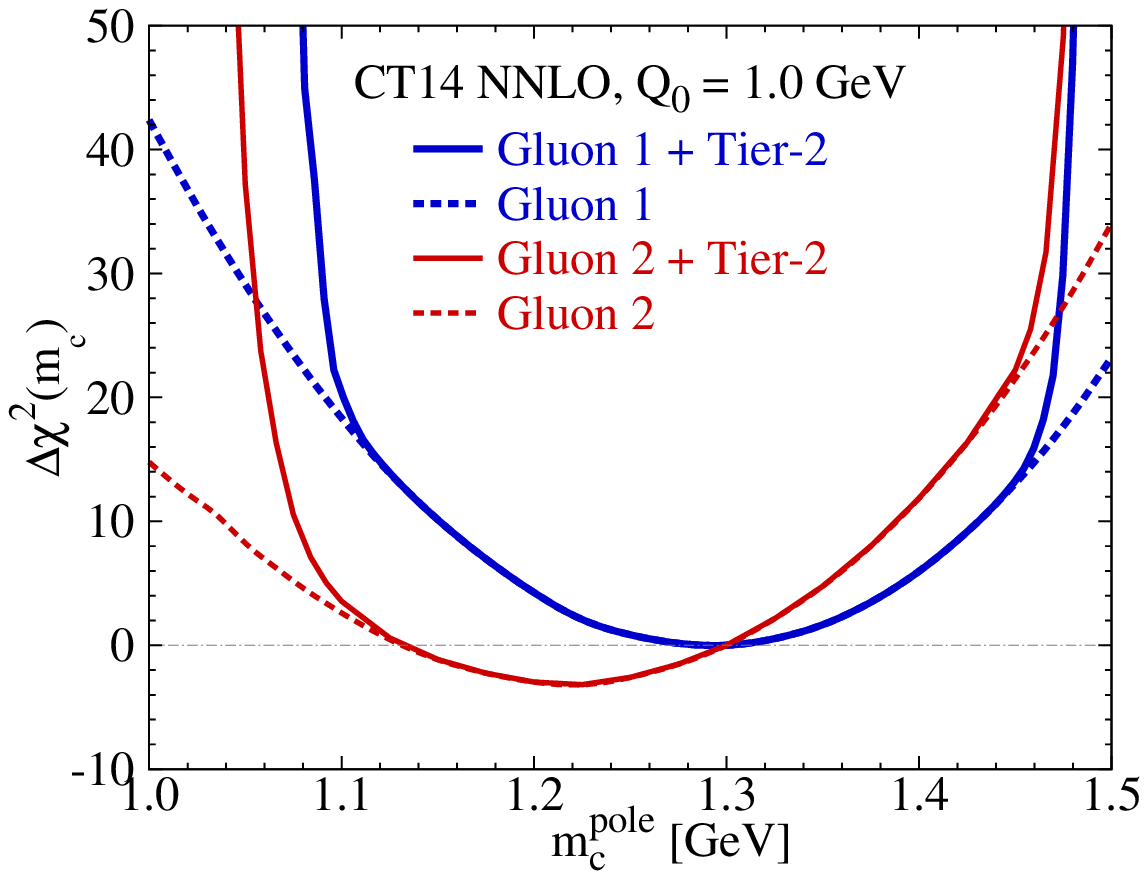}\\
\includegraphics[width=0.45\textwidth]{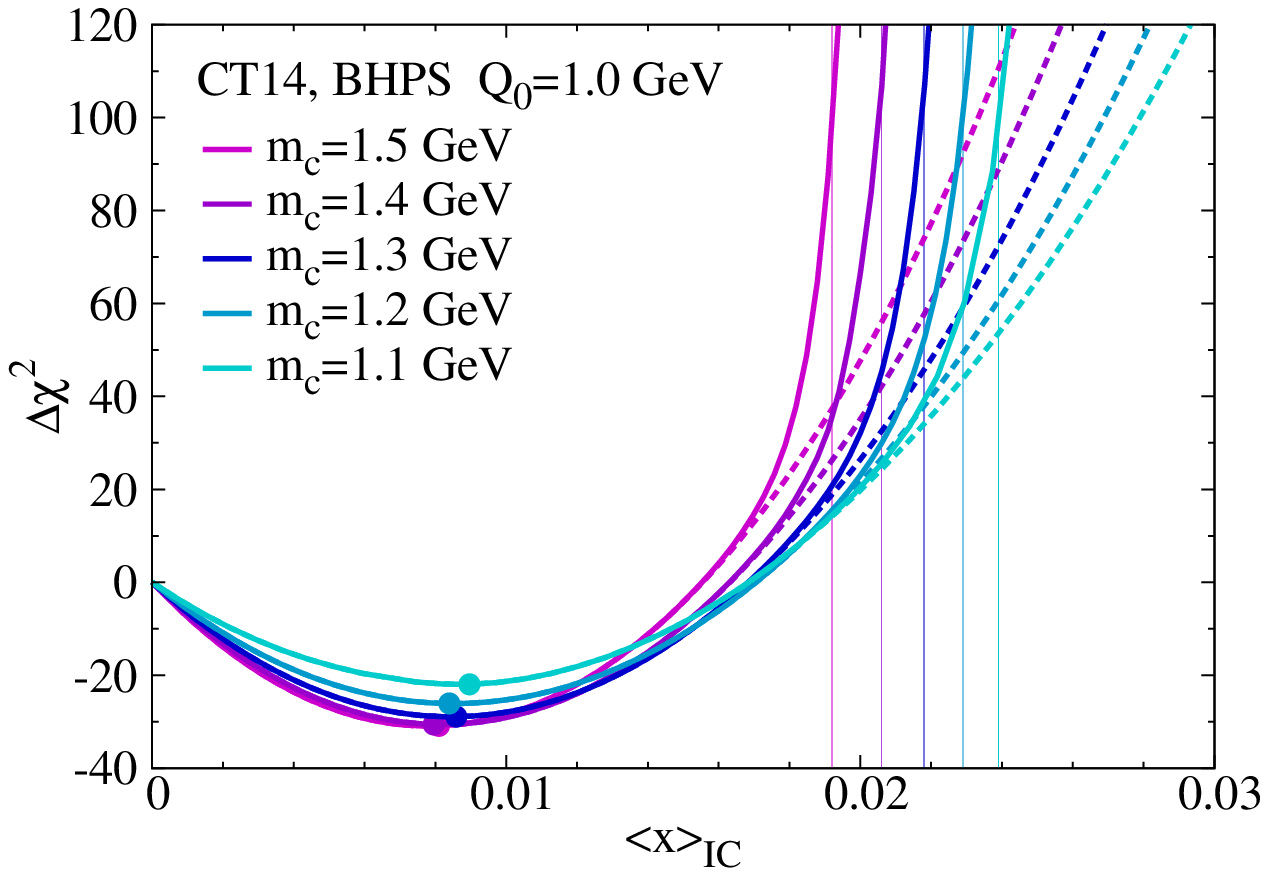}
\includegraphics[width=0.45\textwidth]{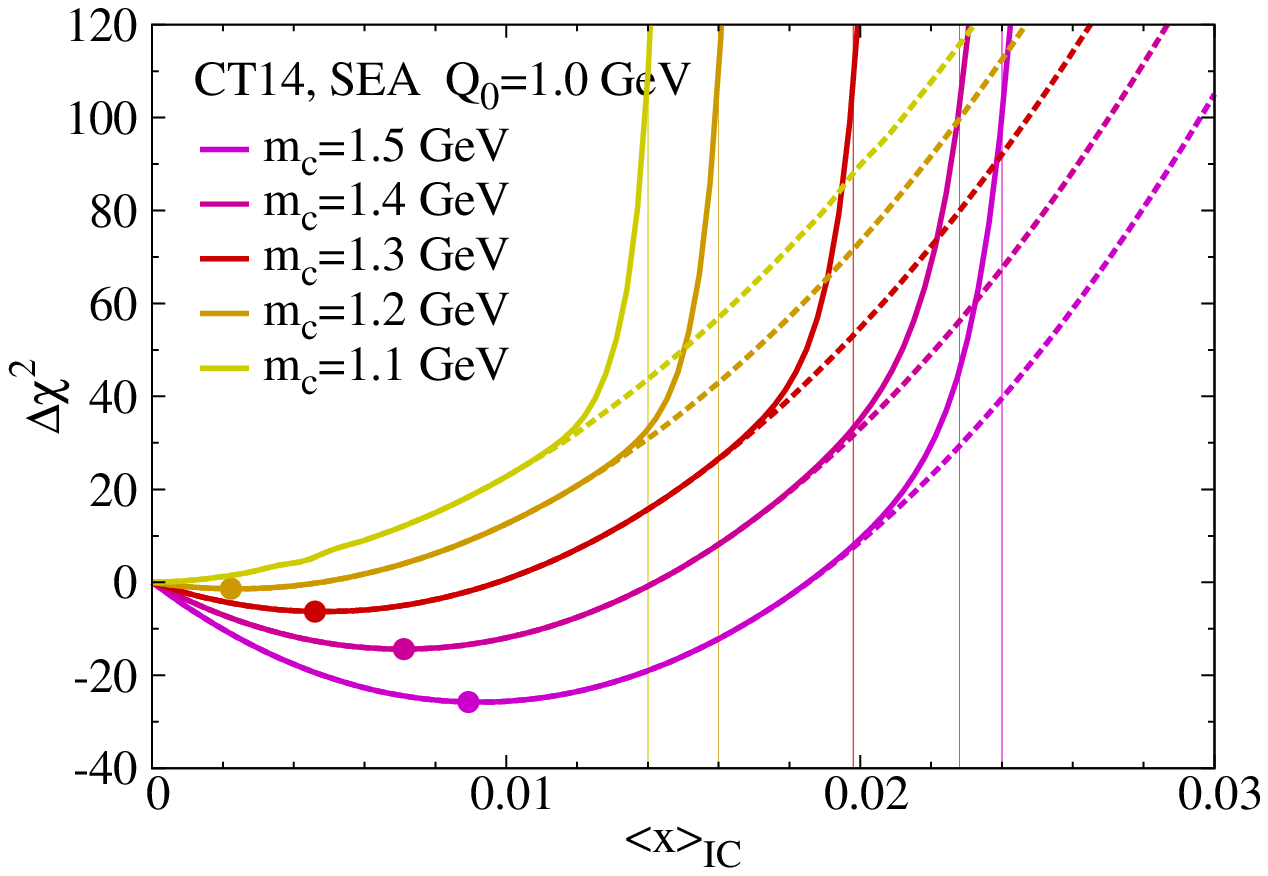}
\end{center}
\caption{Upper: dependence of $\Delta \chi^2$ in the CT14 NNLO fit
  (without the nonperturbative charm) on the charm mass $m_{c}^{pole}$
  for two possible gluon parametrization forms. Lower:
  Dependence of $\Delta \chi^2$ on the intrinsic charm momentum
  fraction for CT14 candidate
  fits with different values of the charm-quark pole mass $m_c^{pole}$.
  $\Delta\chi^2$ is defined as
  $\chi^2 - \chi^2\left(m_c^{pole}=1.3\mbox{\rm GeV}\right)$
  and
  $\chi^2 - \chi^2\left(\, \langle x\rangle_{\rm IC}=0\, \right)$ in the upper and lower insets, respectively.
\label{fig:Mc_dchi2Vxic}}
\end{figure}

Exploratory fits testing
the dependence of $\Delta\chi^2$ on $\langle x\rangle_{\rm IC}$,
for a selection of pole masses,  $m_c^{pole}=\{$1.1, 1.2, 1.3, 1.4, 1.5$\}$
GeV, are illustrated by Fig.~\ref{fig:Mc_dchi2Vxic}.
The general setup of
these $\chi^2$ scans follows the fits to the CT14 data.
To access the masses below 1.3 GeV, we reduced the
initial scale $Q_0$ to 1 GeV and examined
alternative forms for the gluon PDF parametrization,
because DIS charm production is sensitive to the gluon PDF $g(x,Q)$.
Dependence of $\Delta \chi^2$ for CT14 NNLO on $m_c^{pole}$
for two representative gluon parametrizations at $Q_0=1$ GeV,
dubbed ``gluon 1'' and ``gluon 2'', is shown
in the upper inset of Fig.~\ref{fig:Mc_dchi2Vxic}.
With the ``gluon 1'' parametrization, used in the default
CT14 fit with $Q_0 = 1.3$  GeV, $g(x, Q_0)$
 is constrained to be positive at all $x$;
 while for ``gluon 2'', it is allowed
 to be negative at the smallest $x$ and $Q$,
 provided that the negative gluon does not lead to unphysical predictions.
In the latter case, an additional theoretical
constraint was enforced to ensure positivity of the longitudinal structure
function $F_L(x, Q)$ measured
by the H1 Collaboration~\cite{Collaboration:2010ry}.
The more flexible ``gluon 2'' parametrization results
in a marginally better $\chi^2$ with respect to the nominal CT14,
or ``gluon 1'', at a slightly lower $m_c^{pole}=1.22$ GeV, and with a
large uncertainty. This best-fit $m_{c}^{pole}$ value in this range
is consistent with the previously observed tendency of the DIS data
to prefer smaller $\overline{MS}$ masses at ${\cal O}(\alpha_s^2)$,
{\it e.g.}, $m_c(m_c) = 1.15^{+0.18}_{-0.12}$ GeV obtained in the CT10
setup~\cite{Gao:2013wwa}.

The two lower insets of Fig.~\ref{fig:Mc_dchi2Vxic} illustrate
the variations in $\Delta\chi^2$,
with the  more flexible ``gluon 2'' parametrization,
 when the IC component is
included for five values of $m_c^{pole}$. The circles on the curves
mark the $\chi^2$ minima; the thin vertical lines indicate the exclusion
limits on $\langle x \rangle_{\rm IC}$ for each $m_c^{pole}$ value.

For the BHPS model in the left inset,
the position of the $\chi^2$ minimum is relatively stable
as $m_c^{pole}$ is varied, while the upper limit on $\langle x
\rangle_{\rm IC}$ decreases to 1.9\% as $m_c$ increases.
The overall conclusion is that the preferred
$\langle x \rangle_{\rm IC}$ at scale $Q_0$ is not strongly sensitive to the
variations of $m_c$ in the case of the BHPS parametrizations. On the
other hand, as we will see in a moment, the total momentum fraction
$\langle x \rangle_{c+\bar c}(Q)$ at scales
above $Q_c =m_{c}^{pole}$ is sensitive to $m_c^{pole}$
due to the growing perturbative charm component.

The situation is very different for the SEA model shown
in Fig.~\ref{fig:Mc_dchi2Vxic} (right), where the dependence on
$m_c^{pole}$ is more pronounced.
In this case, $\Delta \chi^2$ develops a pronounced minimum for
$m_c^{pole} > 1.3$ GeV, while the minimum totally disappears, and
$\langle x \rangle_{\rm IC} \gtrsim 0.015$ is totally excluded, for
$m_{c}^{pole}=1.1$ GeV.

This can be understood as follows: when $m_c^{pole}$ increases, the
twist-2 $\gamma^*g$ fusion contribution in the inclusive
DIS structure functions is reduced due to phase-space suppression.
This suppression is compensated
by allowing a larger magnitude of intrinsic $c(x,Q)$,
which enhances the $\gamma^*c$ scattering contribution.
An opposite effect occurs when $m_c$ decreases (i.e., less
phase-space suppression for $\gamma^*g$ fusion, a smaller
intrinsic charm momentum allowance in $\gamma^*c$ scattering).
But the $\bar{u}$ and $\bar{d}$ quark PDFs are well constrained by the
data, especially from novel cross section measurements for vector
boson production in $pp$ and $p\bar{p}$ in the intermediate/small $x$
region. The net effect is the $\Delta\chi^2$ enhancement in the
sea-like scenario for $m_c^{pole} < 1.3$ GeV and
also for larger $\langle x \rangle_{\rm IC}$ fractions.

\begin{figure}[p]
\centering
\includegraphics[width=0.49\textwidth]{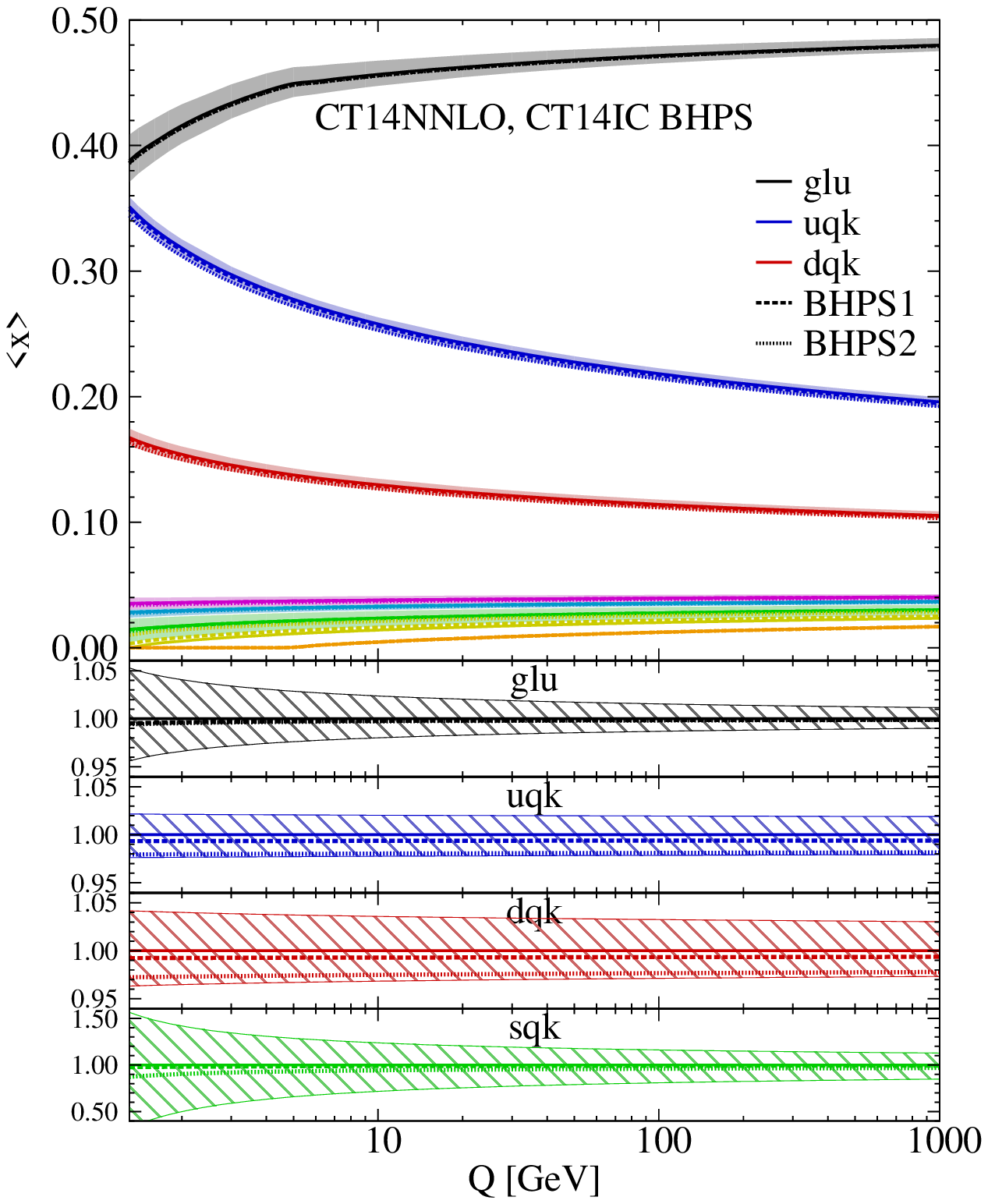}
\includegraphics[width=0.49\textwidth]{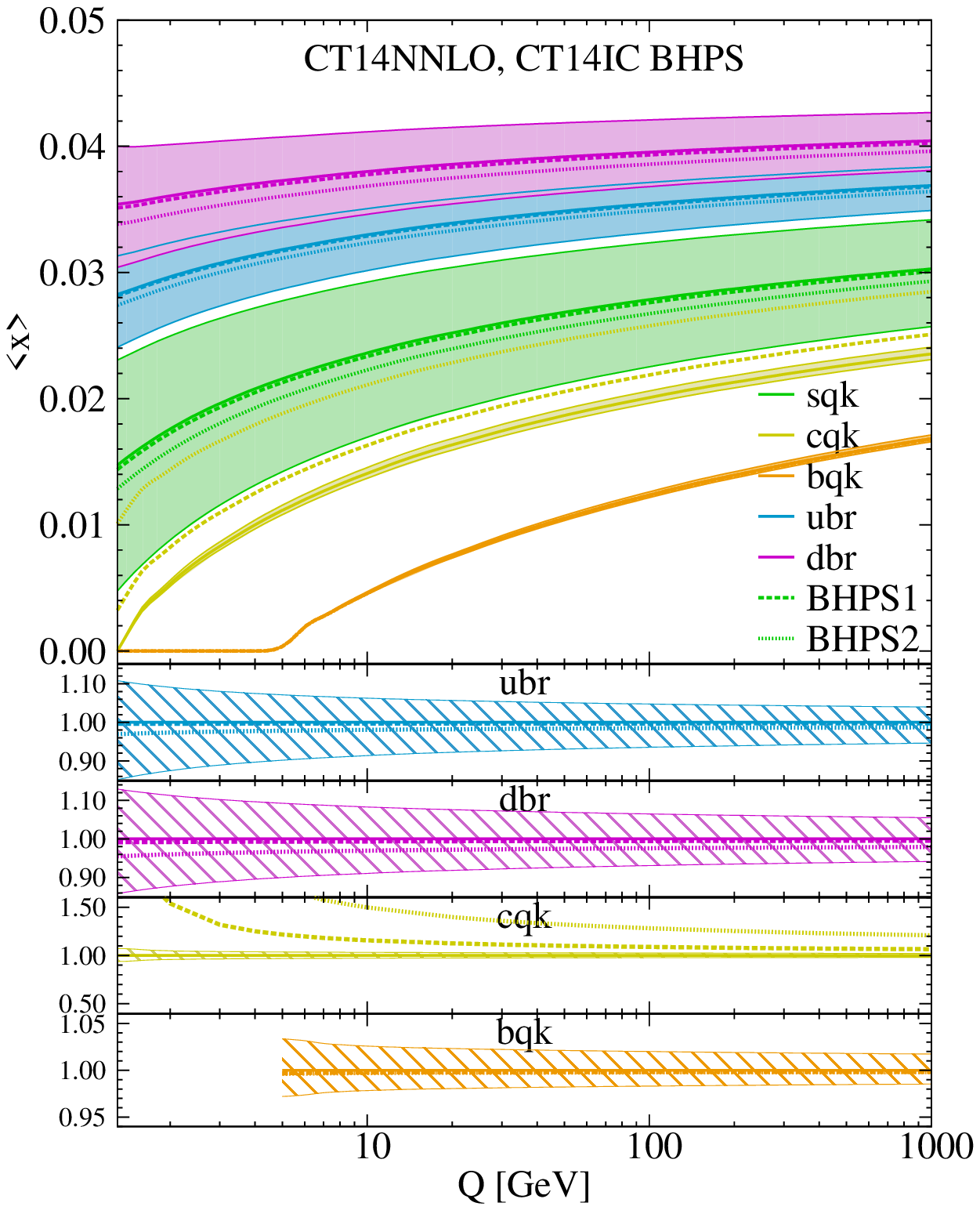}
\includegraphics[width=0.49\textwidth]{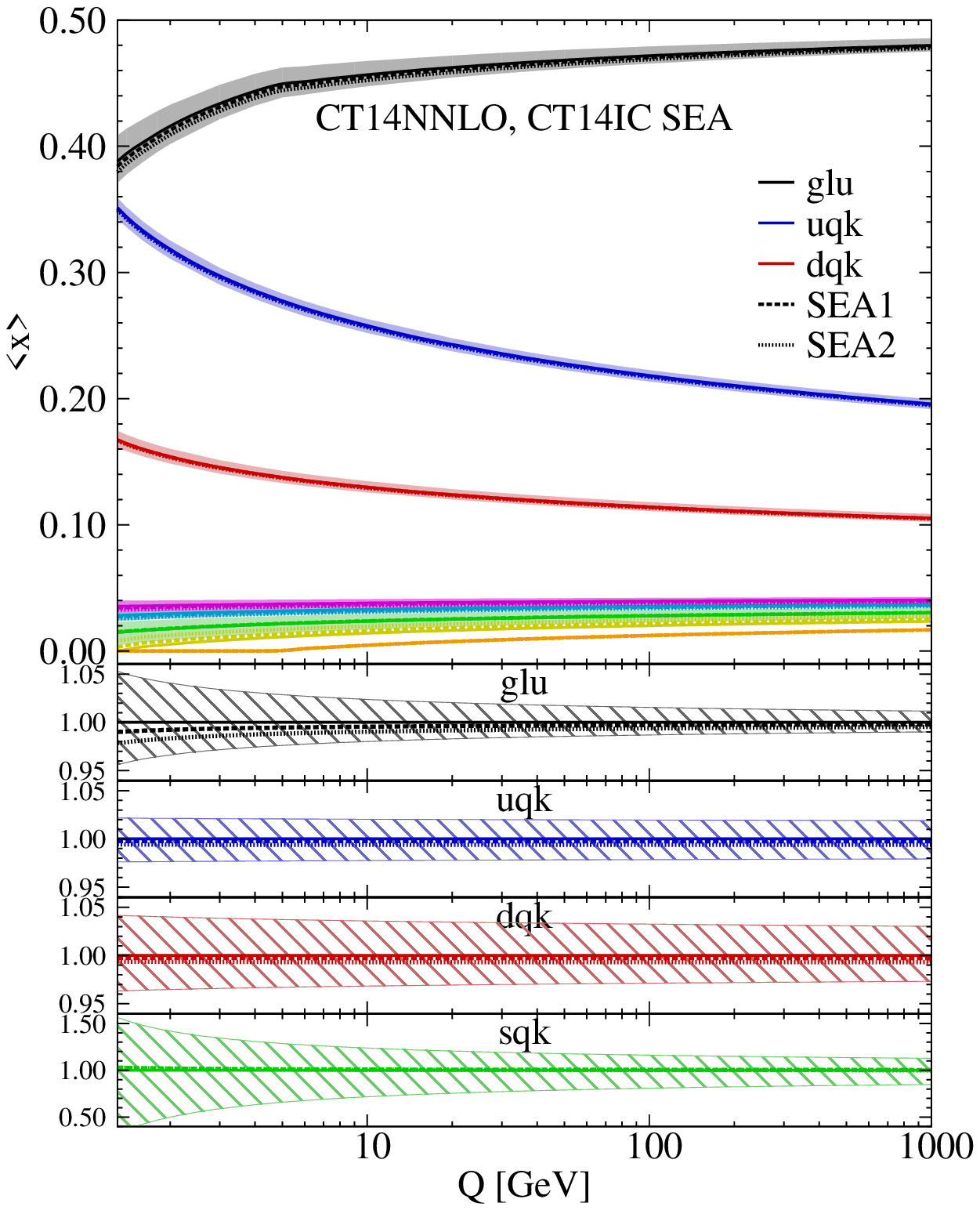}
\includegraphics[width=0.49\textwidth]{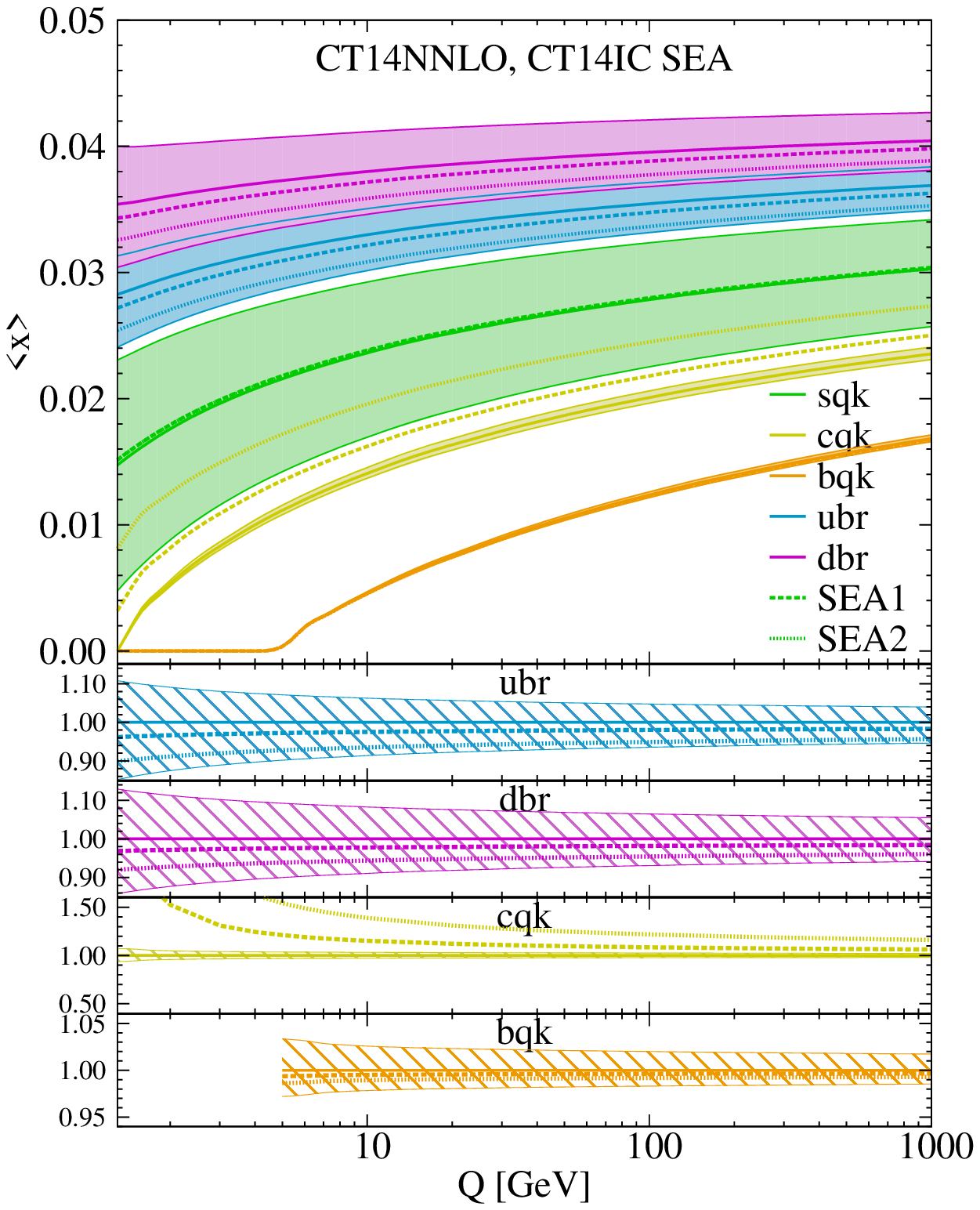}
\caption{Momentum fractions $\langle x\rangle(Q)$ for CT14 and CT14 IC
  vs. $Q$, shown independently for gluons, quarks and antiquarks.
The momentum fractions of PDFs for BHPS1 (SEA1) are denoted by the
dashed curves, while those for BHPS2 (SEA2) are denoted by dotted
curves. (Here, the label ``cqk'' indicates that only charm quark is
counted, and ``ubr'' is for up antiquark only, etc.)
The uncertainty bands are for CT14 with no intrinsic charm.}
\label{fig:xQ}
\end{figure}

To conclude the discussion of the partonic momentum fractions,
Fig.~\ref{fig:xQ} illustrates the first
moments $\langle x\rangle(Q)$
of the other parton flavors as a function of the factorization scale $Q$.
The momentum fractions are computed separately for quarks, antiquarks,
and gluons in the context of the CT14 setup.
In the two upper subfigures, the PDF first moments
are shown for the BHPS model, while those from the SEA model are shown
in the lower two subfigures. The dashed curves  represents BHPS1 (SEA1),
the dotted ones represent BHPS2 (SEA2).

The lower part of each figure shows $\langle x\rangle$
normalized to its CT14 central value. The BHPS2
model curve lies on the edge of the allowed CT14 $u$ and $d$ quark
uncertainties, while the SEA2 is on the boundary of the $\bar{u}$ and
$\bar{d}$ uncertainties.  This corroborates the earlier statement that
BHPS2 and SEA2 are the extreme choices
for the valence-like and sea-like charm distributions,
respectively. Next, we will consider the full $x$ dependence of the
PDFs provided by our models.

\subsection{Impact of IC on the PDFs \label{sec:PDFs}}

\begin{figure}[t]
\centering
\includegraphics[width=0.49\textwidth]{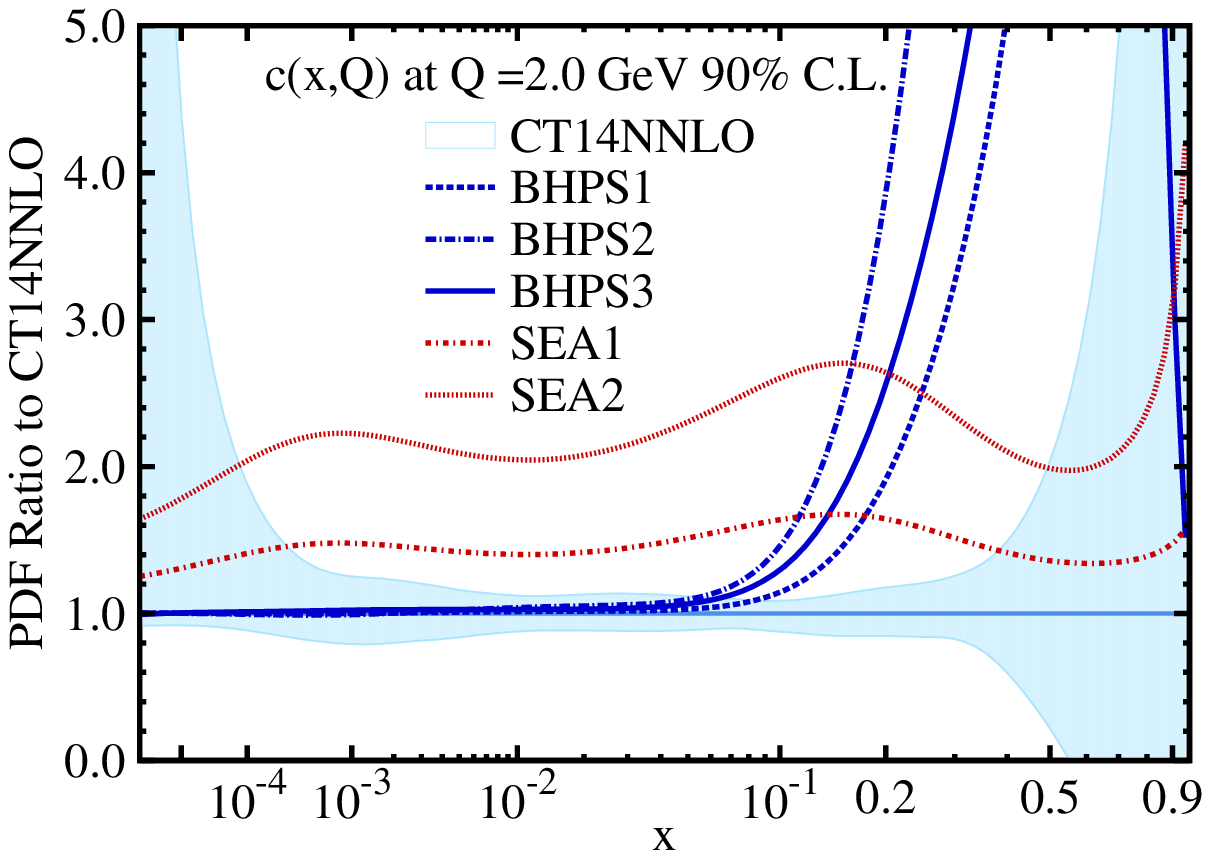}
\includegraphics[width=0.49\textwidth]{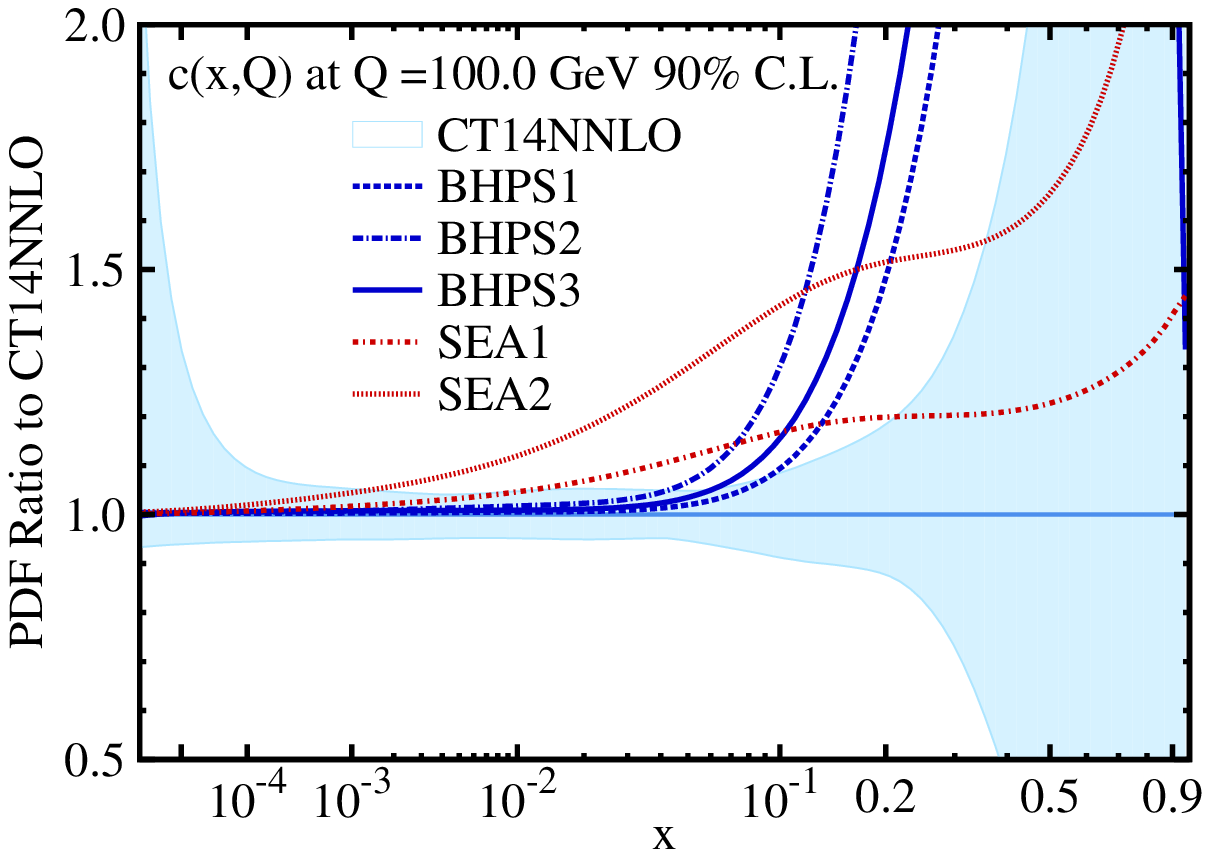} \\
\caption{
Ratio of $c(x,Q)_{\rm IC}/c(x,Q)_{\rm CT14}$ within the CT14 uncertainties at 90\% C.L.
at the scale $Q = 2\ {\rm GeV}$ (left) and $Q = 100\ {\rm GeV}$ (right).
\label{fig:cratios}}
\end{figure}
To complement the visualization in Fig.~\ref{fig:fourmodels}
of the $x$ dependence of the BHPS/SEA charm quark PDFs, in Fig.~\ref{fig:cratios}
these PDFs are shown
normalized to the CT14 charm PDF with no IC contribution.
The blue shaded region represents the CT14 uncertainty for $c(x,Q)$
at the 90\% C.L.

At low scales ($Q=2$ GeV),  the charm quark in the SEA models, especially the SEA2 model, appears to be larger, with respect to the CT14 central charm,
over a wide range of momentum fraction $x$. The charm quark
distributions from both of these models are clearly outside the CT14
uncertainty bands. Of course, this is not a contradiction, since the
CT14 charm PDF is purely radiative, and so it depends on the
theoretical assumptions in addition to the constraints from the
experimental data. The inclusion of nonperturbative sources of charm
relaxes the theoretical assumptions, and so allows a larger charm PDF.
The SEA models exhibit minor shape distortions;
two bumps are present in both the SEA1 and SEA2 models at
$x\approx 10^{-3}$ and $x=0.1$.

The charm-quark distributions in the BHPS models at low scales are basically coincident with CT14 below $x\approx 5\times 10^{-2}$, while
a rapid growth is observed at high $x$, of the largest rate for the BHPS2 model.
We note that there is no qualitative difference in the behavior of $c(x,Q)$
between the BHPS3 model and the other BHPS models below $x\approx 5\times 10^{-2}$, while the differences at larger $x$ can  be ascribed to
the exact solution for mass dependence in BHPS3.
At a higher scale ($Q=100$ GeV), the excesses for all models are suppressed
for $x \lesssim 10^{-2}$ due to the effects of DGLAP evolution.
The results for the ratio of $c(x,Q)_{\rm IC}/c(x,Q)_{\rm CT14HERA2}$
are analogous to those shown in Fig.~\ref{fig:cratios} and are omitted.

\begin{figure}[tb]
\centering
\includegraphics[width=0.49\textwidth]{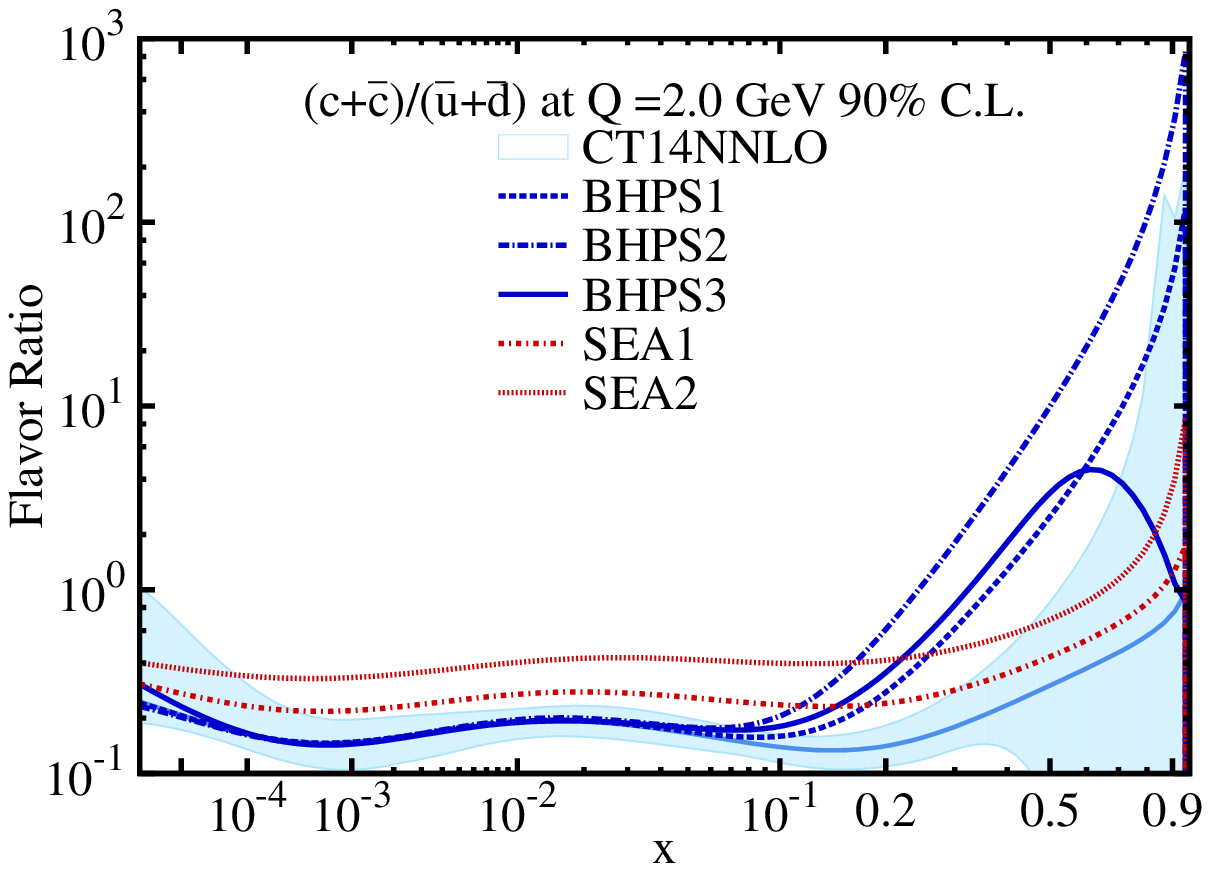}
\includegraphics[width=0.49\textwidth]{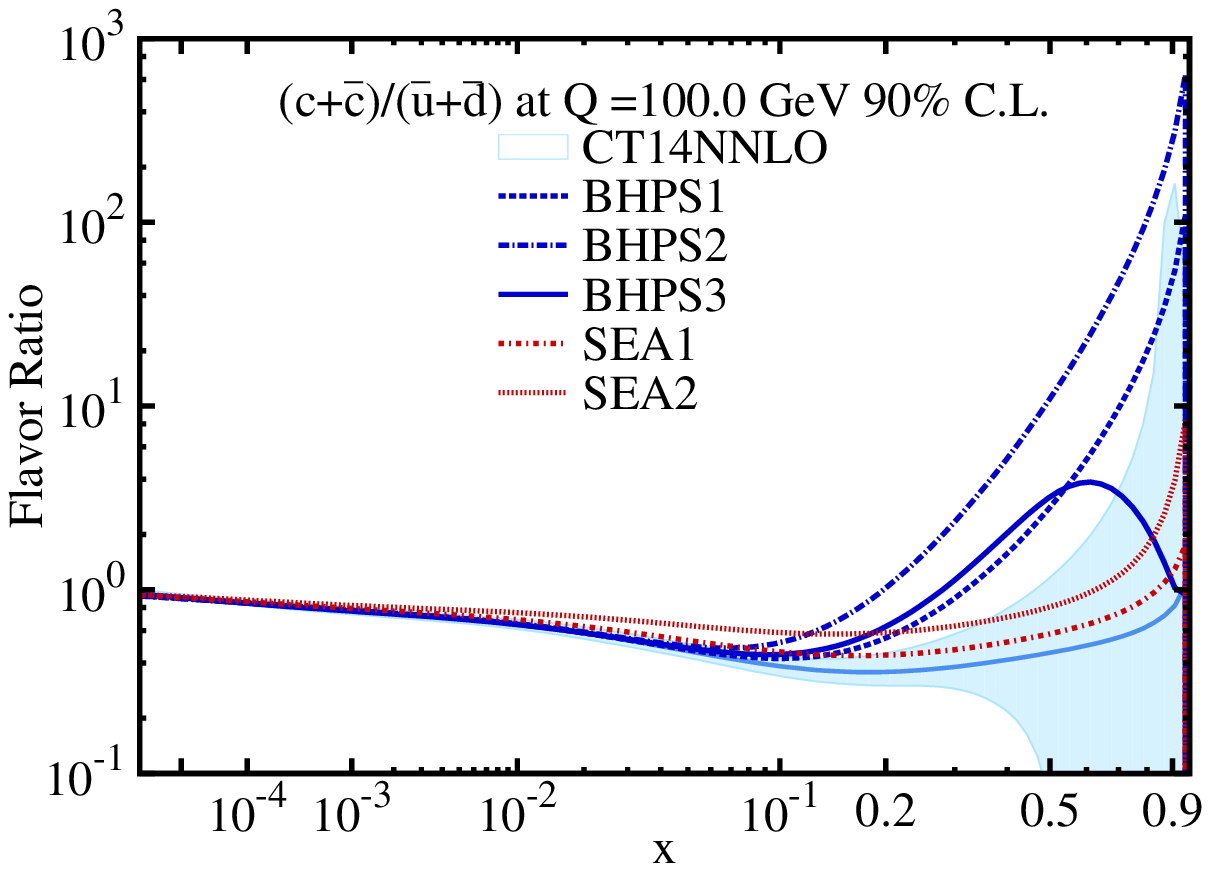} \\
\includegraphics[width=0.49\textwidth]{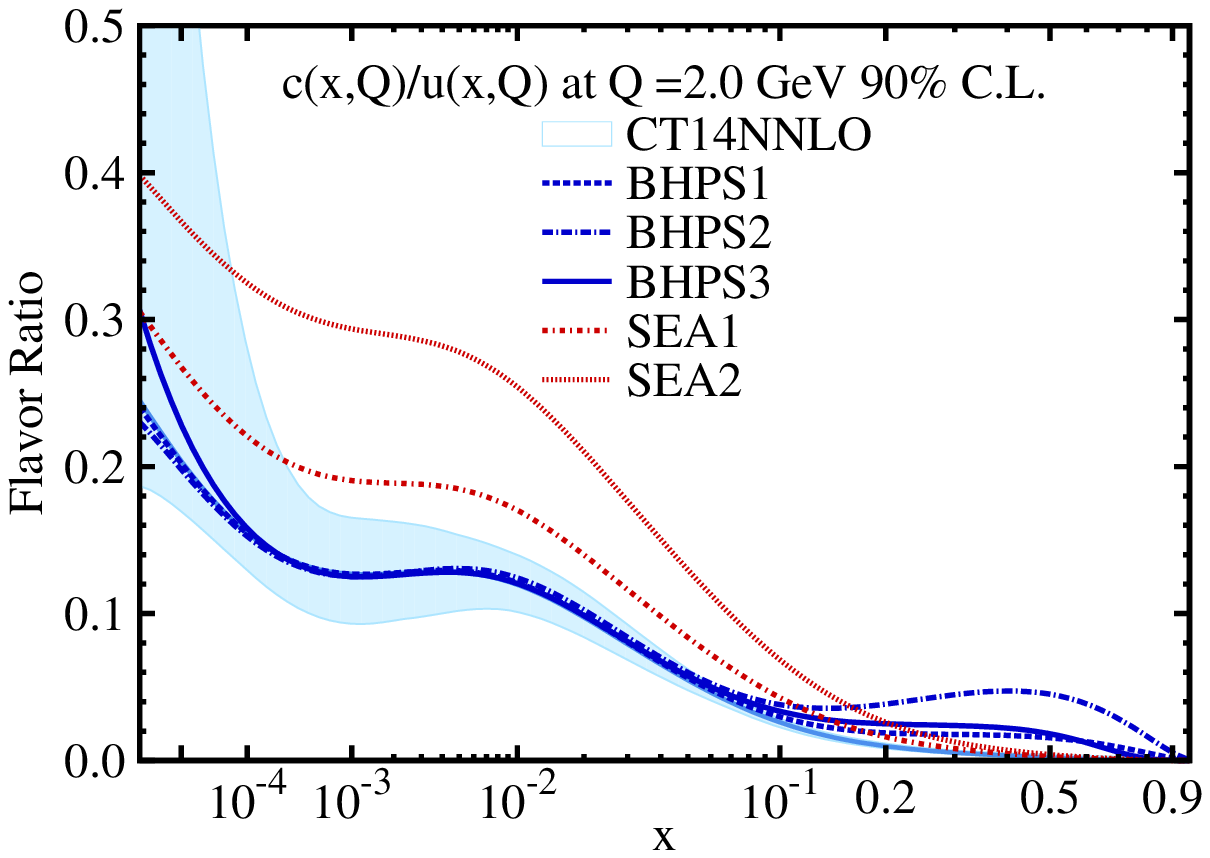}
\includegraphics[width=0.49\textwidth]{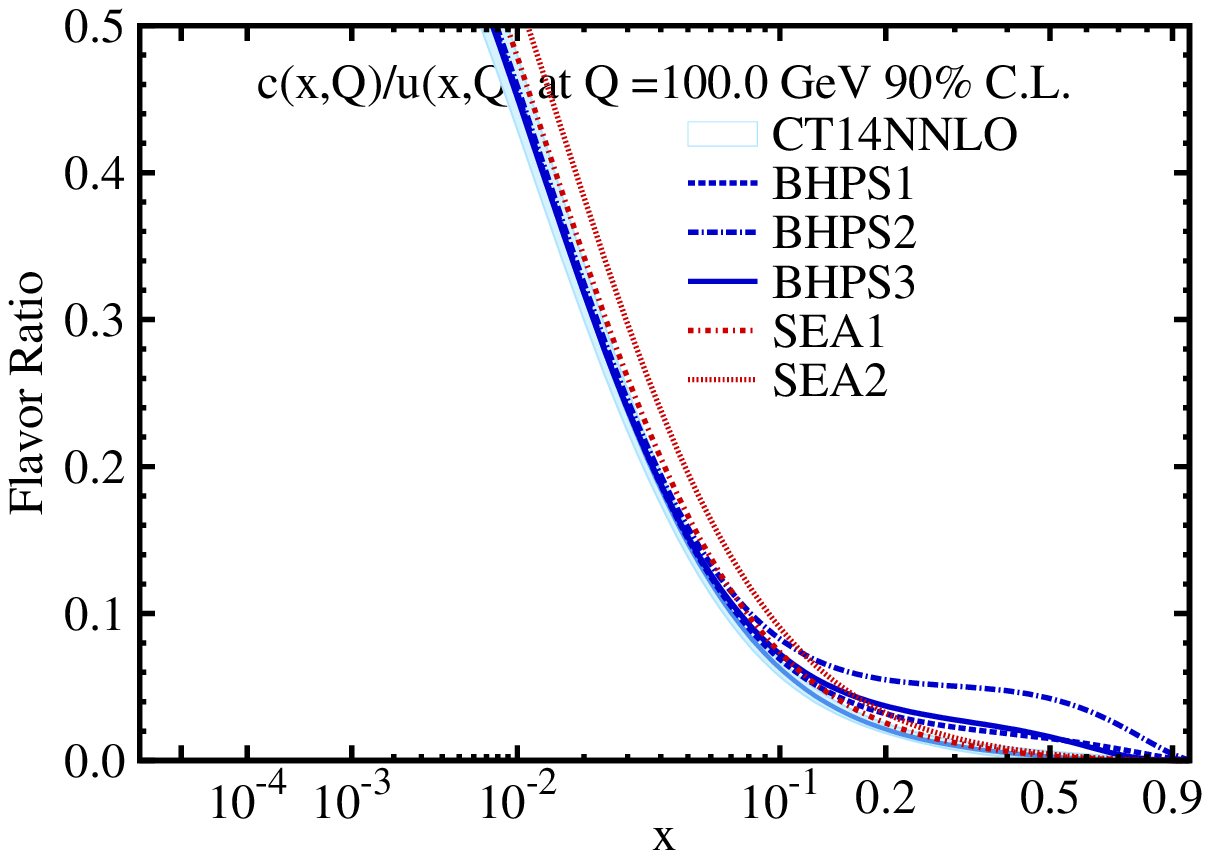} \\
\includegraphics[width=0.49\textwidth]{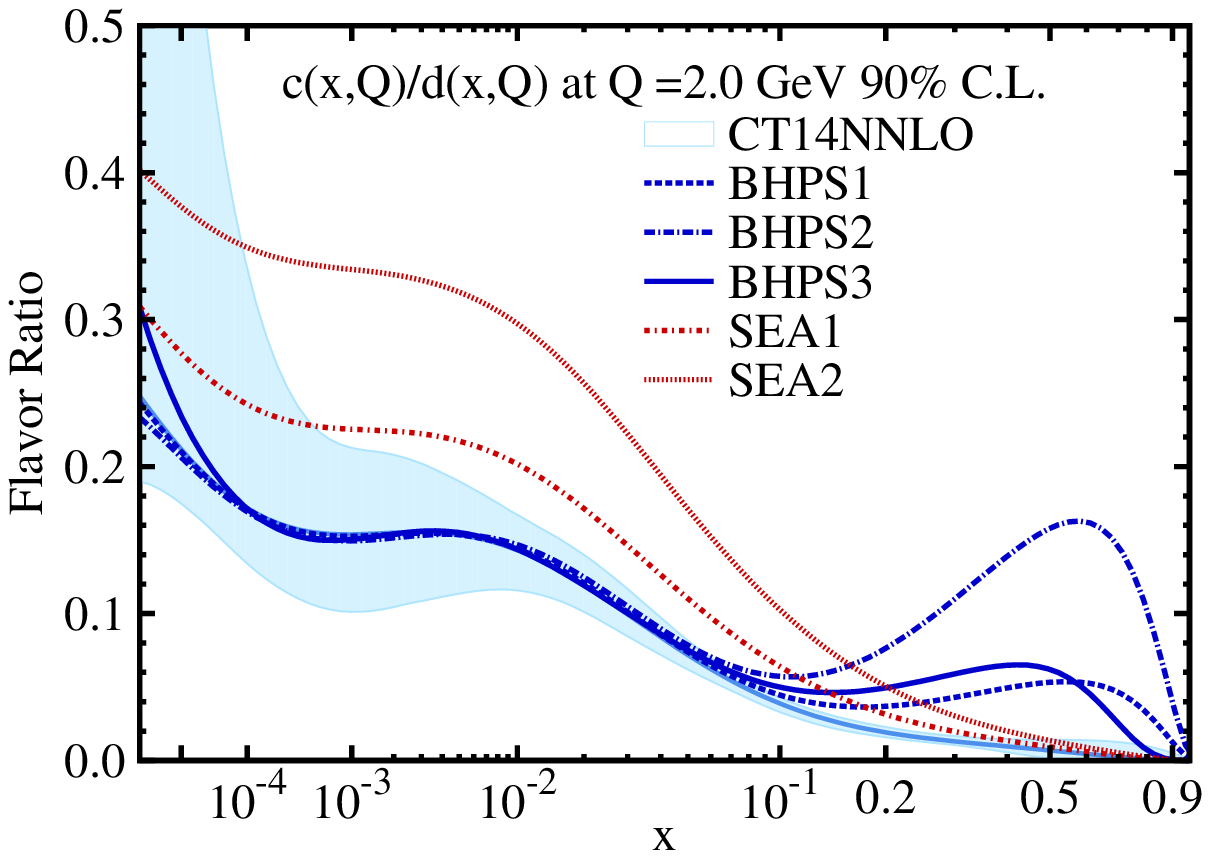}
\includegraphics[width=0.49\textwidth]{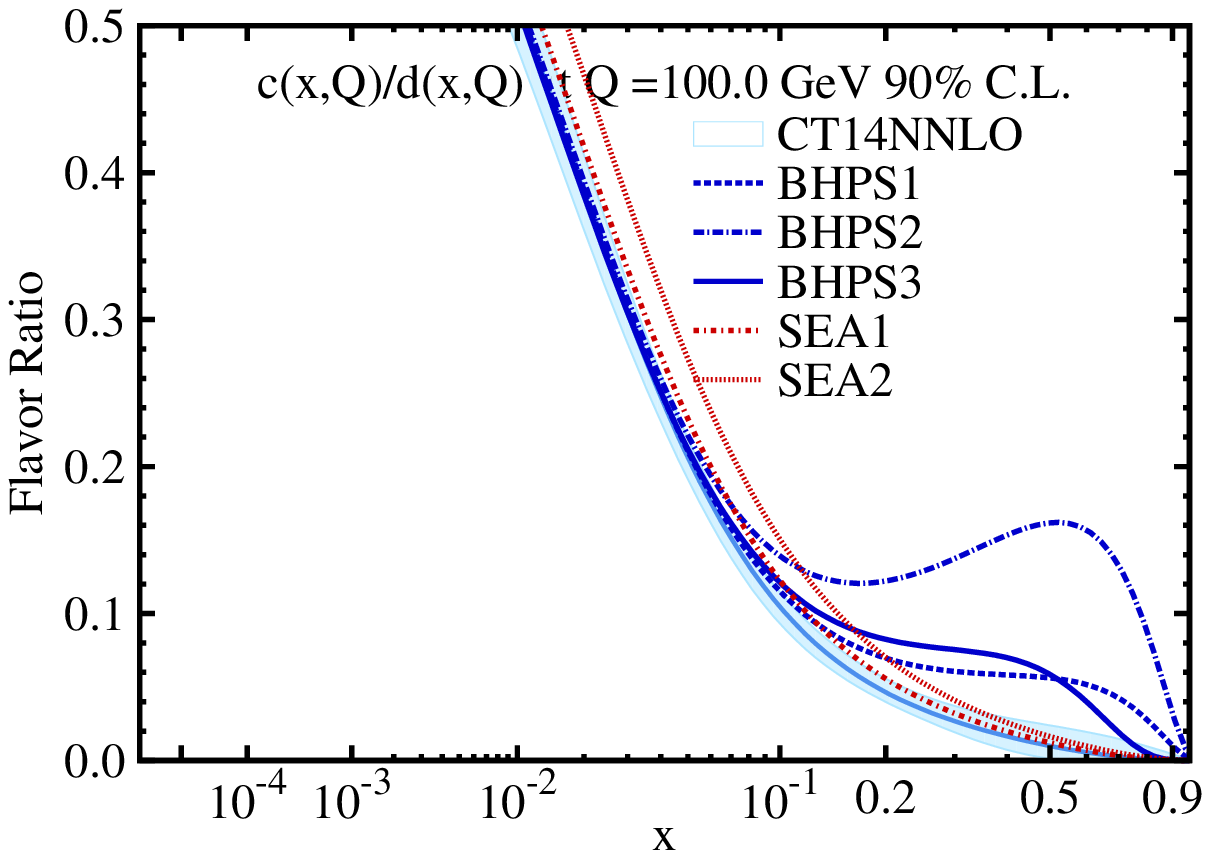}
\caption{Left column: BHPS and SEA models within the CT14 PDF uncertainty at 90\% C.L. in the
charm-quark fraction $\left(c(x,Q)+\bar c(x,Q)\right)/\left(\bar u(x,Q)+\bar d(x,Q)\right)$ (upper), $c(x,Q)/u(x,Q)$ (middle),
and $c(x,Q)/d(x,Q)$ (lower), at $Q=2$ GeV.
Right column: same as left, but at $Q=100$ GeV.
\label{fig:other_ratios}}
\end{figure}

Additional insights can be gathered by examining the ratios of the
charm-quark PDF to other flavors: $\left(c(x,Q)+\bar
  c(x,Q)\right)/\left(\bar u(x,Q) +\bar d(x,Q)\right)$,
$c(x,Q)/u(x,Q)$,  and
$c(x,Q)/d(x,Q)$. These ratios are plotted versus
$x$ in Fig.~\ref{fig:other_ratios},
for two different values of the $Q$ scale. Also shown for a comparison
are the corresponding CT14 PDF uncertainty bands.

For $(c+\bar c)/(\bar u + \bar d)$,
all the BHPS and SEA models reproduce the shape of CT14 at low $x$,
with the ratios in the SEA models shifted upwards. The SEA models
retain the shape of CT14  (but with a larger normalization)  at higher
$x$ as well. All BHPS ratios start to
rise quickly in the range $0.1 \lesssim x\lesssim 0.2$. This rise is
essentially unabated at $x>0.2$ for the BHPS1 and BHPS2 models, because
their respective parametrizations for $\bar{u}$ and $\bar{d}$
fall off as $(1-x)^d$ and are more strongly suppressed at
$x\rightarrow1$ than the BHPS charm quark PDF.  Inclusion of
the intrinsic $\bar{u}$ and $\bar{d}$ components in the BHPS3 model,
together with the numerical estimation of the BHPS integrals for the
$\bar c$, $\bar u$, and $\bar d$ intrinsic parametrizations,
results in a softer BHPS3 $(c+\bar c)/(\bar u + \bar d)$ ratio at large $x$ with
a bump residing at $x\approx 0.5$.
The exact amount of suppression at $x>0.5$ can be
determined, {\it e.g.}, by a fit to the numerical solutions of the
BHPS3 model. In particular, we find that a 6-parameter fit using $f(x)\propto x^{p_1}(1-x)^{p_2}(1+p_3x^{p_4}+p_5 x +p_6 x\ln{(x)})$,
gives a large-$x$ suppression power $p_2\approx 8,9,10$ for intrinsic ${\bar c}$, ${\bar d}$, and ${\bar u}$, respectively.

The $c(x,Q)/u(x,Q)$ ratios in all BHPS models agree with CT14
over the range $10^{-5} \lesssim x\lesssim 0.1$ and exhibit
a bump (most prominent for BHPS2) at $x\approx 0.5$.
The SEA model ratios are notably larger than CT14 in the range $10^{-5} \lesssim x\lesssim 0.3$ and approach CT14 for larger $x$-values.
At higher scale, $Q=100$ GeV, all models are closer to CT14 over the range $10^{-5} \lesssim x\lesssim 0.1$ with the exception of SEA2,
while the bump in the BHPS models at $x\approx 0.5$ are slightly suppressed.
The $c(x,Q)/d(x,Q)$ ratio plot shows essentially the same features as the
$c(x,Q)/u(x,Q)$ plot, with the difference that the bumps present in the
BHPS1, BHPS2 and BHPS3 models, at $x\approx 0.5$, are much more pronounced.

\begin{figure}[tb]
\centering
\includegraphics[width=0.49\textwidth]{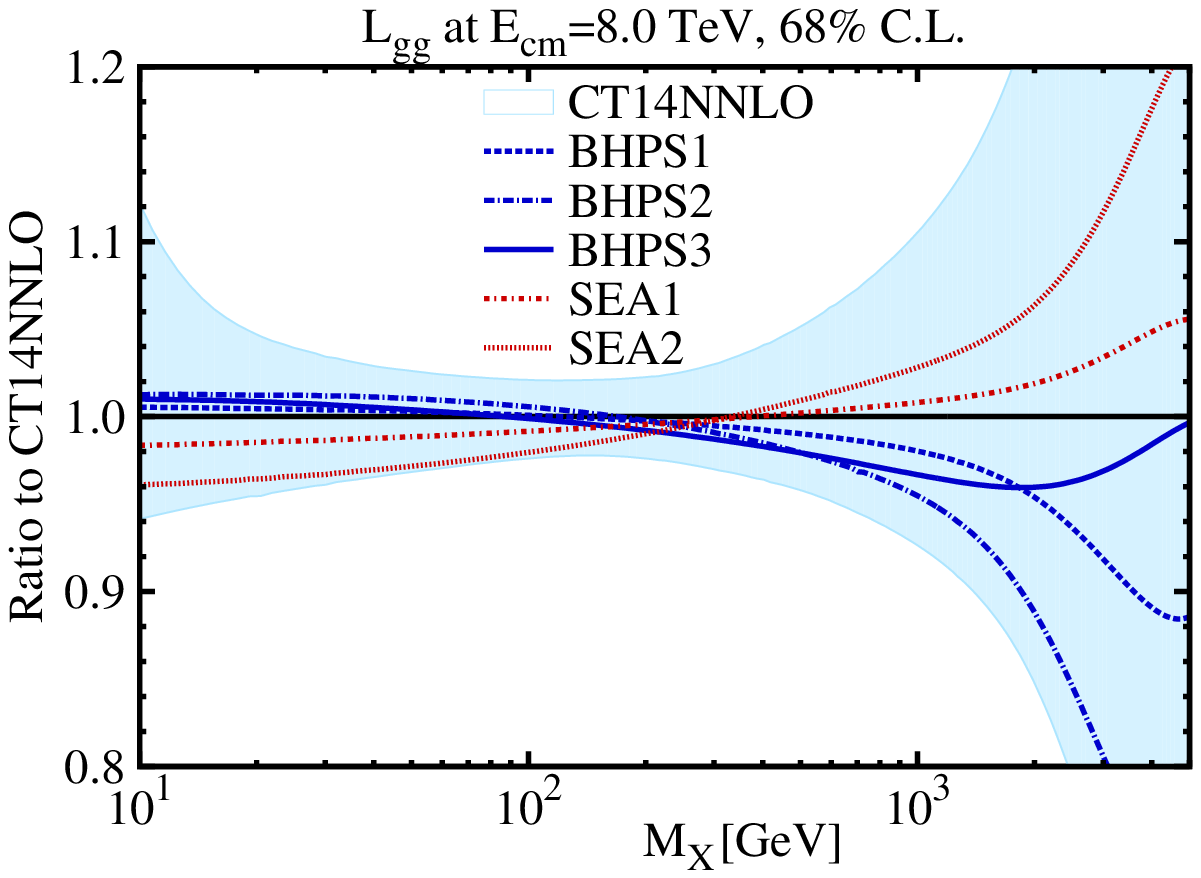}
\includegraphics[width=0.49\textwidth]{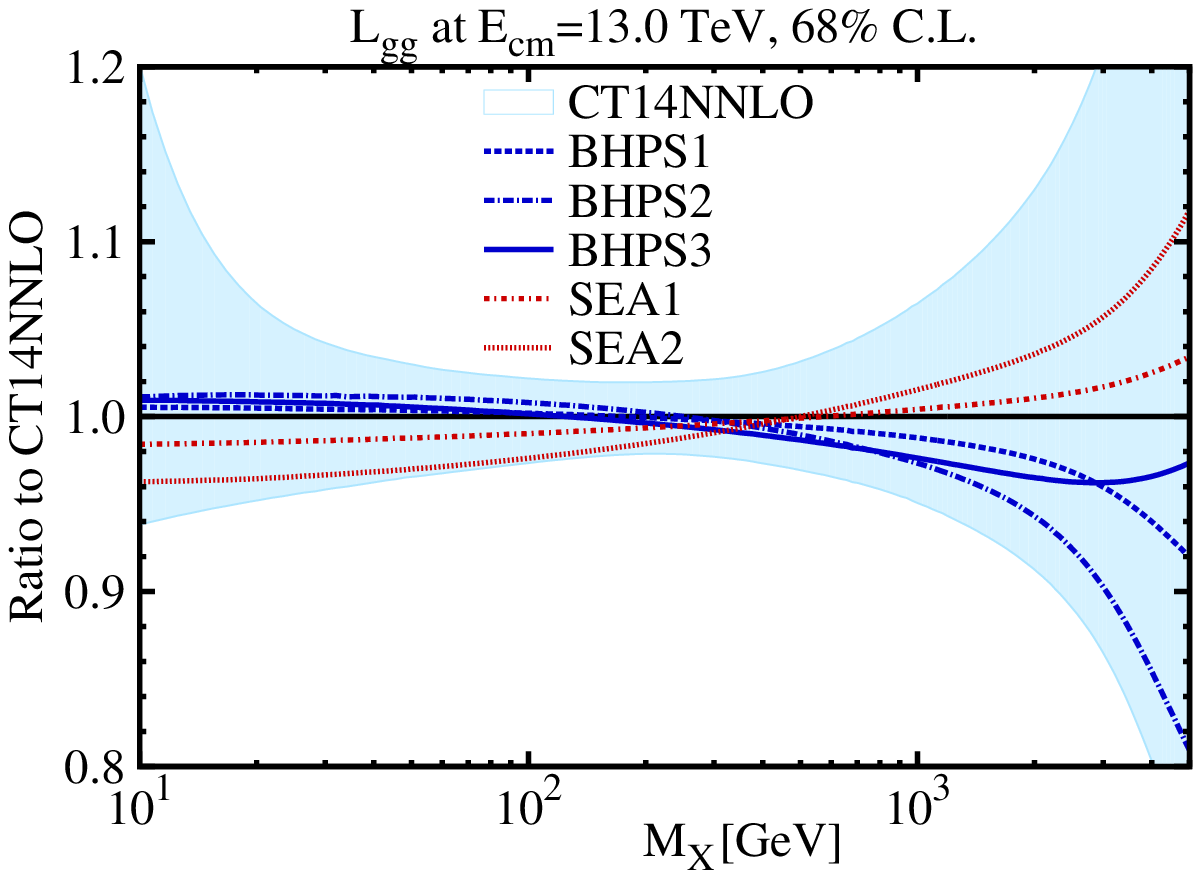} \\
\caption{Impact of the BHPS and SEA models on the gluon-gluon
  luminosity as a function of the invariant mass $M_X$ of a
  hypothetical massive final state $X$. The predictions are
  normalized to the CT14NNLO central PDF set. The shaded bands
  indicate the CT14 uncertainty at 68\% C.L.
\label{ggLumi}}
\end{figure}

An additional charm component (either a sea-like or valence-like one)
affects both those LHC predictions that directly
involve charm quarks in the initial state, and those that do not.
In Fig.~\ref{ggLumi} we show how the gluon-gluon luminosity is affected by BHPS and SEA models
at LHC run I and II energies in the $x$ range sensitive to Higgs production. The parton luminosity is defined
as in Ref.~\cite{Campbell:2006wx}. The various models, shown as ratios
to CT14NNLO, are well within the 68\% C.L. PDF uncertainty.
At $\sqrt{s}=8$ TeV the most prominent deviations are for the SEA2
model, which is suppressed
at lower $\textrm{M}_X$ and is notably larger than CT14 for
$\textrm{M}_X$ in the TeV range.
The BHPS models are almost coincident with CT14 for the invariant mass
$\textrm{M}_X<200$ GeV:
BHPS1 and BHPS2 are highly suppressed above $\textrm{M}_X>300$ GeV,
while BHPS3 is suppressed for $0.3<\textrm{M}_X<3$ TeV and enhanced
above this energy by approximately 3\%.
The impact on the Higgs cross section is small, the influence on the
high-mass $gg$ PDF luminosities is more pronounced,
but still within uncertainties.

\subsection{Agreement with experimental data sets \label{sec:DataImpact}}

In this section we focus on the data sets whose goodness-of-fit values are
affected by the introduction of the intrinsic charm component.
These are selected by computing an effective Gaussian variable, $S_n$,
for each experiment $n$, according to the method
introduced in Refs.~\cite{Lai:2010vv,Gao:2013xoa,Dulat:2013hea}.

For specifications of $S_n$, we refer the reader
to the appendix of Ref.~\cite{Dulat:2013hea}.
$S_n$ maps the goodness-of-fit $\chi^2_n$
for a particular data set, assumed to obey the chi-square probability
distribution with $N_{\rm pts}$ data points, onto a variable $S_n$, which
obeys a standard normal distribution independently of $N_{\rm pts}$.
More precisely, $S_n$ is defined so that
the cumulative standard normal distribution evaluated at $S_n$ equals
the cumulative $\chi^2(\chi^2_n, N_{\rm pts})$ distribution evaluated at
$\chi^2_n$.
We adopt an accurate approximation for $S_n$ given by
\begin{eqnarray}
&&S_n\approx L(\chi^2_n,N_{\rm pts}), \nonumber
\\
&&L = \frac{\left(18 N_{\rm pts}\right)^{3/2}}{18 N_{\rm pts} +1}
\left\{\frac{6}{6-\ln(\chi^2_n/N_{\rm pts})} - \frac{9 N_{\rm pts}}{9 N_{\rm pts}-1}\right\}.
\label{Sndef}
\end{eqnarray}
The $S_n$ distribution over the individual data set characterizes the
agreement with the totality of the fitted experiments,
regardless of their numbers of data points.
Conversely, a naive use of the global $\chi^2$ as the only
discriminating variable may give too much weight to the
data sets with large numbers of data points, even if the correlations with
the fitting parameters are not very significant.

The values of $S_{n}$ can easily be interpreted in terms
of the probabilities associated with a normal distribution.
Fits with $S_{n}$ between -1 and 1 are accepted as reasonable,
within the 68\% C.L. uncertainties. That is, an increase of $S_n$ by 1
has about the same significance (68\%) as the increase of
$\chi^2_n/N_{pts}$ by $\sqrt{2/N_{\rm pts}}$.
Fits with $S_{n} > 3$ are
considered poor, while those with $S_{n} < -3$
actually fit the data much better than one would expect from the
regular statistical analysis: for some reason they have anomalously small residuals.

\begin{figure}[!ht]
\begin{center}
\includegraphics[width=0.49\textwidth]{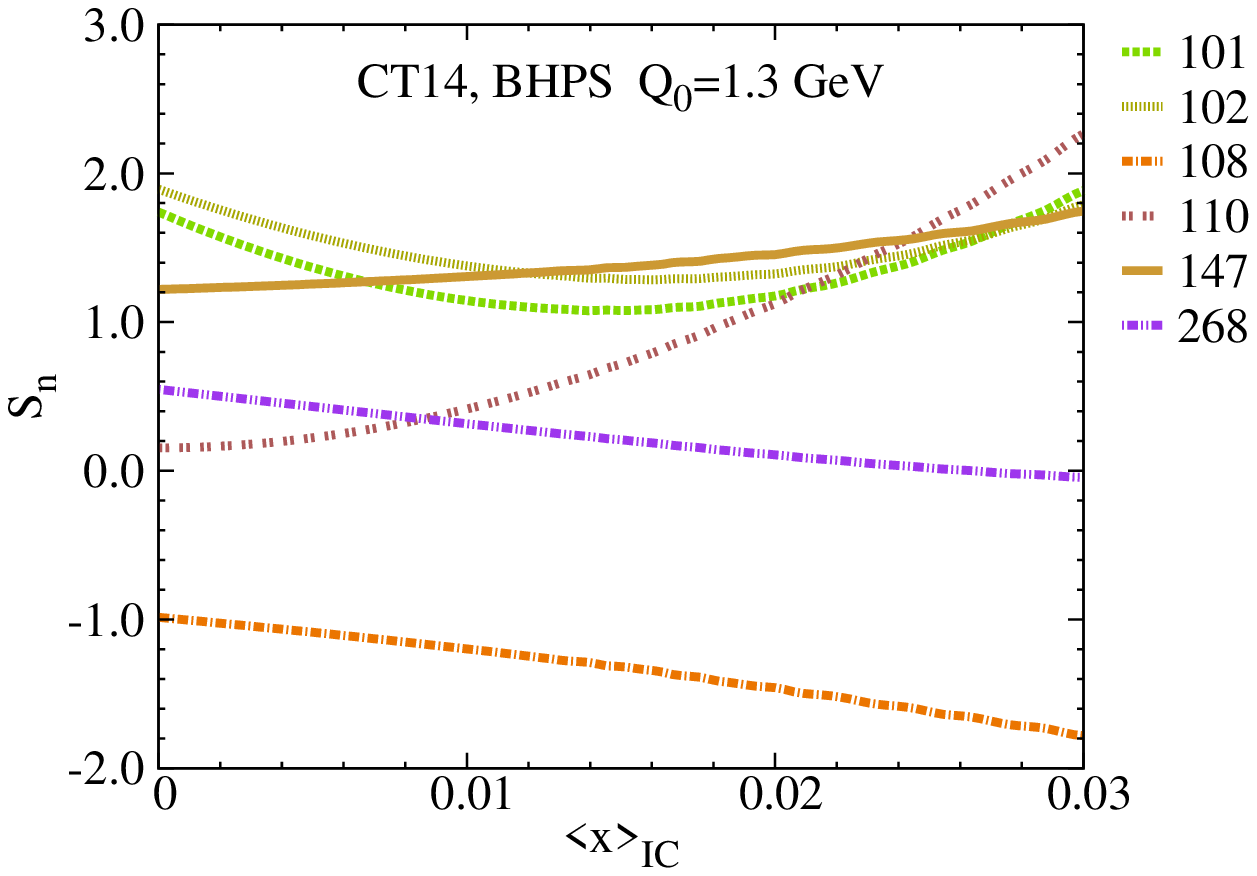}
\includegraphics[width=0.49\textwidth]{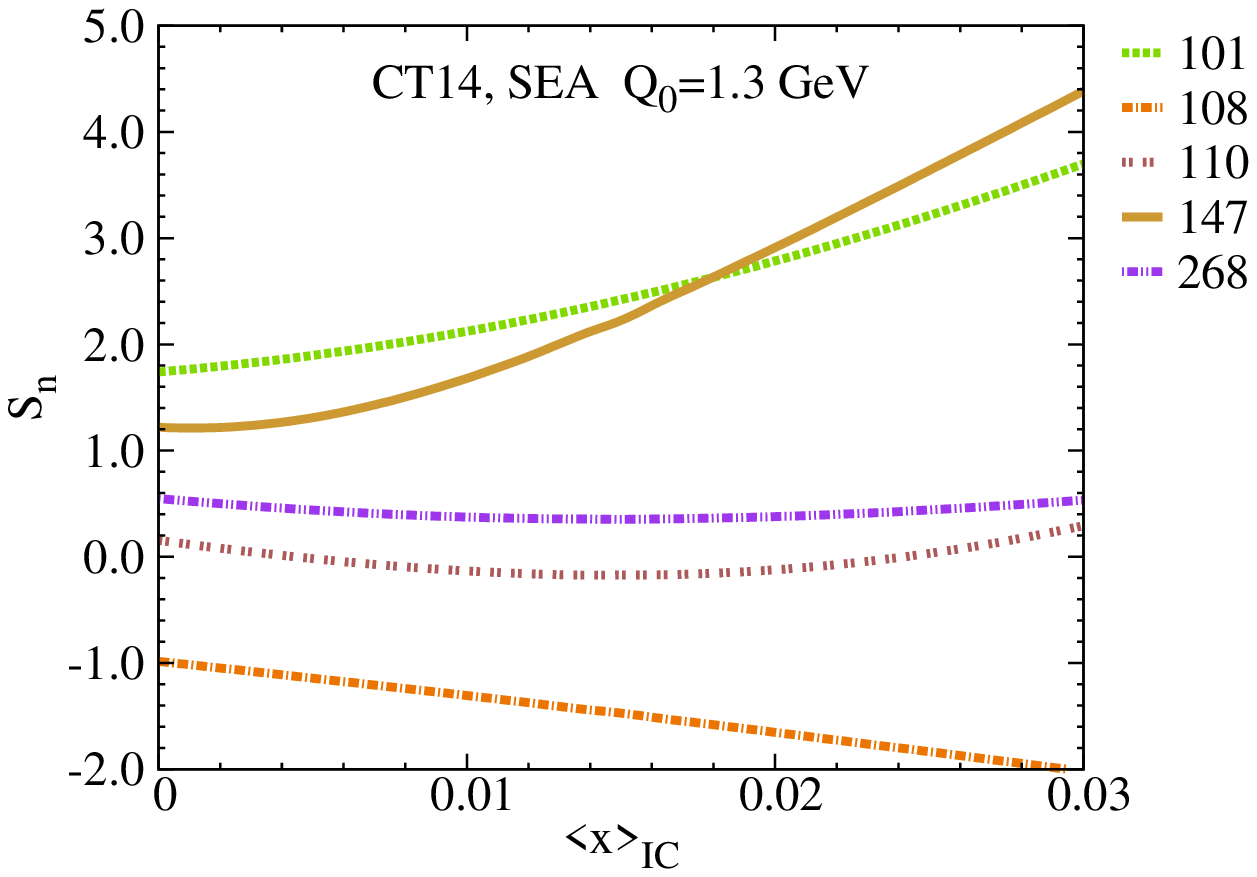}
\caption{The Gaussian variable $S_{n}$ for select experiments as a
  function of $\langle{x}\rangle_{\rm IC}$.
Left: BHPS model; Right: SEA model.
  The curves correspond to BCDMS for $F_2^p$ and $F_2^d$ (ID 101, 102);
ATLAS 7 TeV $W/Z$ cross sections (ID 268);  CC DIS measurements (ID 110);
combined HERA charm production (ID 147); and charged-current neutrino
interactions on iron CDHSW $F_2^p$ (ID 108).
\label{fig:Effective_Sn}}
\end{center}
\end{figure}

In Fig.~\ref{fig:Effective_Sn}, we selectively plot $S_{n}$ for those
data sets whose agreement with theory is most affected by the IC
in the CT14 fit with $m_c^{pole}=1.3$ GeV. $S_n$
is plotted as a function of $\langle{x}\rangle_{\rm IC}$
for both the BHPS (left) and SEA (right) models.

For the BHPS model, the most visible dependence is found for the fixed target measurements from
BCDMS for $F_2^p$ and $F_2^d$ (ID 101,
102)~\cite{Benvenuti:1989fm,Benvenuti:1989rh} and the ATLAS 7 TeV $W/Z$
cross section measurements~\cite{Aad:2011dm} (ID 268).
The E866 Drell-Yan dimuon cross section measurement~\cite{Webb:2003ps}
also shows some variation, however, its $S_n$ is always larger than 3 and
not shown in Fig.~\ref{fig:Effective_Sn}(left).
These experiments, mostly sensitive to $u$ and $d$ quarks at large $x$,
(slightly) favor a non-zero intrinsic charm component. Although the
improvement for the BCDMS $S_n$ is relatively mild, the two data sets contain a
large number of data points ($N_{\textrm{\rm pts}}=339$ for $F_2^p$ and
$251$ for $F_2^d$). The shallow minimum of 20-30 units
occurring in $\chi^2$ for the
BHPS model in Fig.~\ref{fig:delta_chisqVxic} is attributed primarily
to these two experiments; it is not clear whether it
originates from the charm component or reflects a
small admixture of the N3LO contributions
or even some residual $1/Q^2$ terms that may be present
at relatively low $Q$ and large $x$.

Continuing with BHPS, the charged-current (CC) DIS
measurement~\cite{Yang:2000ju} $F_2^p$ by
CCFR (ID 110) has $0<S_n<1$ for
$0\leq \langle{x}\rangle_{\rm IC} \leq 0.02$,
then $S_n$ increases faster for even larger
$\langle{x}\rangle_{\rm IC}$. The combined HERA charm production
~\cite{Abramowicz:1900rp} (ID 147) exhibit $1<S_n<2$ over the whole
range of $\langle{x}\rangle_{\rm IC}$.

The $S_n$ dependencies of various experiments for the SEA model are
shown in the right-side of Fig.~\ref{fig:Effective_Sn}. The HERA charm
production and BCDMS ($F_2^p$) data are very sensitive  to
$\langle{x}\rangle_{\rm IC}$ in the SEA model. A fast growth for $S_n$ is observed
for $\langle{x}\rangle_{\rm IC} > 0.01$, paralleling the increase in
$\chi^2$ observed in Fig~\ref{fig:delta_chisqVxic}. Experiment 108 (charged-current neutrino DIS on iron by CDHSW \cite{Berge:1989hr}) does not impose strong constraints in either model, as it is already fit very well ($S_n \approx -1$). Its $\chi^2$ exhibits mild improvement for larger values of $\langle x\rangle_{IC}$.
Similar conclusions can be drawn for the CT14HERA2 fits and
when $m_c$ is varied as in Sec.~\ref{sec:McDependence}.

\subsection{A global analysis including the EMC charm DIS measurements
\label{sec:EMCdata}}

The measurement of
semi-inclusive dimuon and trimuon production in DIS on an iron target
by the  the European Muon Collaboration
(EMC)~\cite{Aubert:1982tt} has been investigated by various groups
for indications of BHPS-like contributions from the IC.
This data set, published in 1983, did not follow the stringent criteria
on the documentation of systematic uncertainties adopted in
more recent studies; therefore, there is a lack of
the control on the constraints that these data may
impose. This is why the EMC measurements are not included in the
CTEQ PDF analyses, whose policy is to include only data
with documented systematic errors. Moreover, the EMC analysis has been
done at the leading order of QCD,
clearly insufficient for accurate conclusions at NNLO.
Despite the tensions\footnote{See, for example, discussions in the early CTEQ analyses~\cite{Lai:1994bb,Morfin:1990ck}.} stated
between the EMC measurement and its contemporary experiments
in the case of inclusive DIS~\cite{Aubert:1982hn,Arneodo:1996qe,Benvenuti:1989fm,Benvenuti:1989rh}
and semi-inclusive charm DIS production cross
sections~\cite{Aubert:1982tt,Clark:1980pe}\footnote{Keep in mind that EMC
  employed non-identical detection techniques in the measurement
  of the inclusive
 structure functions $F_2$~\cite{Aubert:1982hn} and semi-inclusive
 $F_2^{c}$~\cite{Aubert:1982tt}.},
various studies~\cite{Hoffmann:1983ah,Vogt:1995tf,Vogt:1995fsa,Vogt:1994zf,Harris:1995jx} have interpreted the excess seen in a few
high-$x$ bins of the EMC $F_{2c}(x,Q)$ data as evidence for some
nonperturbative charm contribution, while yet
other studies concluded the opposite~\cite{Steffens:1999hx, Jimenez-Delgado:2014zga, Jimenez-Delgado:2015tma}.
Our special series of the CT14 IC
fits included the EMC $F_{2c}(x,Q)$ data to investigate the above
conclusion. We observe that the EMC $F_{2c}$ data do not
definitively discriminate between the purely perturbative and intrinsic
charm models, hence we do not include them in the final
CT14 BHPS and SEA fits. However, it is still useful
to examine how the EMC data could possibly affect the
amount of the intrinsic charm-quark content,
especially given their emphasis in a recent NNPDF study~\cite{Ball:2016neh}.

\begin{table}[tb]
\begin{center}
\begin{footnotesize}
\begin{tabular}{ |l|l|l|l|l|}
  \hline
  \multicolumn{2}{|c|}{Candidate NNLO PDF fits}{~~~~~$\chi^2/N_{\rm pts}$}\\
  \hline
  \hline
   & All Experiments &  HERA inc. DIS & HERA $c\bar{c}$ SIDIS & EMC $c\bar{c}$ SIDIS \\
  \hline
  CT14 + EMC (weight=0), no IC  &  1.10    &   1.02    &    1.26      &            3.48\\
  \hline
  CT14 + EMC (weight=10), no IC &  1.14    &   1.06    &    1.18       &           2.32\\
  \hline
  CT14 + EMC in BHPS model              & 1.11   &  1.02    &    1.25          &        2.94\\
  \hline
  CT14 + EMC in SEA model               & 1.12   &   1.02    &    1.28        &          3.46\\
  \hline
  \hline
  CT14 HERA2 + EMC (weight=0), no IC & 1.09  & 1.25 & 1.22 & 3.49\\
  \hline
  CT14 HERA2 + EMC (weight=10), no IC & 1.12  & 1.28 & 1.16 & 2.35 \\
  \hline
  CT14 HERA2 + EMC in BHPS model              & 1.09  & 1.25 & 1.22 & 3.05  \\
  \hline
  CT14 HERA2 + EMC in SEA model              & 1.11  & 1.26 & 1.26 & 3.48  \\
  \hline
\end{tabular}
\end{footnotesize}
\caption{$\chi^2/N_{\rm pts}$ for all experiments, the HERA inclusive DIS data, HERA $c\bar{c}$ SIDIS data, and EMC $F_{2c}$ data in representative fits.
\label{TAB1}}
\end{center}
\end{table}

Our findings concerning the fit to the EMC data can be summarized as
follows.

\subsubsection{$\chi^2$ values for the EMC data  set\label{sec:chi2EMC}}
Either by fitting to the EMC $F_{2c}$ data or not, we obtain
$\chi^2/N_{\rm pts}$ between 2.3 and 3.5 for the EMC data set
in various candidate fits.
So, for their nominal experimental errors,
the EMC data is in general not fit well in either CT14 or CT14HERA2 setup,
regardless of the charm model. On the other hand, these $\chi^2/N_{\rm pts}$
values are not dramatically high, it may be argued that allowing for a modest
systematic error would improve the agreement to tolerable
levels. One way or another, the unknown systematics of this
measurement prevents us from concluding for or against the
preference of the EMC $F_{2c}$ data for a particular charm model.
To show an example of this, Table~\ref{TAB1} reports the values of
$\chi^2/N_\textrm{\rm pts}$ for all experiments, HERA inclusive DIS, HERA
charm SIDIS, and EMC charm SIDIS in the CT14 (CT14HERA2) NNLO IC candidate
fits in the upper (lower) half of the table.

The first two lines in
each half present the fits without the nonperturbative charm. When
$\chi^2$ for the EMC $F_{2c}$ data is included with weight 0 (so that
the EMC $F_{2c}$ data has {\it no} effect on the PDFs), we obtain
$\chi^2/N_{\rm pts} \approx 3.5$ -- it is quite poor. When the EMC weight
is increased to 10 to emphasize its pull, $\chi^2/N_{\rm pts}$ decreases
to 2.4, at the cost of a worse $\chi^2$ for the inclusive HERA I+II data and
other experiments, and somewhat better $\chi^2$ for charm DIS hadroproduction.
Again the quality of the fits is poor, yet it is also compatible with the
possibility of moderate unaccounted systematic errors,
as those are unknown in the EMC case.

We can also see from Table~\ref{TAB1} that including the BHPS
intrinsic charm does not qualitatively change the fit to the EMC data.
Without the IC, the $\chi^2$ for all experiments
slightly grows if we increase the weight of the EMC data set;
with the BHPS intrinsic charm, there seem to be no effect with and
without the EMC data, as $\chi^2$ does not change in either case.
In the SEA model fit, inclusion of the EMC data results in a larger $\chi^2$
with respect to the fits without the intrinsic charm;
description of both HERA inclusive DIS and HERA combined charm SIDIS
production deteriorates.
To summarize, in all considered intrinsic charm models (BHPS,
SEA, and the mixed model that produces a similar outcome),
the intrinsic charm has no decisive effect on improving the fit to the EMC data.

\subsubsection{Constraints from EMC on the IC momentum
  fraction \label{sec:EMCxIC}}

\begin{figure}[tb]
\begin{center}
\includegraphics[width=0.49\textwidth]{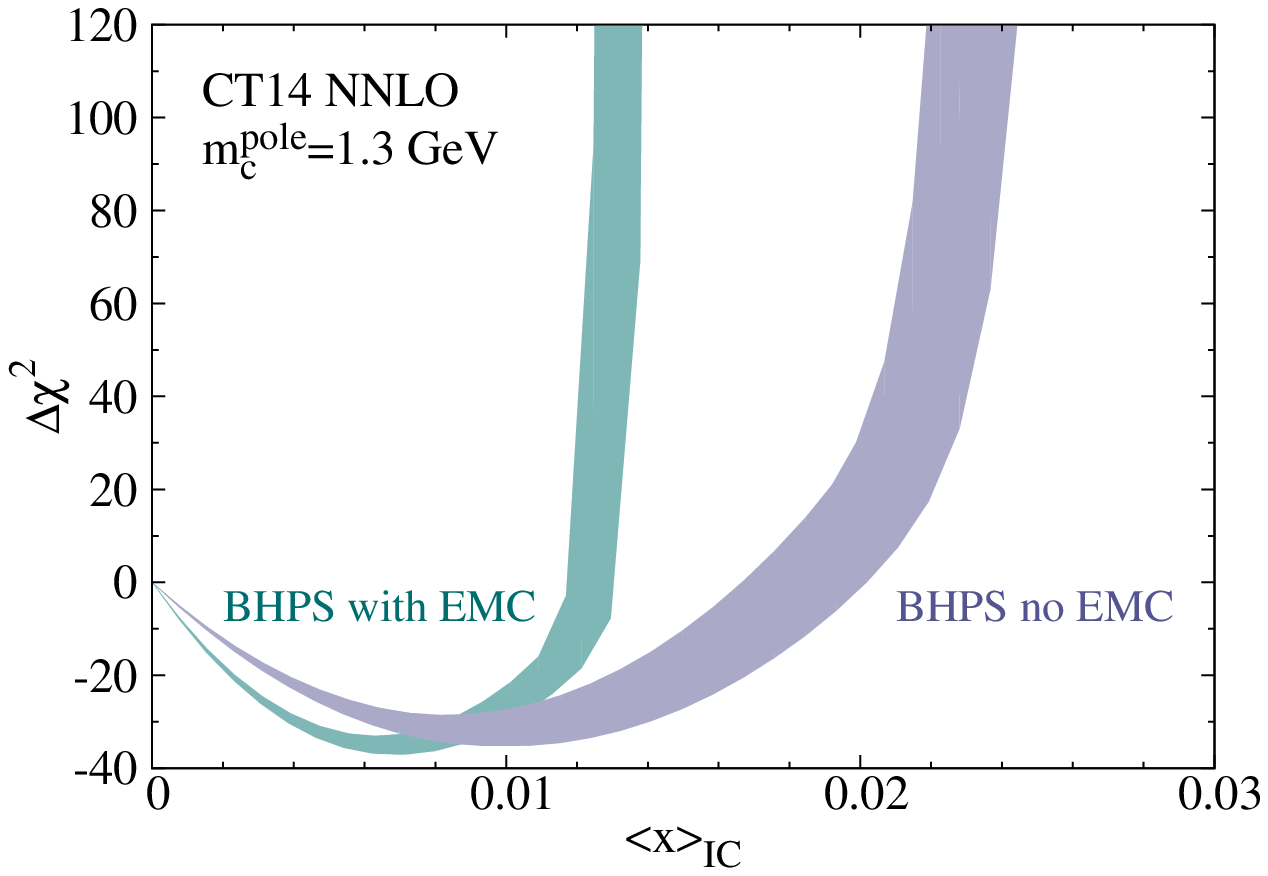}
\includegraphics[width=0.49\textwidth]{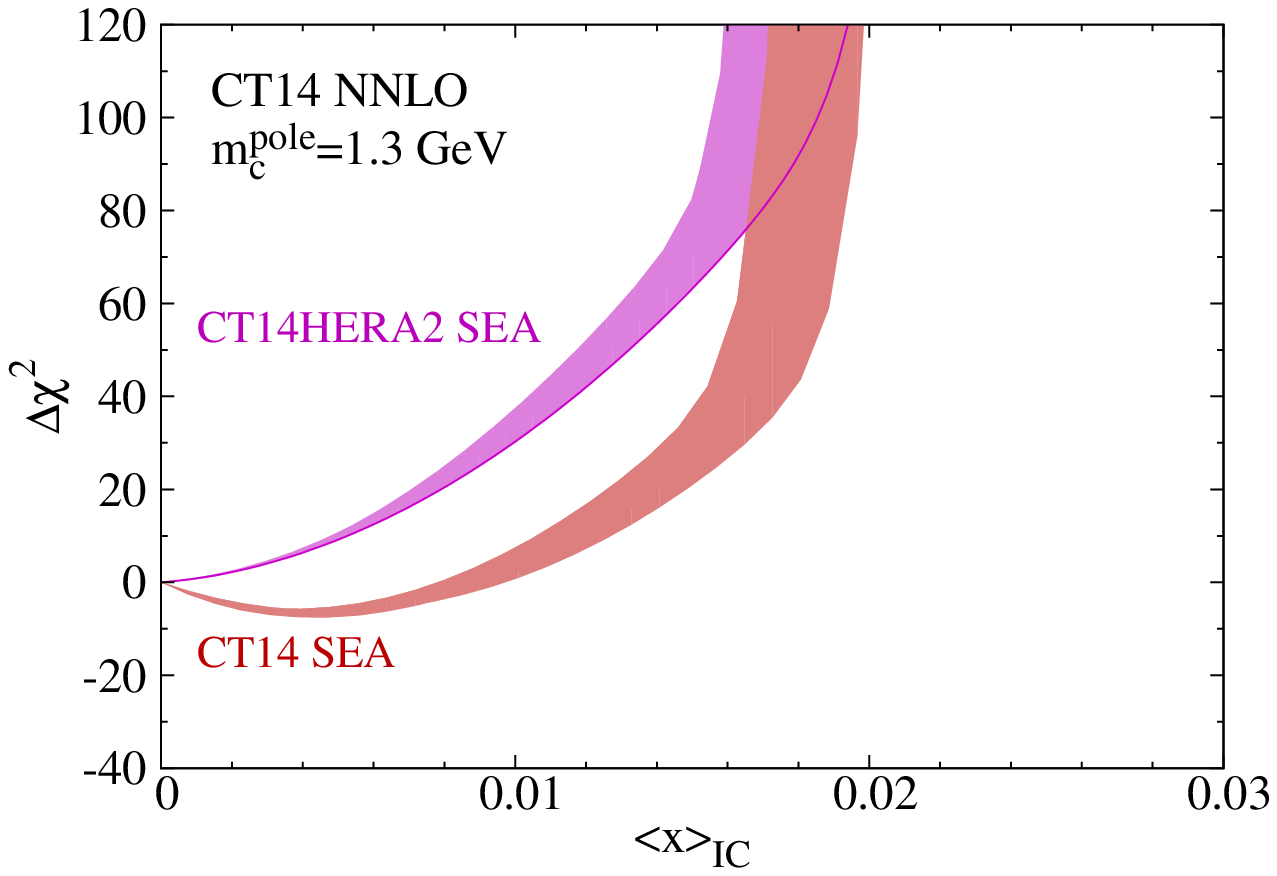}
\end{center}
\caption{$\Delta\chi^2$ as function of $\langle x \rangle_{\rm IC}$
in fits with and without the EMC data for both the BHPS
and SEA models for $m_c^{pole}=1.3$ GeV. For the BHPS model (left), the two bands are from the fits with and without the EMC data.
For the SEA model (right), the bands are from the CT14 and CT14HERA2 fits with the EMC data.
\label{fig:scan-EMC}}
\end{figure}

Figure~\ref{fig:scan-EMC} compares
the dependence of $\Delta\chi^2$ on $\langle x \rangle_{\rm IC}$ in
the context of the CT14 and CT14HERA2 global analyses with and
without EMC data. It must be noted upfront that, since the EMC
$F_{2c}(x,Q)$ data set are not well described, these $\Delta \chi^2$
scans do not establish clear-cut constraints on $\langle x \rangle_{\rm
  IC}$, contrary to the CT14 IC fits without the EMC data set
that were presented earlier.

The outcomes shown here are for
$m_c^{pole}=1.3$ GeV and remain analogous for the other
$m_c^{pole}$ values. The bands of various shades
illustrate the spread in $\Delta\chi^2$ values
induced by: (a) the choice of different data sets and strangeness
parametrization used in CT14 and CT14HERA2, and (b) various gluon
PDF parametrizations utilized.

For the BHPS model in Fig.~\ref{fig:scan-EMC} (left), we observe two
distinct trends in the fits with and without the EMC data. The spread
in the $\Delta\chi^2$ band without the EMC data is mostly driven by the
differences in the data sets and in the strangeness parametrization
between CT14 and CT14HERA2 (the dependence on the gluon
parametrization is weak). Meanwhile, after including the EMC data, the
spread due to the gluon parametrization dependence is much larger
and gives the major contribution to the band.
The BHPS model is affected more by the EMC data,
the $\Delta \chi^2$ band narrows near the minimum when these data are included.
The $\chi^2$ minimum with the EMC data moves to a lower value of $\langle
x \rangle_{\rm IC}\approx 0.006$, with substantially the same $\chi^2$
(same depth) at the minimum. The nominal upper limit on
$\langle x \rangle_{\rm IC}$ moves to about 0.012; its exact
location is debatable because of the overall poor quality of the EMC fit,
see above.

To contrast with the BHPS case, in the SEA model in
Fig.~\ref{fig:scan-EMC} (right), the $\Delta \chi^2$ behavior
is only mildly impacted by the EMC data. As already discussed
in Section~\ref{sec:xICDependence} and shown in
Fig.~\ref{fig:delta_chisqVxic}, the $\Delta\chi^2$ trend in the SEA
model is mostly affected by the differences
between the CT14 and CT14HERA2 fits. The EMC data do not change this
trend. Both minima are shallow and higher than in the BHPS case.

\begin{figure}[tb]
\begin{center}
\includegraphics[width=0.49\textwidth]{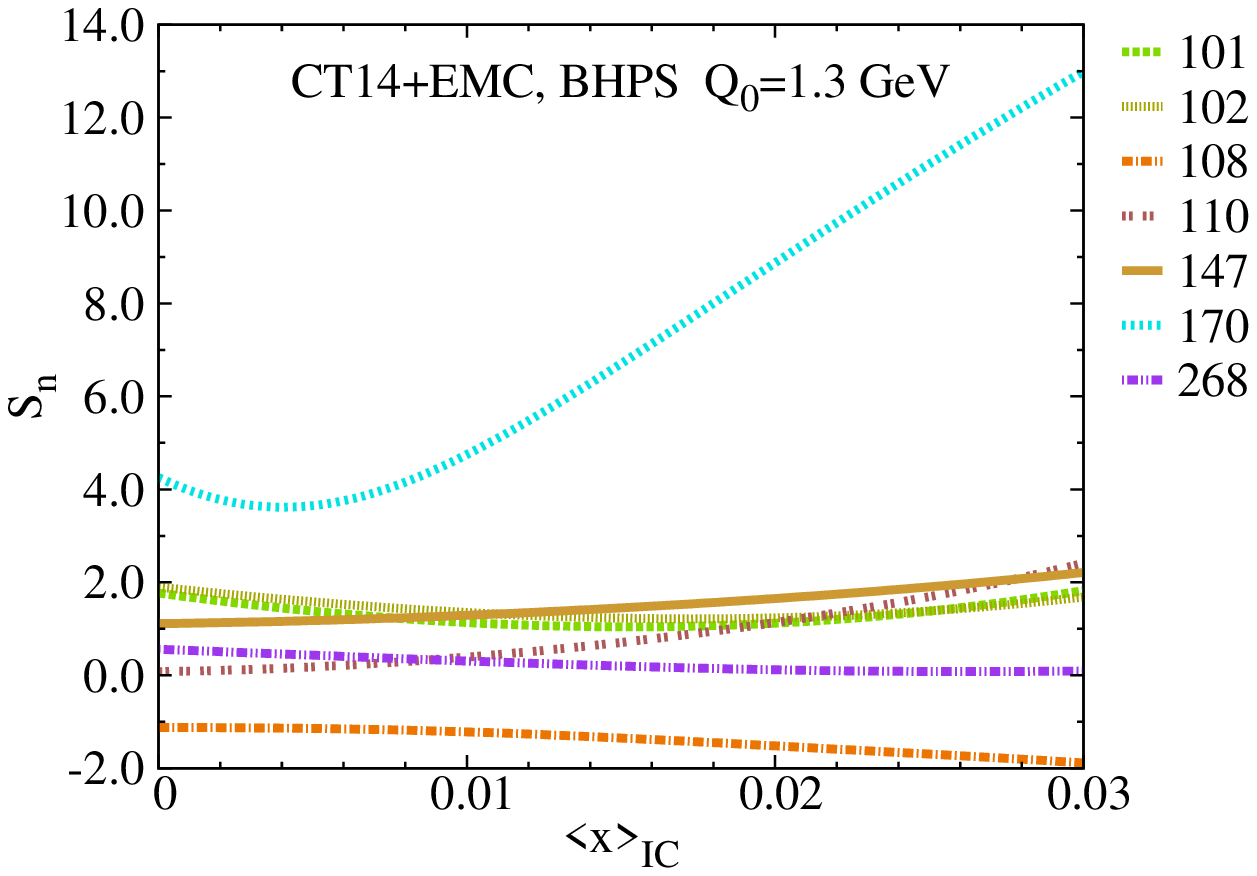}
\includegraphics[width=0.49\textwidth]{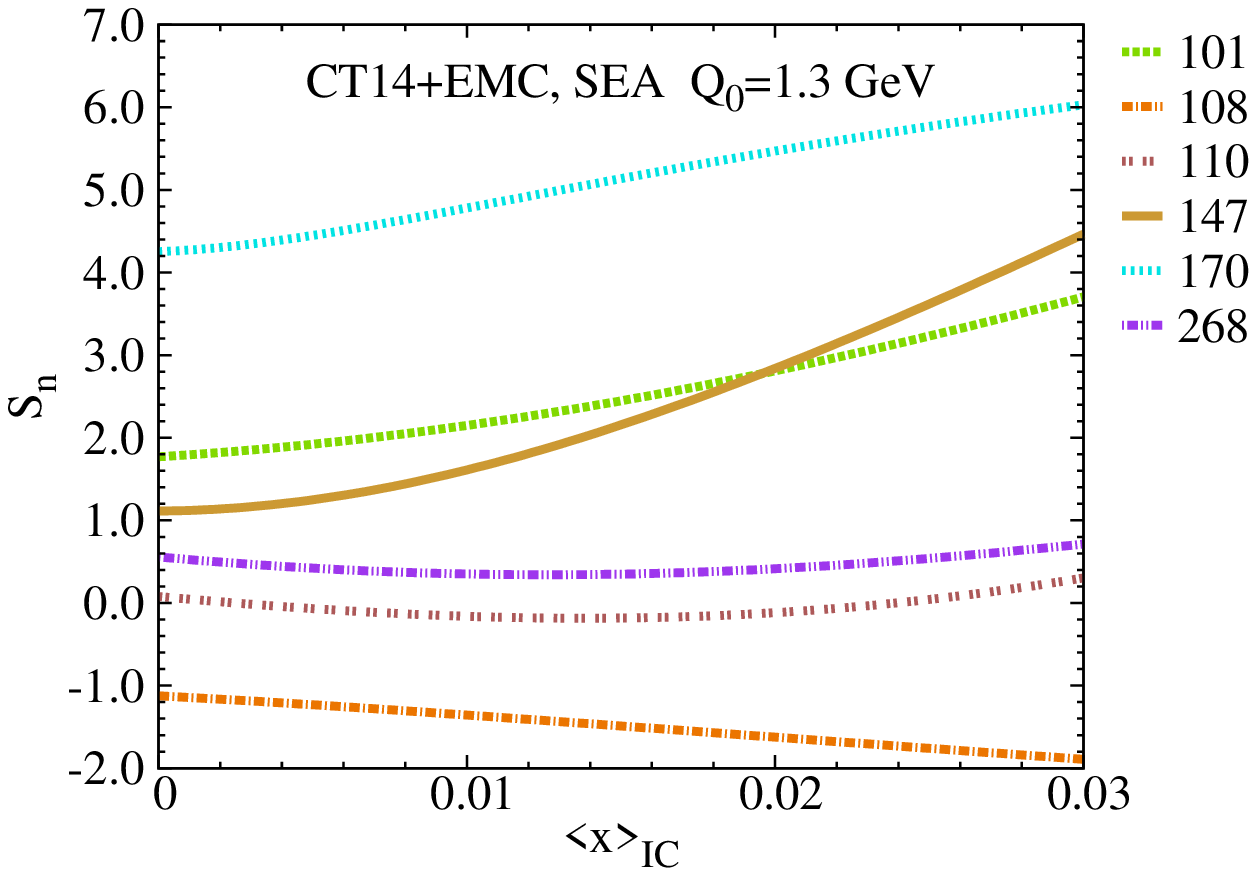}
\end{center}
\caption{The $S_{n}$ variable as a function of $\langle{x}\rangle_{\rm IC}$ for the
  BHPS (left) and SEA (right) models. The curves correspond to $S_n$ for
  EMC $F^c_2$ (data set ID 170);
BCDMS for $F_2^p$ and $F_2^d$ (ID 101, 102);
ATLAS 7 TeV $W/Z$ cross sections (ID 268); CC DIS measurements (ID 110);
combined HERA charm production (ID 147); charged-current neutrino interactions on iron CDHSW $F_2^p$ (ID 108).
\label{fig:EMC_Sn}}
\end{figure}

The Gaussian variables $S_n$ quantifying the agreement with the
individual data sets are shown for the CT14 fits and
for various $\langle x\rangle_{\rm IC}$ values in  Fig.~\ref{fig:EMC_Sn}.
[The behavior of $S_n$ in the CT14HERA2 fit is largely analogous.]
In this figure we selected only the experiments that have pronounced
dependence on $\langle x\rangle_{\rm IC}$.

Comparing Fig.~\ref{fig:EMC_Sn} with Fig.~\ref{fig:Effective_Sn} in
which the EMC data are not included, one sees that the dependence of
$S_n$ for the non-EMC experiments on $\langle x\rangle_{\rm IC}$  does
not qualitatively change upon the inclusion of the EMC. The $S_n$
value for the EMC $F_{2c}$, indicated as ``experiment ID 170'', is very
high for any $\langle x\rangle_{\rm IC}$. In the BHPS
model in Fig.~\ref{fig:EMC_Sn} (left), the $S_n$ variable for the EMC
experiment increases rapidly past $\langle x \rangle_{\rm
  IC}$ of about 0.005, up to very high values at $\langle x
\rangle_{\rm IC}=0.03$. The tier-2 contribution associated with the
rapid increase of this $S_n$ above 6 produces the rapid rise of the global
$\Delta \chi^2$ for $\langle x \rangle_{\rm IC} > 0.01$ in
Fig.~\ref{fig:scan-EMC}. In the SEA model in Fig.~\ref{fig:EMC_Sn}
(right), we observe $S_n > 4$ for the EMC regardless of $\langle
x\rangle_{\rm IC}$.

\begin{figure}[tb]
\begin{center}
\includegraphics[width=0.48\textwidth]{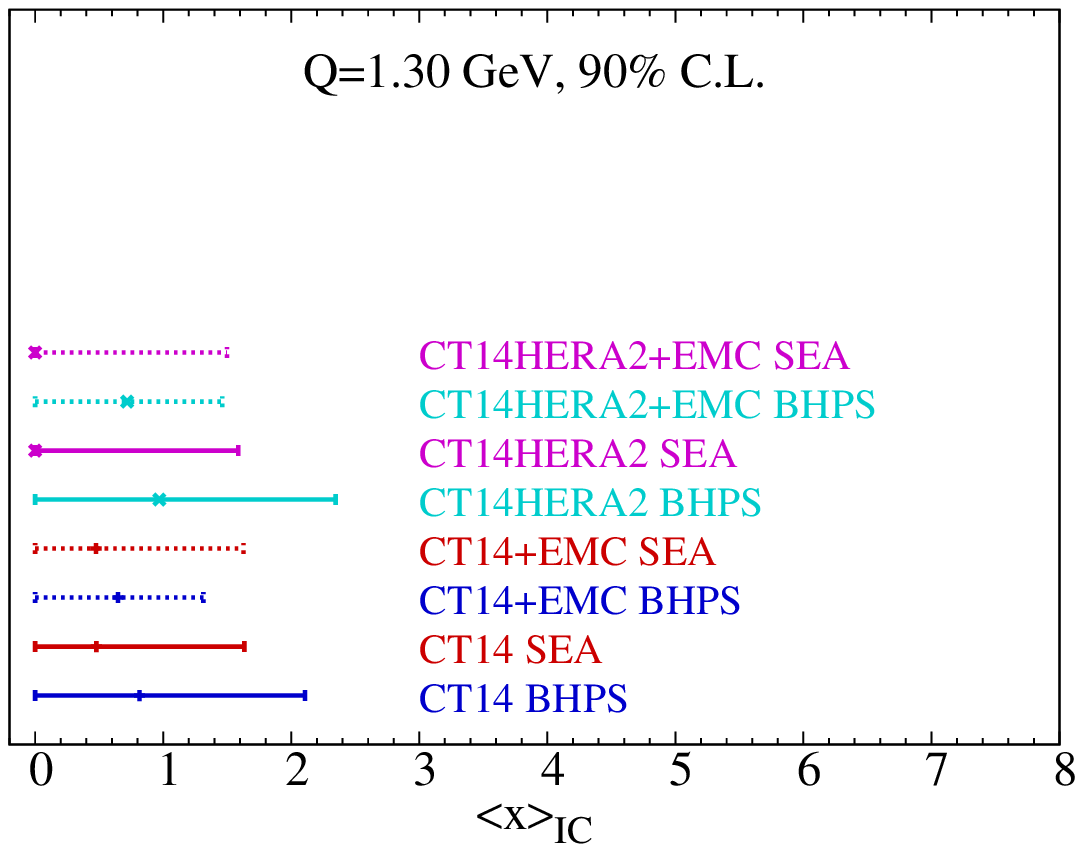}
\includegraphics[width=0.48\textwidth]{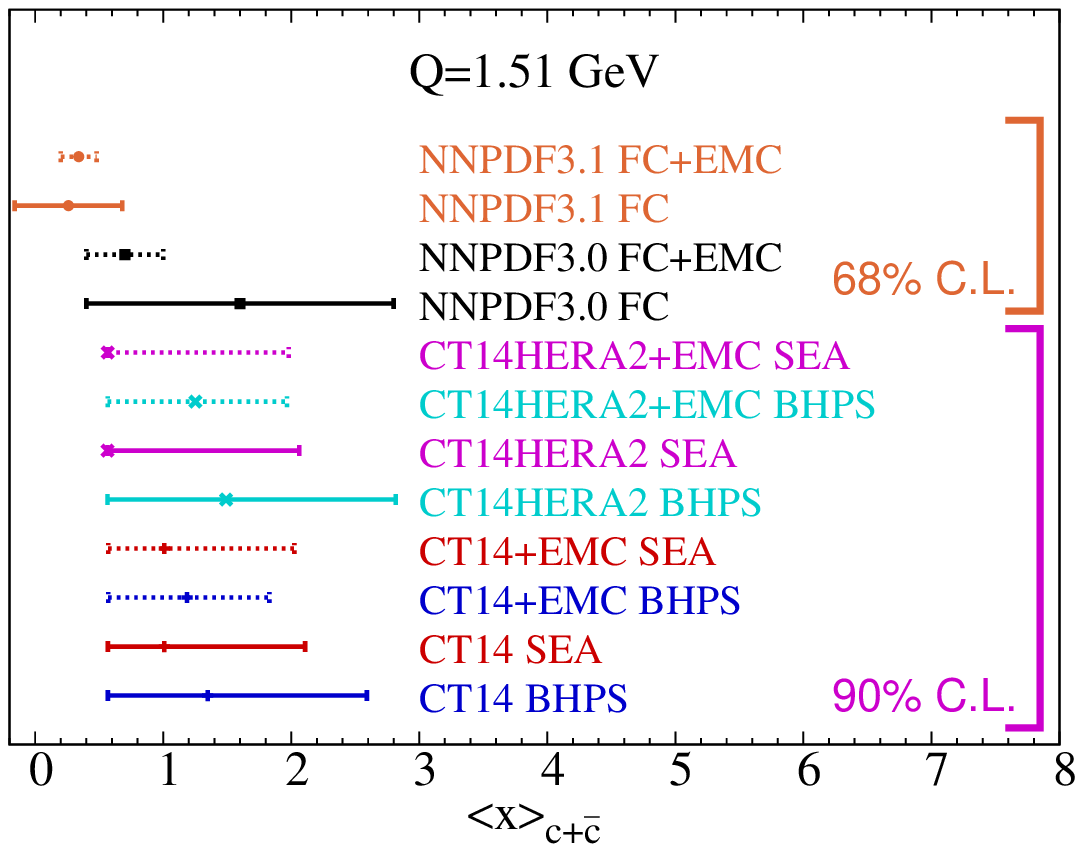}
\end{center}
\caption{The 90\% C.L. intervals on the charm momentum fraction evaluated
at $Q=1.3$ GeV and $Q=1.51$ GeV. For $Q=1.51$ GeV, the 68\%
C.L. intervals from the NNPDF3.0 \cite{Ball:2014uwa,Ball:2016neh} and NNPDF3.1
  \cite{Ball:2017nwa} are superimposed.
\label{fig:xICCTNNPDF}}
\end{figure}

To recap, the EMC data has a weak impact on fitting the rest of
the CT14/CT14HERA2 data. Increasing the weight of the EMC data to 10
without the IC improves the description of the HERA charm
production data at the expense of a worse fit to the inclusive
DIS data and to the full data set. Including the nonperturbative charm
contribution of the BHPS, SEA, or mixed type does not improve the fit
to the EMC $F_{2c}(x,Q)$, in contrast to the findings in
\cite{Ball:2016neh}.

It might be argued that a larger set of
parametrization forms for the IC needs to be explored, as in the NNPDF method,
to see if a better fit to the
EMC $F_{2c}(x,Q)$ could be reached.
In the absence of control of
experimental and (N)NLO theoretical systematic effects in the EMC $F_{2c}$
data set, such an exercise again appears to be excessive.
Indeed, when using a purely
perturbative charm only, the NNPDF3.1 study \cite{Ball:2017nwa}
obtains a considerably worse $\chi^2_n/N_{\rm pts} = 4.8$
for the EMC $F_{2c}$ data set
than our results quoted in Table~\ref{TAB1}. After including a flexible
``fitted charm'' parametrization they arrive at a much better agreement
with the EMC data sample, with $\chi^2_n/N_{\rm pts} = 0.93$  and $\langle
x\rangle_{c+\bar{c}} =0.34\pm 0.16\%$ at $Q_c=m_c^{pole}=1.51 $ GeV
at 68\% C.L. Their $\chi^2/N_{pts}$ values in Table 4.3 of \cite{Ball:2017nwa}
are somewhat better for the inclusive HERAI+II data set (1.16) and somewhat worse for the HERA charm SIDIS data set (1.42), compared to our 1.25 and 1.22 in Table~\ref{TAB1}.

Some of these disparities are
explained by non-identical PDF parametrization forms (positive-definite BHPS/SEA models
in the case of CT14 IC, vs. the neural networks of NNPDF3.0), the general-mass schemes, and the choices of
the mass parameters:
$Q_c=m_c^{pole}=$1.3, 1.275, and 1.51 GeV in the CT14, NNPDF3.0, and NNPDF3.1 studies,  respectively.
The preferred $\langle x\rangle_{c+\bar{c}}(Q)$ at $Q=1.51$ GeV
are smaller in the NNPDF3.1 framework than for CT14 IC
in part because the evolved perturbative charm PDF is absent at this $Q$
in NNPDF3.1. The S-ACOT-$\chi$ scheme that we use
is at present the only ACOT scheme in
which the massive coefficient functions are fully
available to NNLO, or ${\cal O}(\alpha_s^2)$ \cite{Guzzi:2011ew}. NNPDF3.1 used a different mass scheme \cite{Ball:2015tna,Ball:2016neh} and set to zero some
${\cal O}(\alpha_s^2)$/NNLO massive terms that are not available in that scheme \cite{Ball:2017nwa}. We thus expect some differences between the schemes.

Another difference arises from the definitions of
uncertainties. The current paper quotes 90\%
probability intervals obtained by scanning $\Delta \chi^2$ with
respect to $\langle x\rangle$, as explained  in
Sec.~\ref{sec:xICDependence}. The NNPDF works quote their errors as
symmetric standard deviations obtained from averaging over many
replica fits, each of which is not a perfect fit and may deviate from
the central fit by hundreds of units of $\chi^2$ \cite{Hou:2016sho}.

As an illustration, Fig.~\ref{fig:xICCTNNPDF} compares the probability intervals
on the momentum fractions from the CT14/CT14HERA2 and
NNPDF3.0/NNPDF3.1 NNLO analyses.
The left frame shows the CT14/CT14HERA2 90\% probability intervals for
$\langle x\rangle_{\rm IC}$ at $Q=1.3$ GeV.
The right frame shows the CT14/CT14HERA2 intervals
for $\langle x\rangle_{\rm c+\bar{c}}(Q)$ at $Q=1.51$ GeV and
superimposes the 68\% C.L. uncertainties on the fitted charm (FC)
copied from the NNPDF3.0 and 3.1 publications.
Apart from the constant horizontal shift due to
the $Q_c$ choice, without the EMC data, the CT14 and
NNPDF probability intervals for $\langle x\rangle_{c+\bar{c}}$ are reasonably
compatible, minding their non-equivalent definitions.
 [The upward shift in $\langle x\rangle_{c+\bar{c}}(Q)$ by $\approx 0.5$\% due
  to the choice of $Q_c$, an auxiliary scale in a general-mass scheme,
  is of little physical significance,
  it is canceled up to ${\cal O}(\alpha_s^3)$ in the complete DIS cross section
  because of the compensating shift in ACOT subtraction terms.]
Inclusion of precise LHC data sets
helped to reduce the uncertainty in NNPDF3.1.
 The symmetric definition of the NNPDF3.1 errors allows a negative value
of uncertain interpretation
for $\langle x\rangle_{c+\bar{c}}$ at 68\% C.L. if the EMC data are
not included. A very small uncertainty on $\langle
x\rangle_{c+\bar{c}}$ quoted by the NNPDF3.1+EMC fit is accompanied
by the reduction in the global $\chi^2$ by less than 13 units for
4300 data points when the EMC data are added into the fit,
cf.~Table 4.3 in Ref.~\cite{Ball:2017nwa}. Needless
to say, the impact of the new experiments and assumptions on the
uncertainty of $\langle x\rangle_{c+\bar{c}}$ warrants a further investigation.

\section{Impact of IC on electroweak $Z$ and $H$ boson production cross sections at the LHC run II \label{sec:key-obs}}

Next, we will analyze the impact of the fitted/intrinsic charm (or the ``IC'',
for short) on key observables at the LHC,
assuming that the fitted charm does not strongly depend on the hard
process at NNLO. [We argued in Section~\ref{sec:QCDFactorization} that this
  assumption is not self-evident. We will nevertheless make it to
  investigate sensitivity of the LHC predictions.]

\begin{figure}[p]
\begin{center}
\includegraphics[width=0.49\textwidth]{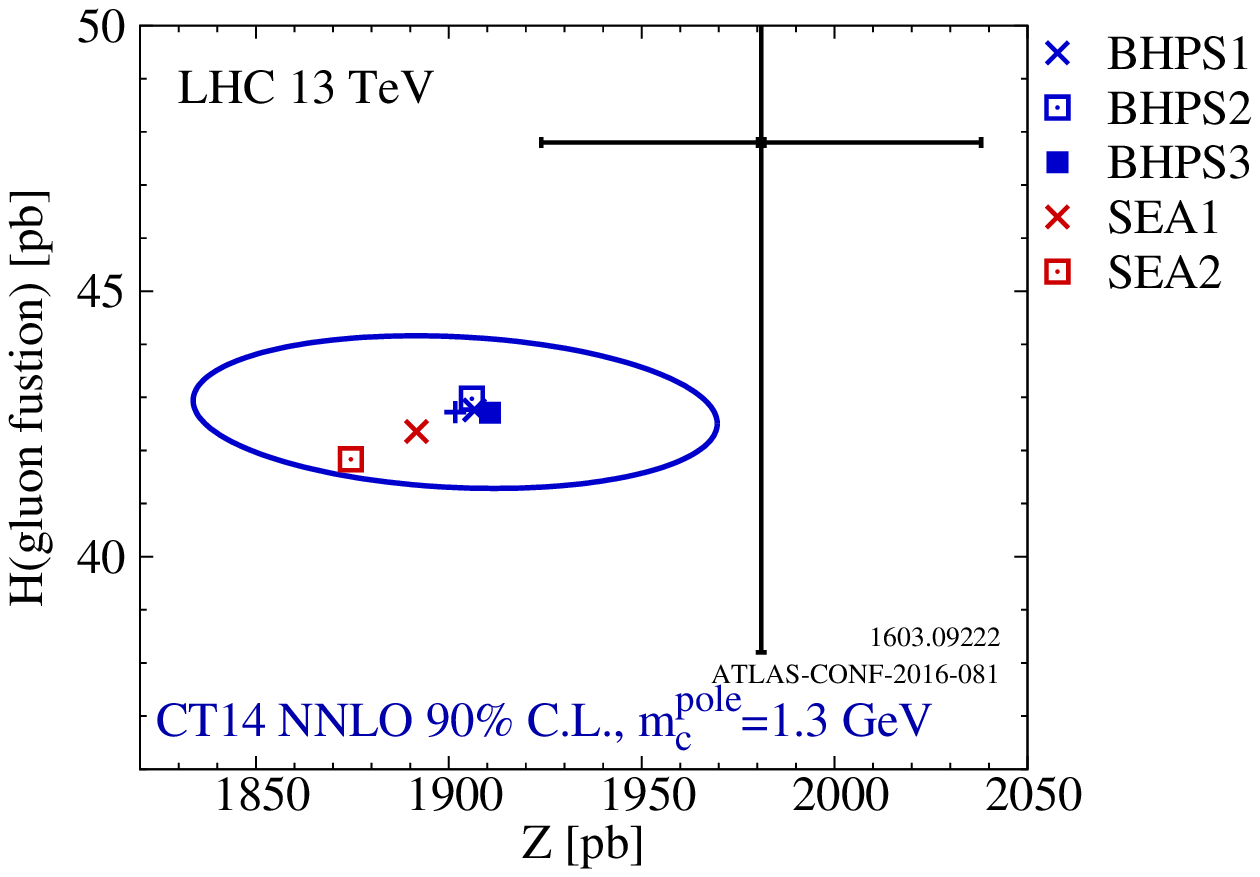}
\includegraphics[width=0.49\textwidth]{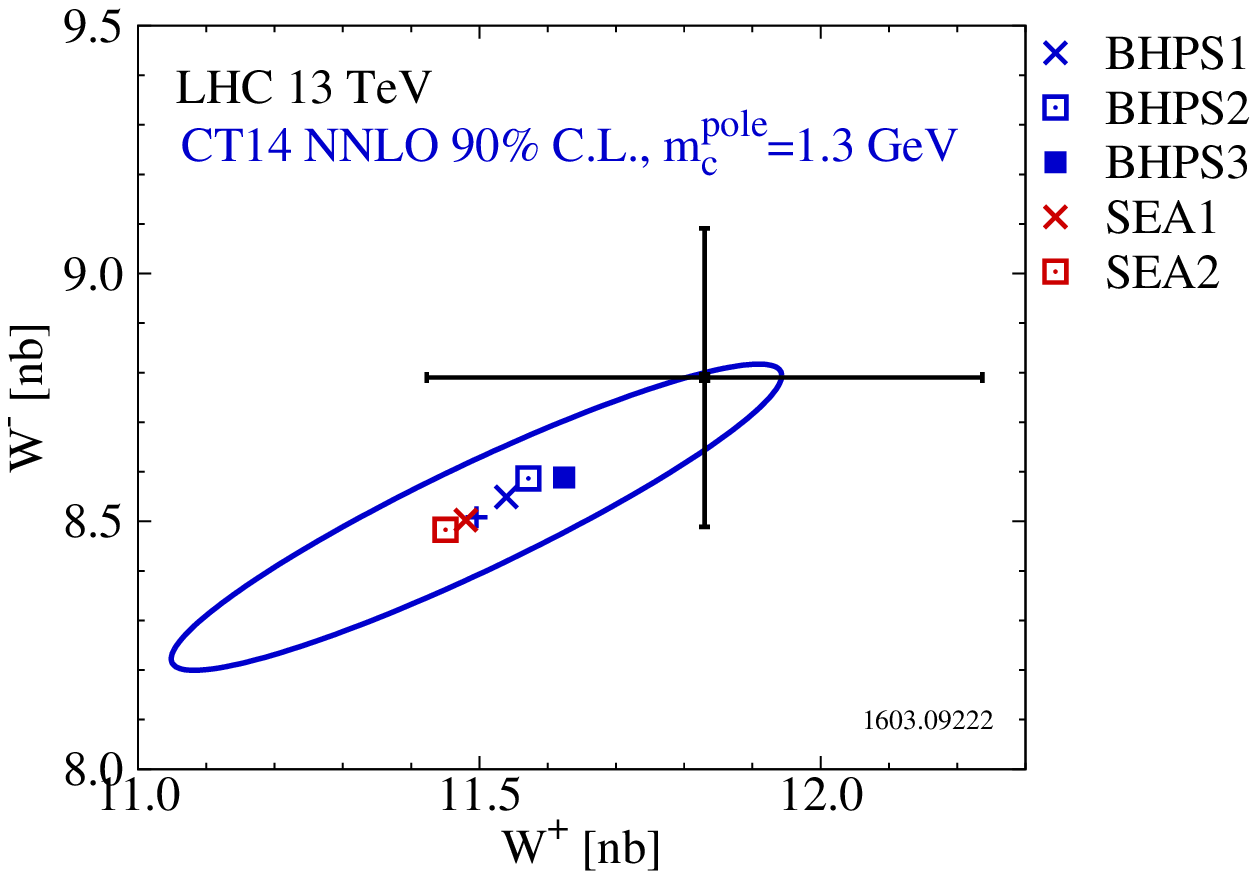}
\includegraphics[width=0.49\textwidth]{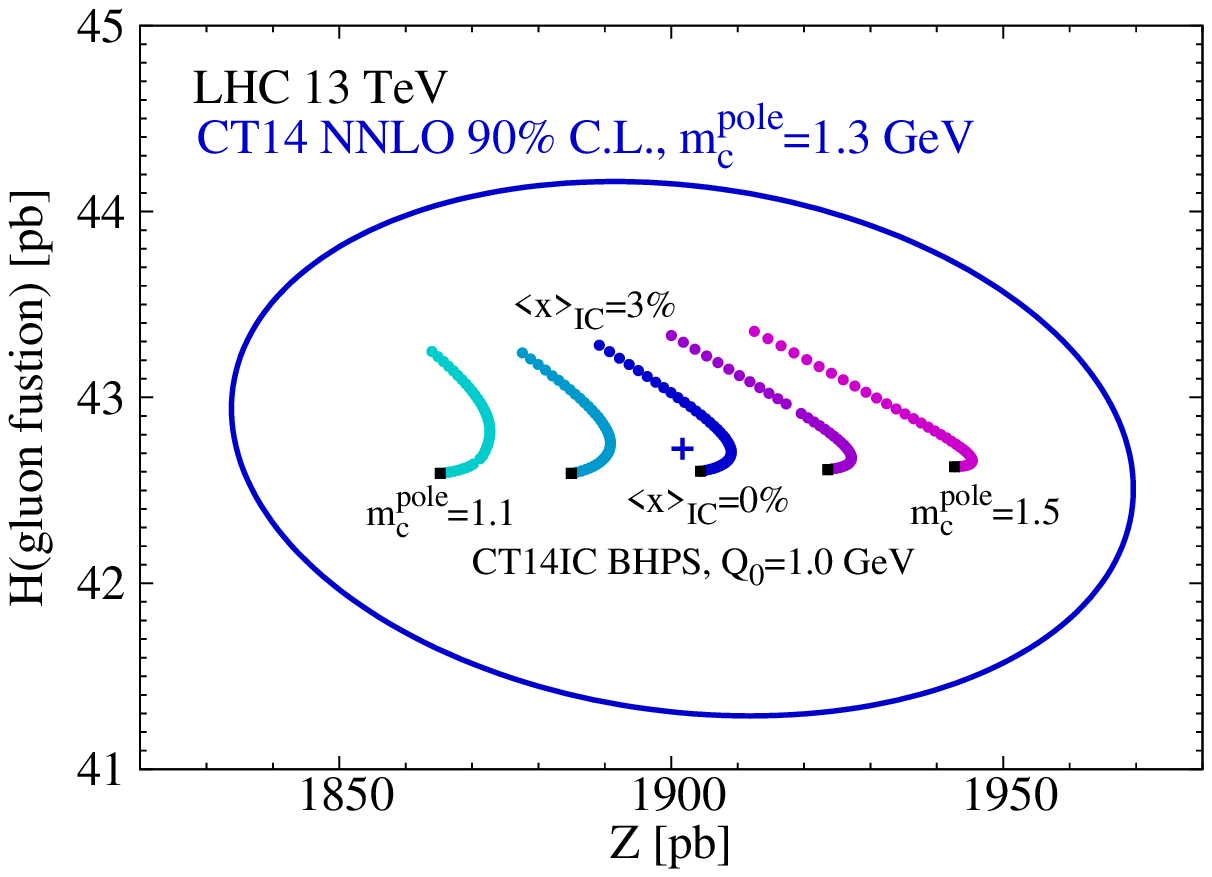}
\includegraphics[width=0.49\textwidth]{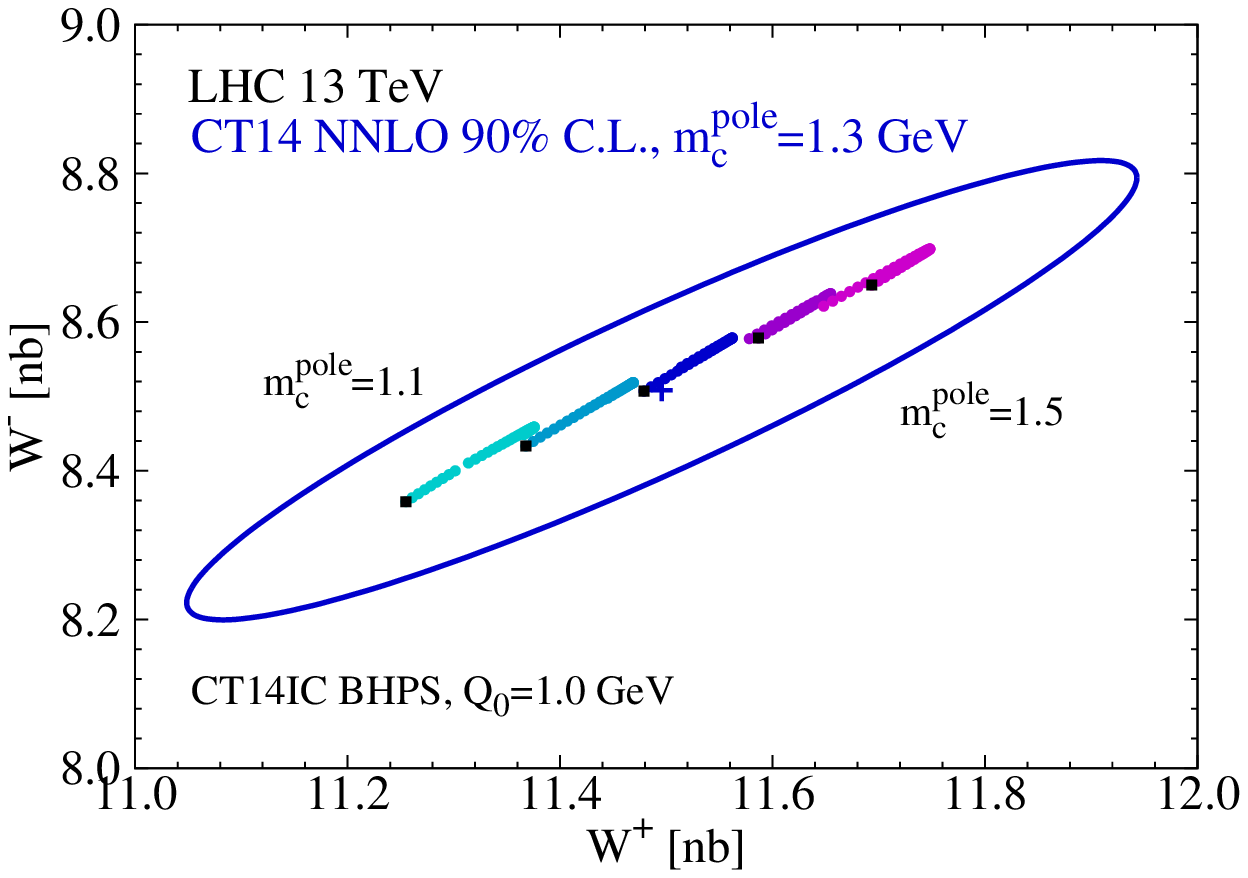}\\
\includegraphics[width=0.49\textwidth]{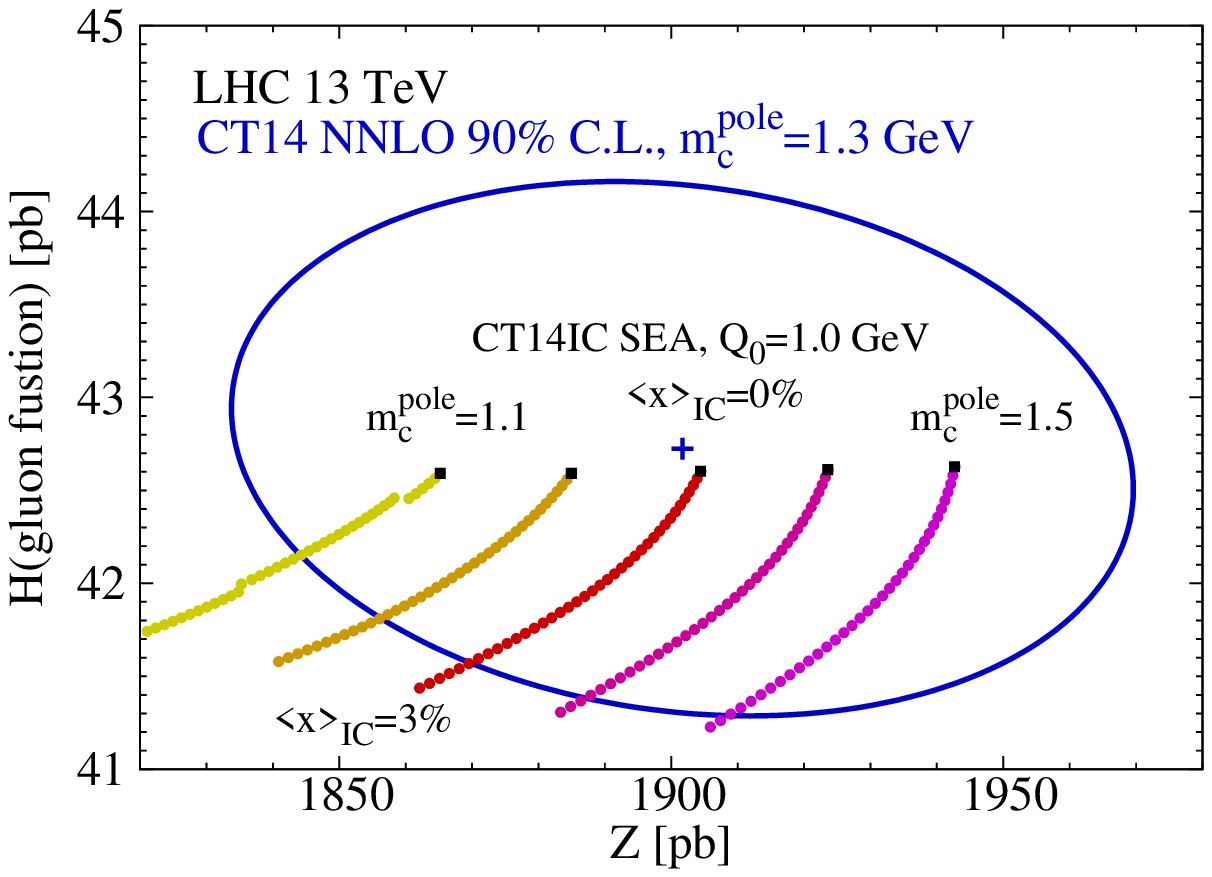}
\includegraphics[width=0.49\textwidth]{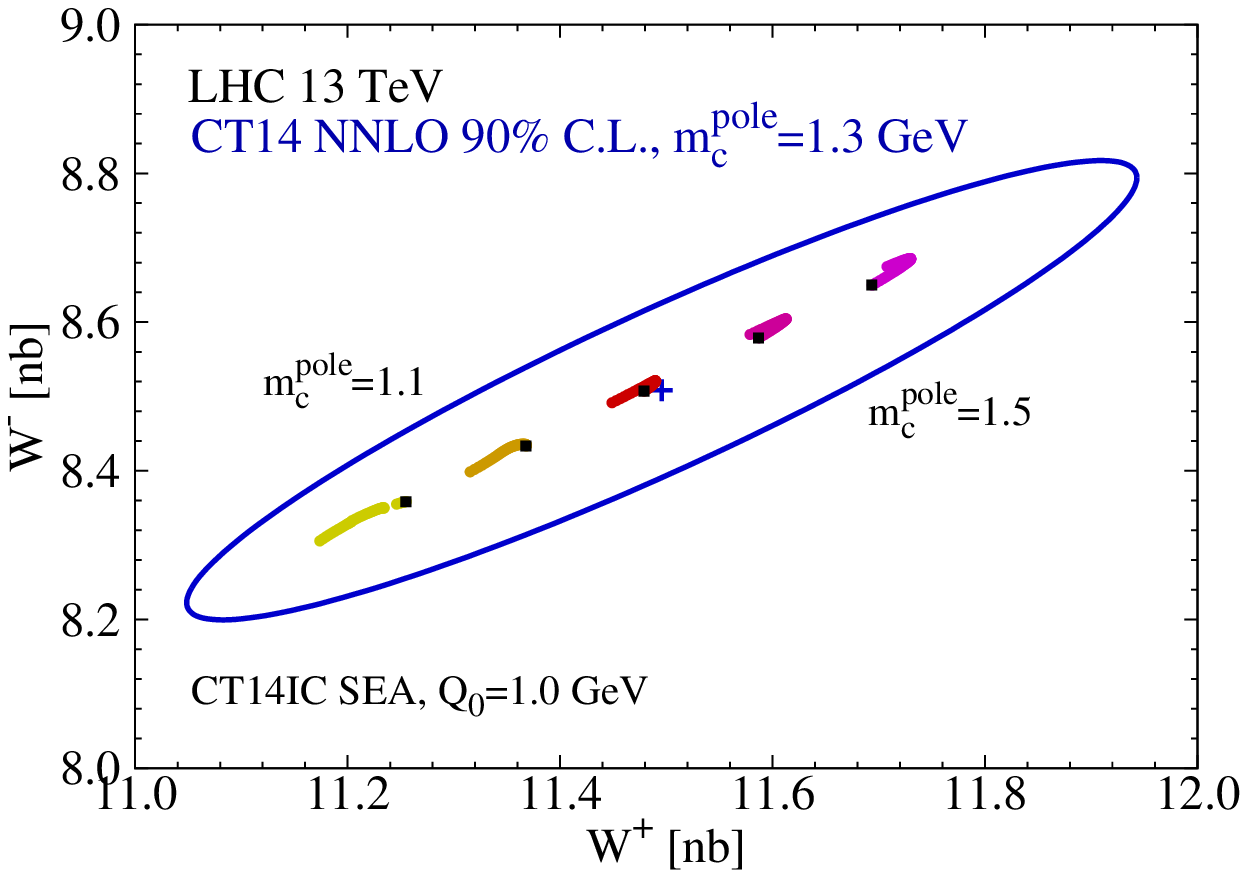}
\end{center}
\caption{CT14 NNLO $H$ (gluon-gluon fusion), $Z$, $W^+$ and $W^-$
  production cross sections at the LHC
  $\sqrt{s}$ = 13 TeV, for various charm models, as a function of the pole mass
  $m_c^{pole}=1.1-1.5$ GeV and
  charm momentum fraction $\langle x \rangle_{\rm IC}=0-3\%$.
  The 90\% C.L. uncertainty regions
  for CT14 at NNLO and experimental points~~\cite{Aad:2016naf,ATLAS:2016hru}
  are also shown.
\label{fig:ZH-Xsec}}
\end{figure}

Figure~\ref{fig:ZH-Xsec} illustrates dependence of the total cross
sections for inclusive production of electroweak bosons $W^\pm$,
$Z^0$, and $H$ (via gluon-gluon fusion) on the IC model
and charm quark mass at the LHC $\sqrt{s}=$ 13 TeV.
To provide a visual measure of the CT14NNLO uncertainty, each figure shows an error ellipse  corresponding to CT14 NNLO at the 90\% C.L.
The $W$ and $Z$ inclusive cross sections
(multiplied by branching ratios for the decay into one charged lepton flavor),
are calculated by using the \textsc{Vrap} v0.9 program~\cite{Anastasiou:2003ds,Anastasiou:2003yy} at NNLO in QCD, with the renormalization
and factorization ($\mu_R$ and $\mu_F$) scales set equal
to the invariant mass of the vector boson.
The Higgs boson cross sections via gluon-gluon fusion
are calculated at NNLO in QCD by using the \textsc{iHixs} v1.3
program~\cite{Anastasiou:2011pi},
in the heavy-quark effective theory (HQET) with finite top quark mass correction,
and with the QCD scales set equal
to the invariant mass of the Higgs boson.
The first row of Fig.~\ref{fig:ZH-Xsec} shows predictions for $W^\pm$, $Z^0,$ and $H^0$
production cross sections in the five BHPS and SEA fits for $m_c^{pole}=1.3\mbox{ GeV}$.
Predictions for
different values of the IC momentum fraction $0\% < \langle x \rangle_{\rm IC} <3\%$ and charm-quark mass $1.1< m_c^{pole}<1.5$ GeV, obtained with the initial scale $Q_0=1$ GeV, are  illustrated in the second and third rows of Fig.~\ref{fig:ZH-Xsec}.
The varied $\langle x\rangle_{IC}$ values are indicated by the point color for each $m_c^{pole}$ value.

The central value predictions for the BHPS and SEA models are all within the CT14 NNLO uncertainties, with BHPS very close to the CT14
nominal fit. The impact of IC on these key LHC observables is mild. For BHPS, increasing $\langle x\rangle_{IC}$ generally increases, and then reduces the $W^\pm$, $Z^0$ cross sections, and increases the Higgs cross sections. For SEA, increasing $\langle x\rangle_{IC}$ reduces all cross sections.

The intrinsic charm may partially offset the variations in the
electroweak cross sections due to the pole charm mass. As
$m_{c}^{pole}$ is increased from 1.1 to 1.5 GeV, the light-quark PDFs
in CT14/CT14 HERA2 are mildly increased at $x > 10^{-3}$ and $Q \sim
M_Z$, while the gluon is reduced at $x>0.1$. As mentioned before,
$m_{c}^{pole} \approx 1.5$ GeV results in a worse fit
to the CT14HERA2 data set, cf. the upper
Fig.~\ref{fig:Mc_dchi2Vxic}. For the LHC $W/Z$ cross sections,
increasing $m_{c}^{pole}$ to 1.5 GeV results in two competing
trends. On the one hand, 1.5 GeV leads to a somewhat better
description of the total $W$ and $Z$ cross sections in
Fig.~\ref{fig:ZH-Xsec}, even though the changes are well within the
CT14 uncertainty. This increase reflects larger $u$ and $d$
(anti)quark PDFs for $m_{c}^{pole}=1.5$ GeV.

On the other hand, the LHC data on high-$p_T$  $Z$-boson
production~\cite{Aad:2014xaa,Aad:2015auj,Khachatryan:2015oaa}
show contradictory preferences for the $m_c$ and $\langle x \rangle_{\rm IC}$,
 depending on the collider energy [7 or 8 TeV] and the format of the data [absolute or normalized cross sections].
Our conclusion at the moment is that the LHC inclusive $W$ and $Z$ production cross sections may provide helpful correlated constraints on $m_c$ and $\langle x \rangle_{\rm IC}$ in the future. We may also consider
more exclusive scattering
processes~\cite{Kniehl:2009ar, Kniehl:2012ti,Stavreva:2009vi, Stavreva:2010mw, Bednyakov:2013zta, Lipatov:2016feu, Bailas:2015jlc, Rostami:2015iva,Boettcher:2015sqn}
to look for evidence of the IC in the LHC environment.

\clearpage

\section{$\boldsymbol{Z}$ + charm-jet production in pp collisions at
  the LHC\label{sec:PREDICTIONS}}

A suitable test scenario is given by the production of a
$Z$ boson in association with a charm jet, for which a CMS measurement
at $\sqrt S=8\tev$ has been recently published in
Ref.~\cite{CMS:2016dyh}. The corresponding calculation
$pp\to\gamma/Z\,c$ is available at NLO in QCD, building on the
important feature that the LO partonic process
$g\,+\,c\,\to\,\gamma/Z\,+\,c$ (consisting of $\hat s$ and $\hat t$
channel contributions) is directly sensitive to the initial-state
charm distribution. Provided that the charm-quark
transverse momentum is much larger than its mass, the NLO corrections
to this process can be calculated working in the S-ACOT
scheme~\cite{Aivazis:1993pi,Kramer:2000hn,Tung:2001mv}. Using this
scheme enables one to neglect the charm mass throughout, while only
making a small error of the order of
$1/\ln\left(\frac{M_Z}{m_c}\right)\times\frac{m^2_c}{p^2_T}$~\cite{Campbell:2002zm}.
The contributing subprocesses are given by $gc\to Zc$ (one-loop level
production), $q/g\;c\to Zc\;q/g$ (light-flavour parton emission)
($q=u,d,s$) and $gg\to Zc\bar c$ (charm pair production).\footnote{The
  $Z$ mass window constraint of the measurement will ensure the strong
  suppression of any $\gamma+c$ contribution. We therefore neglect
  these contributions.}
Another subprocess leading to charm-quark pairs in the final state is
$q\bar q\to Zc\bar c$. It is not regarded as a correction to
$gc\to Zc$, but it is an additional source of $Zc$ events, and
therefore taken into account at LO. There is one subtlety that
concerns the $Zc\bar c$ final states. They are evaluated by retaining
the charm-quark mass in order to regulate the gluon splitting
singularity that would arise for massless collinear quarks. Taking all
of these subprocesses, one then arrives at an NLO accurate description
for the associated production of a $Z$ boson and a single charm jet,
as has been presented in Ref.~\cite{Campbell:2003dd} and implemented in
the program \textsc{MCFM}~\cite{Campbell:2010ff}.
To compare the impact of the different IC PDF fits, we use the
\textsc{MCFM} calculation to generate the various $Z$+$c$-jet cross
sections in the presence of intrinsic charm at NLO.

The main drawback of the fixed-order predictions is their limitation
in describing effects that arise from multi-particle final states.
One complication is due to the importance of jet production at higher
orders, which enhances the size of the $Z$+charm-jet cross section
especially for high-$p_T$ $Z$ boson production. The inclusive cross
section definition ($Z$+$c$-jet+$X$) employed by the CMS analysis
makes it important to account for the contributions from more complex
topologies like $gq\to Zc\bar cq$ or gluon-jet splitting to $c\bar c$
occurring at higher perturbative orders (i.e.~in Monte Carlo physics
language, later in the event evolution). The fixed-order approach will
miss these multijet contributions, but we can invoke matrix-element
plus parton-shower merging (MEPS) to study these effects. This can be
particularly important if the final state is binned in a variable such
as the $Z$ boson transverse momentum, while a fixed (low) cut is
placed on any jets in the event. We can also investigate at which
point (in terms of the number of multileg MEs included), saturation
(stabilization of the cross section) can be found.

Another complication stems from the fact that in an experimental
environment, we are required to use a cross section definition, which
is based on the detection of charm hadrons/objects in the event,
i.e.~charm tagging is involved one way or another to determine the
inclusive $Z$+charm-jet rate. The theory-driven,
parton-level definition employed in the fixed-order case
cannot be applied here, as it ignores the evolution of the hard event
to energy scales of the order of $1\gev$, where the measurement takes
place. In this context and, especially, for the identification of
specific particles/objects -- as in our case, charm jets -- aspects of
multi-particle production (beyond hard jets) therefore need to be
taken into account to arrive at a more realistic simulation. For our
studies, we will rely on the parton shower to describe the
fragmentation of the charm partons~\cite{Hoeche:2009xc}, and assuming
factorization of the initial-state and final-state QCD radiations
as a reasonable approximation.
The cross section based on charm tagging will be affected
by parton showering. Thus, we have to deal with contributions emerging
from $Z$+non-$c$ partonic processes because the $g\to c\bar c$
splittings have the potential of turning a $Z$ plus light-flavour jet
into a $Z$ plus $c$-jet contribution. This additional source of
$Z$+charm events enhances the size of the measured cross section.
However, this enhancement simply serves to dilute the impact of any
intrinsic charm, since in most cases it emerges from initial states not
involving a charm quark, i.e.~the enhancement comes from final-state
gluon splitting into a $c\bar c$ pair.
The rate for this enhancement depends on both the charm-jet transverse
momentum threshold and the number of jets in the final state.

For these reasons, one cannot ignore the multi-particle aspects when
dealing with realistic scenarios. We therefore generate predictions
using the LO matrix-element plus parton-shower merging (MEPS@LO)
approach~\cite{Hoeche:2009rj}, adding additional jets and subsequent
parton showering, and requiring the presence of a charm jet in the
final state. The MEPS@LO approach allows us to estimate the impact of
the higher-order radiative corrections and charm tagging at the
same time. Using the various IC models, we can examine (on a
quantitative level) to what extent the multi-particle effects alter
the outcome of the NLO calculations provided by \textsc{MCFM}. All
MEPS@LO predictions presented here have been obtained from the Monte
Carlo event generator \textsc{Sherpa-2.2.1}~\cite{Gleisberg:2008ta}.
To perform the charm tagging in the \textsc{Sherpa} simulations, we
rely on the flavorful version of the anti-$k_t$ jet algorithm as
implemented in \textsc{FastJet}~\cite{Cacciari:2011ma}.
We generate $Z$+jets samples in the five-flavour scheme (massless $c$
and $b$ quarks) involving tree-level matrix elements for $Z$+$0,1$
parton up to those for $Z$+$n_\textrm{ME}$ partons where
$n_\textrm{ME}$ denotes the maximum outgoing-parton multiplicity of
these matrix elements. Three \textsc{Sherpa} samples are provided,
namely for $n_\textrm{ME}=1,2,3$, using a merging cut of
$Q_\textrm{cut}=20\gev$. Each \textsc{Sherpa Nj} prediction is then
drawn from the respective $Z$+jets sample with $n_\textrm{ME}=N$.

\begin{table}
  \centering
  \begin{tabular}{lcccc}\hline
    & \multicolumn{4}{c}{\textbf{Calculation}}\\\cline{2-5}\\[-5mm]
    \multirow{2}{*}{\textbf{~PDF}}
    & \multicolumn{2}{c}{\small[ratio to \textsc{MCFM} CT14]}
    & \multicolumn{2}{c}{\small(increase wrt.~CT14)}\\
    & \textsc{~~~~MCFM~~~~} & \textsc{~~Sherpa 1j~~}
    & \textsc{~~Sherpa 2j~~} & \textsc{~~Sherpa 3j~~}\\\\[-5mm]\hline
    ~CT14NNLO~~ &
    $6.04$~~[1.0]   &$5.93$~[0.982]  &$6.59$~[1.091]  &$6.64$~[1.099]\\
    ~BHPS3 &
    $6.18$~(+2.3\%)  &$6.04$~(+1.9\%)  &$6.70$~(+1.7\%)  &$6.76$~(+1.8\%)\\
    ~BHPS2 &
    $6.41$~(+6.1\%)  &$6.21$~(+4.7\%)  &$6.90$~(+4.7\%)  &$6.97$~(+5.0\%)\\
    ~SEA1 &
    $6.51$~(+7.8\%)  &$6.29$~(+6.1\%)  &$6.97$~(+5.8\%)  &$7.03$~(+5.9\%)\\
    ~SEA2 &
    $7.23$~(+19.7\%) &$6.82$~(+15.0\%) &$7.57$~(+14.9\%) &$7.63$~(+14.9\%)\\\hline
    ~NNPDF3.0 &
    $6.09$~[1.008]\\
    ~$\cdot$~fitted charm &
    $5.78$~[0.957]\\
    ~$\cdot$~fitted charm, no EMC &
    $6.00$~[0.993]\\\hline
  \end{tabular}
  \caption{\label{tab:zclhc}
    Total inclusive $Z$+charm-jet cross sections (in pb) at the LHC
    for $\sqrt{S}=8\tev$ for two different standard PDFs (CT14 and
    NNPDF3.0) as well as different fits containing an IC component.
    The predictions were obtained from \textsc{MCFM} at NLO, and from
    $Z$+jet samples generated by \textsc{Sherpa} using the MEPS@LO
    approach at different levels of including higher-order tree-level
    matrix elements. The details of the calculations are given in the
    text. Note that entities in square brackets show ratios
    with respect to (wrt.)~the \textsc{MCFM} result for CT14NNLO, while numbers in
    parentheses quantify the percentage of increase in the cross section
    for the various CT14 IC models in relation to the respective
    CT14 standard result.}
\end{table}

The simplest observable to look at is the inclusive $Z$+charm-jet
cross section. Hence, we start by presenting a summary of
cross section predictions in Table~\ref{tab:zclhc}. In both types of
calculations (fixed order and MEPS@LO), we employ kinematic
requirements that are similar to those utilized by the CMS
analysis~\cite{CMS:2016dyh}.\footnote{Although we aim at a fairly
  close reproduction of the kinematic selections used in the CMS
  analysis (cf.~Ref.~\cite{CMS:2016dyh}), we refrain from comparing
  our results directly to the experimental data for reasons such as
  unapplied/unknown hadronization corrections and neglecting certain
  $\Delta R$ constraints.}
Most notably, we impose the following kinematic requirements on the
two leptons from the $Z$ boson decay:
$p_{T\!,\ell}>20\gev$, $\left|\eta_\ell\right|<2.1$ and
$71 \, {\rm GeV} <m_{\ell\ell} <111\,{\rm GeV}$.
Jets are defined by using the anti-$k_t$ algorithm with a size
parameter of $0.5$ and threshold requirements reading
$p_{T\!,\mathrm{jet}}>25\gev$ as well as $\left|\eta_\mathrm{jet}\right|<2.5$.
Table~\ref{tab:zclhc} shows that the predictions from the two standard
PDFs (CT14NNLO and NNPDF3.0) agree very well. All CT14 IC models
lead to an increase of the $Z$+charm-jet cross section varying from
about 2\% for the specific choice of using BHPS3 to almost 20\% for
the SEA2 model. On the contrary, the fitted charm PDFs of the NNPDF
group~\cite{Ball:2014uwa,Ball:2016neh} lead to a small reduction of
the total cross section, however by no more than 5\%. The results from
Table~\ref{tab:zclhc} also confirm the rise of cross sections owing to
the inclusion of multijet contributions.
This increase can grow as large as 10\%. From a fixed-order point of
view, the \textsc{Sherpa 1j} calculation is LO-like
while the \textsc{Sherpa 2j} calculation is closest to the one
provided by \textsc{MCFM}. The largest differences
with respect to ~\textsc{MCFM}
lie in \textsc{Sherpa}'s neglect of virtual contributions that are
non-Sudakov like and the usage of a dynamical plus local scale setting
prescription.\footnote{We note that the \textsc{Sherpa 2j}
  calculations can be made even more \textsc{MCFM}-like by relying on
  Sudakov reweighting but applying no parton showers at all. These
  modified \textsc{Sherpa} predictions show good agreement with the
  cross sections predicted by \textsc{MCFM} though they are still
  larger by about 2\%.}
The \textsc{Sherpa 3j} computation then goes beyond the \textsc{MCFM}
calculation, resulting in an additional but smaller increase with
respect to the \textsc{Sherpa 2j} cross sections. In other words, we
observe the expected saturation effect that stabilizes the $Z$+$c$-jet
rate with increasing $n_\textrm{ME}$.
As in the fixed-order case, the CT14 IC models enhance the
\textsc{Sherpa} cross sections by different amounts. For a specific
model, the predicted gains are of similar size among the different
\textsc{Sherpa Nj} calculations (as indicated by the numbers in
parentheses in Table~\ref{tab:zclhc}), but turn out to be smaller when
compared to the respective fixed-order result. The MEPS@LO predictions
therefore show the expected dilution of the IC signals as previously
described. Furthermore, we can take this as evidence for similar
mitigation effects applying to experimental signatures for intrinsic
charm.

\begin{figure}[t!]
\centering
\includegraphics[width=0.48\textwidth]{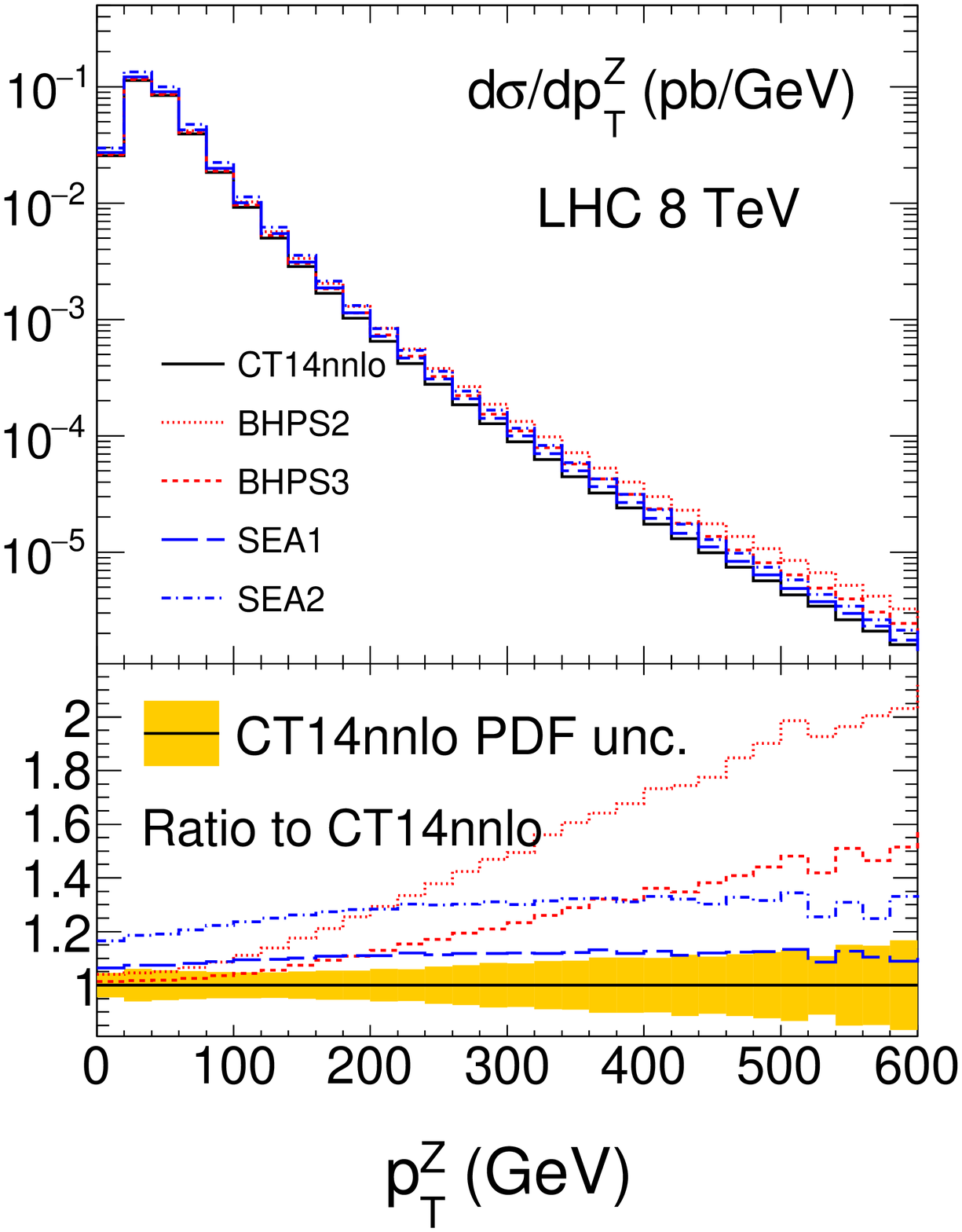}
\hskip5mm
\includegraphics[width=0.48\textwidth]{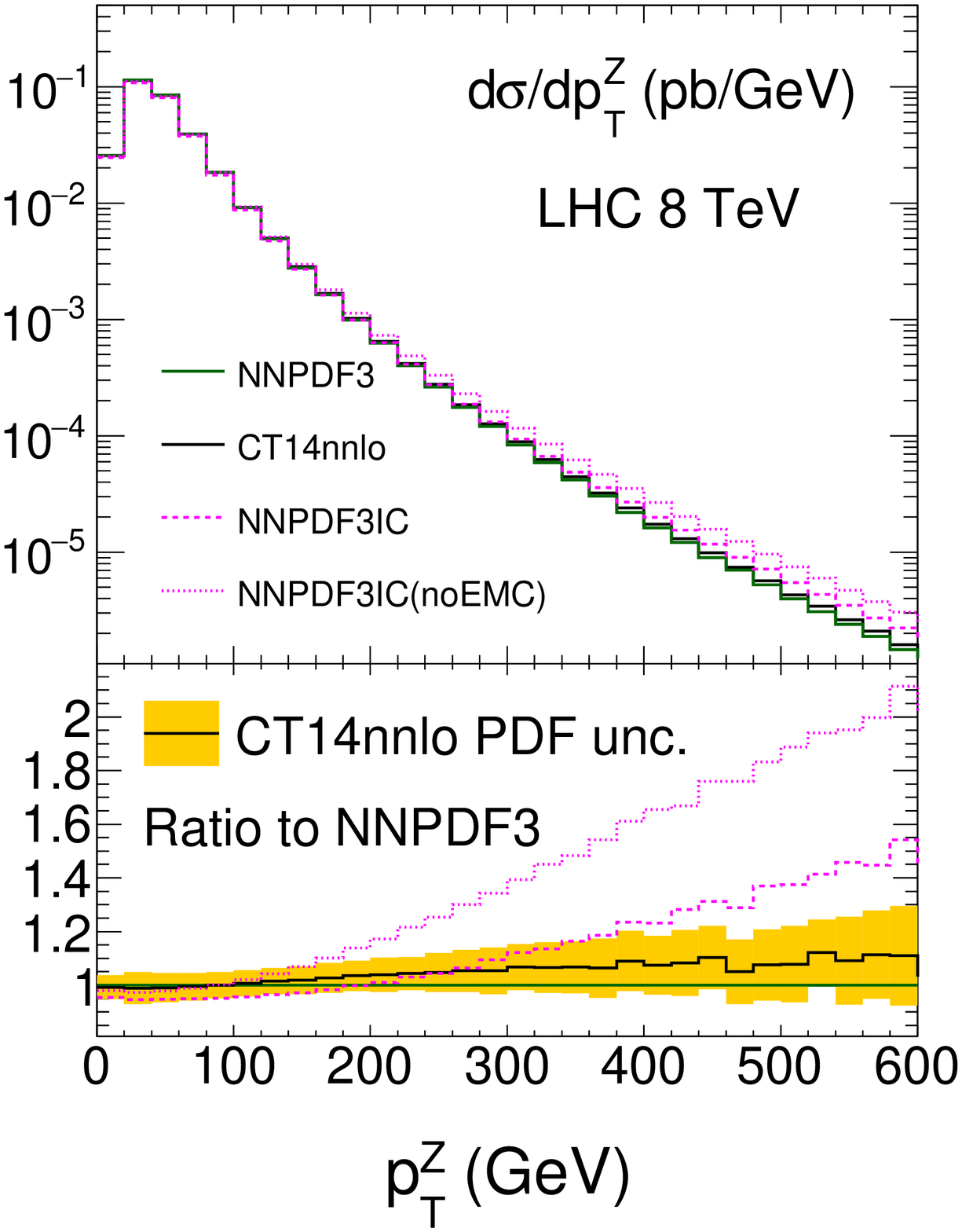}
\vskip-5mm
\caption{\label{fig:zclhc-8}
  Transverse momentum distribution of the $Z$ boson for the production
  of $pp\to Zc$ at the LHC with $\sqrt{S}=8\tev$. Various predictions
  based on different fitted charm models are compared to their
  respective standard predictions, obtained by using the CT14NNLO PDF
  set on the left and the NNPDF3.0 set on the right. Note that the
  CT14NNLO prediction is shown in both plots, together with its
  uncertainty envelope for 90\%~C.L. All results have been generated
  using the program \textsc{MCFM}. The lower panels are used to depict
  the relative changes induced by the different models with respect to the
  CT14NNLO prediction (left) and the NNPDF3.0 prediction (right).}
\end{figure}

\begin{figure}[t!]
\centering
\includegraphics[width=0.48\textwidth]{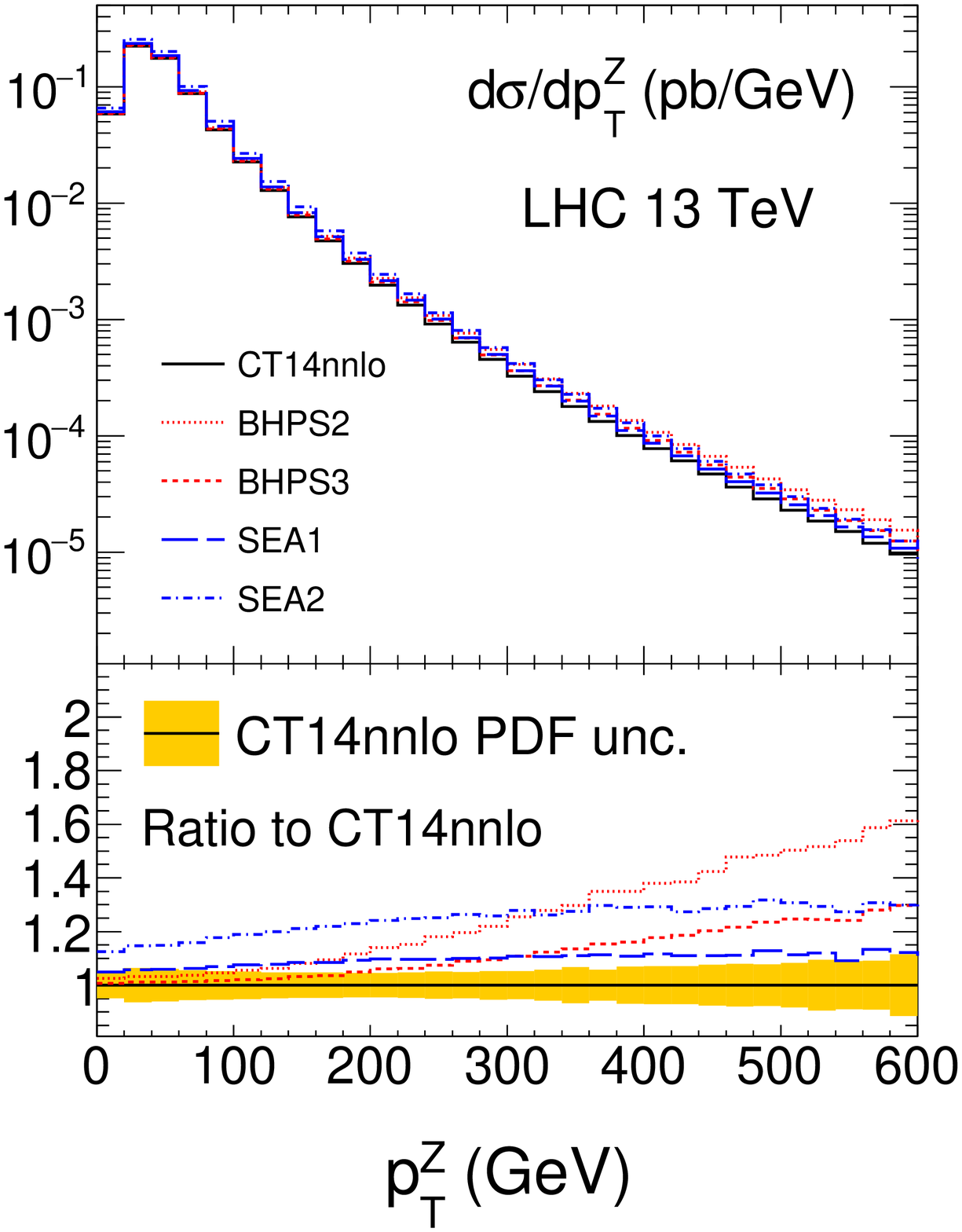}
\hskip5mm
\includegraphics[width=0.48\textwidth]{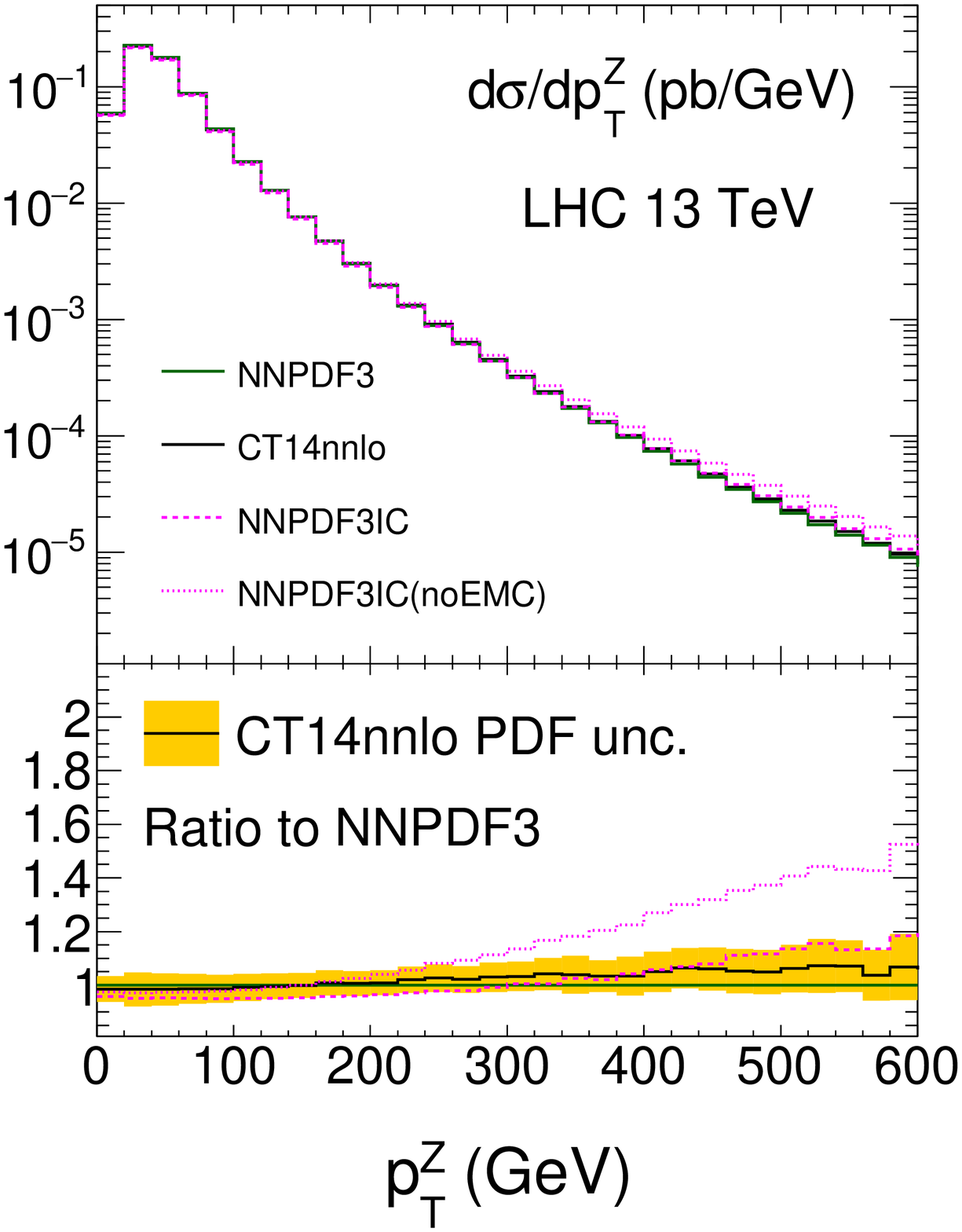}
\vskip-5mm
\caption{\label{fig:zclhc-13}
  Same as Fig.~\protect{\ref{fig:zclhc-8}}, for the LHC with $\sqrt{S}=13\tev$. }
\end{figure}

The total inclusive cross section as measured by CMS,
$\sigma(pp\to Zc+X)\times\mathrm{BR}(Z\to\ell^+\ell^-)=
8.6\pm0.5\textrm{\:(stat.)}\pm0.7\textrm{\:(syst.)}\mathrm{\:pb}$,
comes with an overall relative uncertainty of 10\%.
This cross section is larger than any of the predictions in Table~\ref{tab:zclhc}.
With this rather
large value, we cannot yet draw any conclusion regarding the
preference or exclusion of the various IC models. For example, if we
assume that the baseline CT14 prediction describes the data, the SEA2
model, which predicts the largest relative cross section change among
all IC models, would only occur at the upper edge of the allowed
($2\sigma$) range (neglecting the impact of PDF and theory
uncertainties for a moment). However, the various intrinsic charm
models affect the low and high $x$ regions differently, making it
worthwhile to investigate the effects on differential cross sections
as well. As mentioned earlier, the transverse momentum distribution of
the $Z$ boson
in association with a
charm jet is a suitable candidate because larger $x$ values
predominantly affect the high $p_T$ region. Focusing on different
$p_T$ regions may therefore increase our chances to distinguish
certain IC models from each other.

Figures~\ref{fig:zclhc-8} and~\ref{fig:zclhc-13} show \textsc{MCFM}
predictions of the differential $Z$ boson $p_T$ cross sections at the
LHC, for energies of $8\tev$ and $13\tev$, respectively. Apart from
presenting the $p_T^Z$ distributions themselves, we also depict
the respective ratios taken with respect to~the CT14NNLO result.
We furthermore use the panels on the right in
Figures~\ref{fig:zclhc-8} and~\ref{fig:zclhc-13} to present a similar
set of plots obtained by using various PDFs from the NNPDF group,
namely their current default version, NNPDF3.0, also serving as the
reference curve in the lower part of the right panes, and their
associated fitted charm PDFs with and without accounting for the EMC
data. These NNPDF plots also contain the CT14 baseline predictions
(including their PDF uncertainties) to allow for direct comparison
between both PDF families.

The results of Figures~\ref{fig:zclhc-8} and~\ref{fig:zclhc-13} reveal
the existence of sizable deviations between the predictions
from the standard PDFs and the IC models (for both families).
The BHPS intrinsic charm fits produce larger cross sections for high
$Z$ transverse momenta, while the PDFs using the SEA parametrization
affect the cross sections fairly equally at all values of $p_T^Z$,
and in a similar way at both $8\tev$ and $13\tev$ predictions. In
particular, the SEA2 fit yields increases of the order of 20\%.
Regarding the BHPS models, the critical issue is the reach of
the LHC data into regions of higher $x$ (corresponding to large values
of $p_T^Z$) where the enhancement in the BHPS models
becomes significant.
At $8\tev$, the effects can be up to 100\% higher than the baseline;
they however occur in a region without data
(for $p_T^Z>500\gev$).
At $13\tev$, we deal with smaller $x$ values on average and therefore
observe smaller deviations (dropping by nearly a factor 2) for the
corresponding BHPS predictions. We also note that the relative changes
predicted by the fitted charm PDFs of the NNPDF group resemble those
of the BHPS fits for the CT family. This resemblance is found at both
collider energies, for which we also observe good agreement between
the central predictions of NNPDF3.0 and CT14NNLO.

\begin{figure}[t!]
\centering
\includegraphics[width=0.48\textwidth]{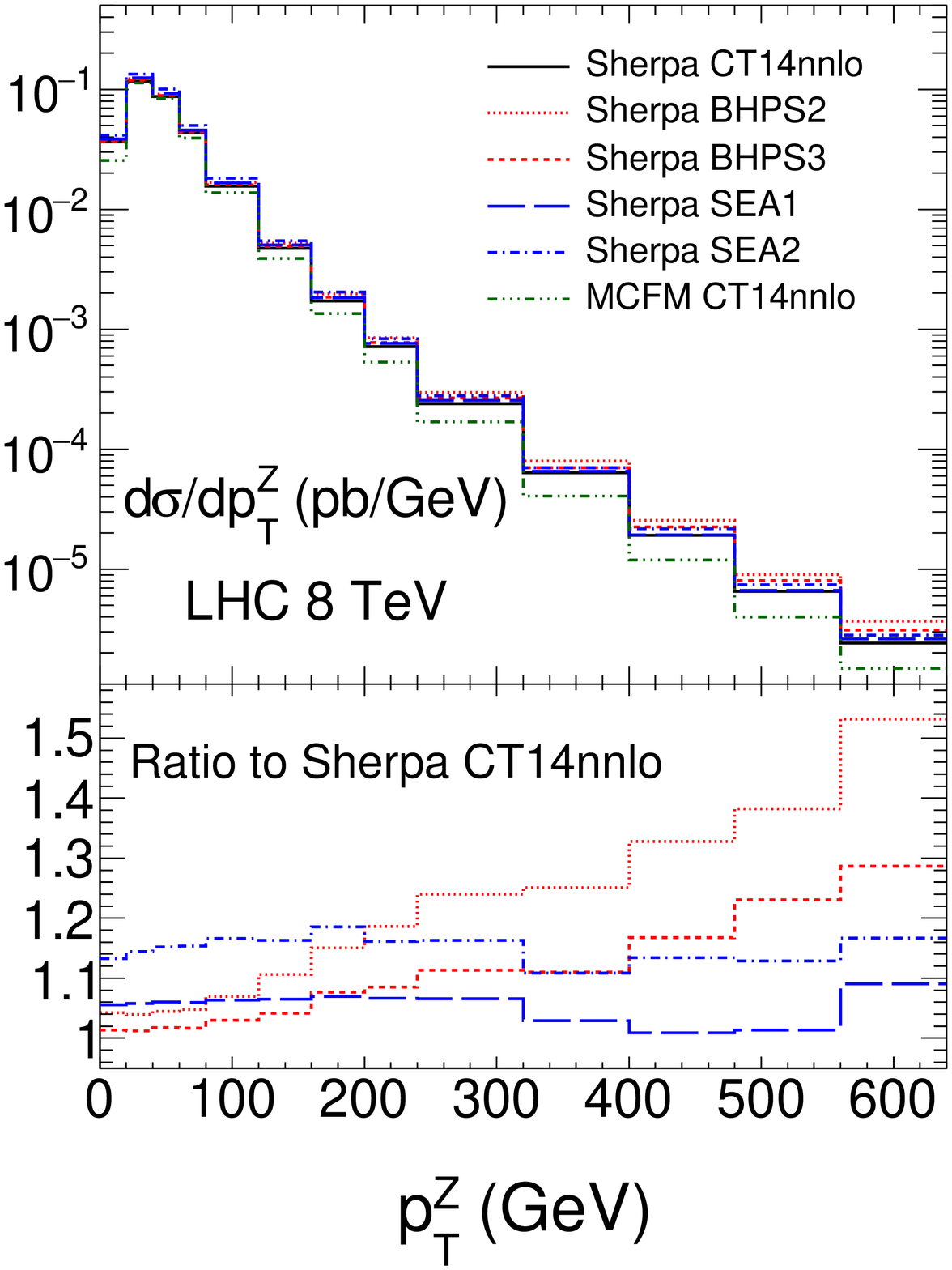}
\hskip5mm
\includegraphics[width=0.48\textwidth]{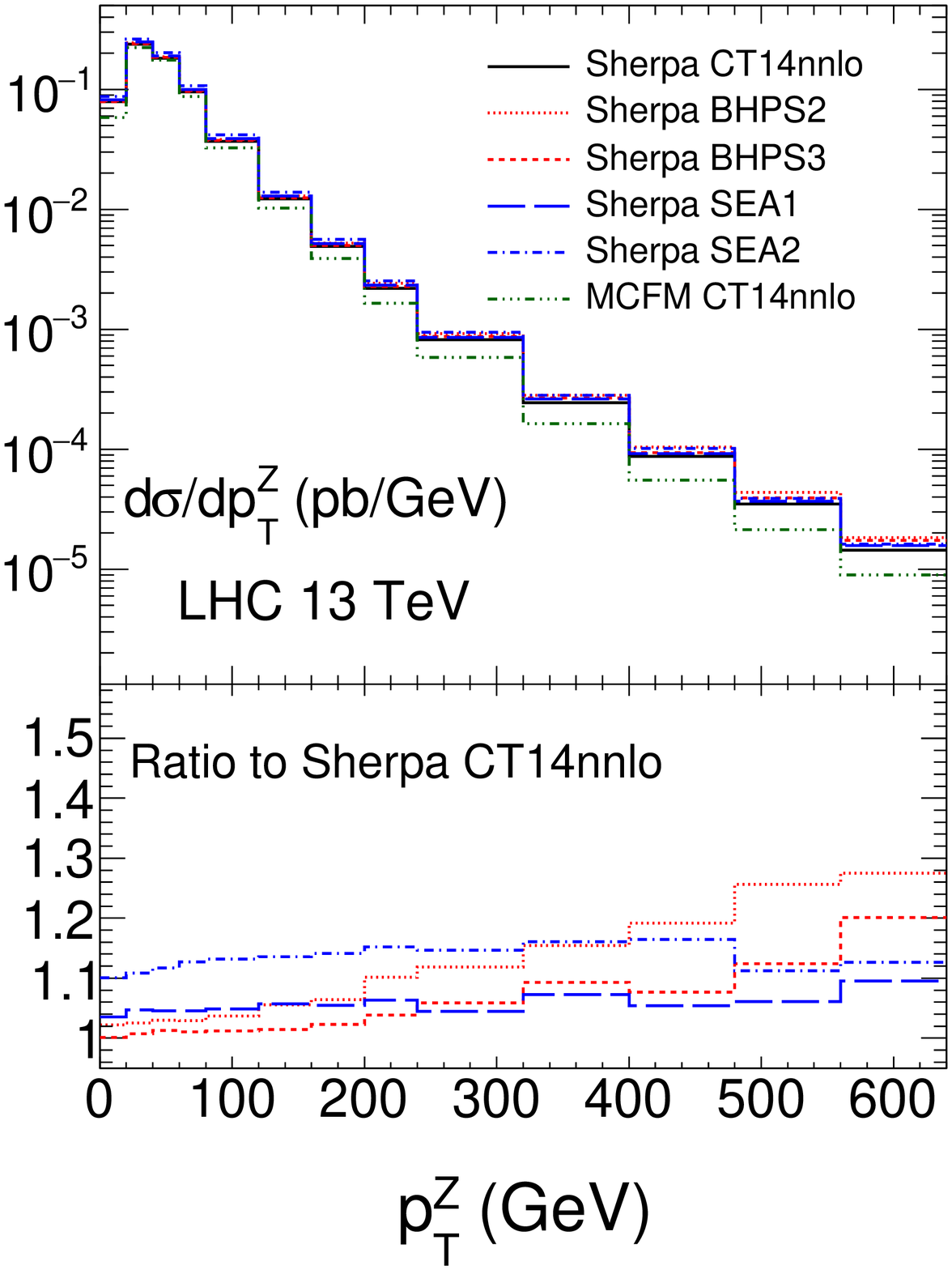}
\vskip-5mm
\caption{\label{fig:zclhc-meps}
  Transverse momentum distribution of $Z$ bosons produced in
  association with at least one charm jet at the LHC, for
  $\sqrt{S}=8\tev$ (left panel) and $\sqrt{S}=13\tev$ (right panel).
  Except for the reference \textsc{MCFM} result, all predictions were
  obtained by using \textsc{Sherpa}'s MEPS@LO algorithm for $Z$+jets
  production with $n_\textrm{ME}=3$ (supplemented by proper charm
  tagging). The bottom panels show the ratios between the
  \textsc{Sherpa 3j} prediction using CT14NNLO and those using the IC
  models.}
\end{figure}

\begin{figure}[t!]
\centering
\includegraphics[width=0.48\textwidth]{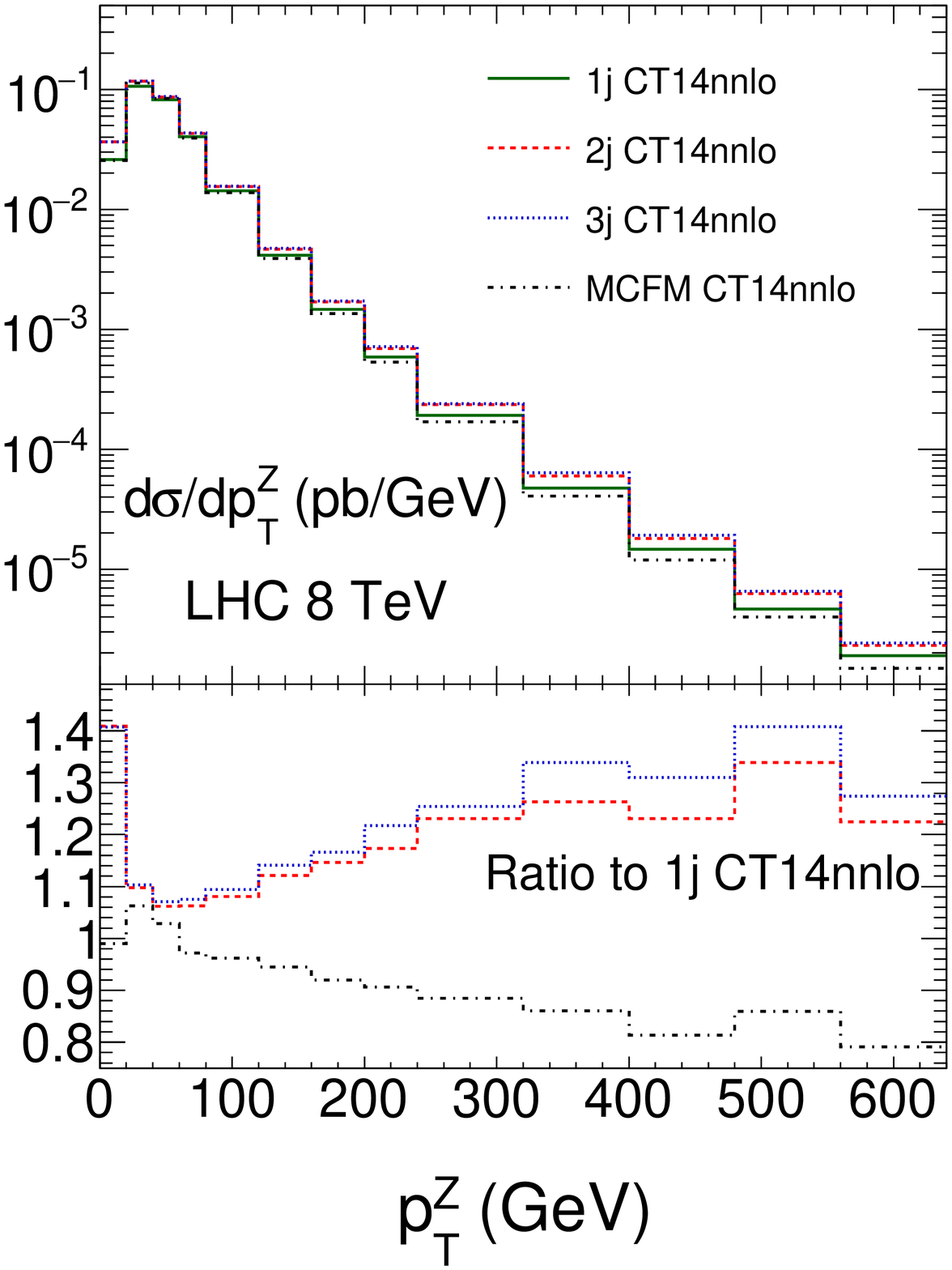}
\hskip5mm
\includegraphics[width=0.48\textwidth]{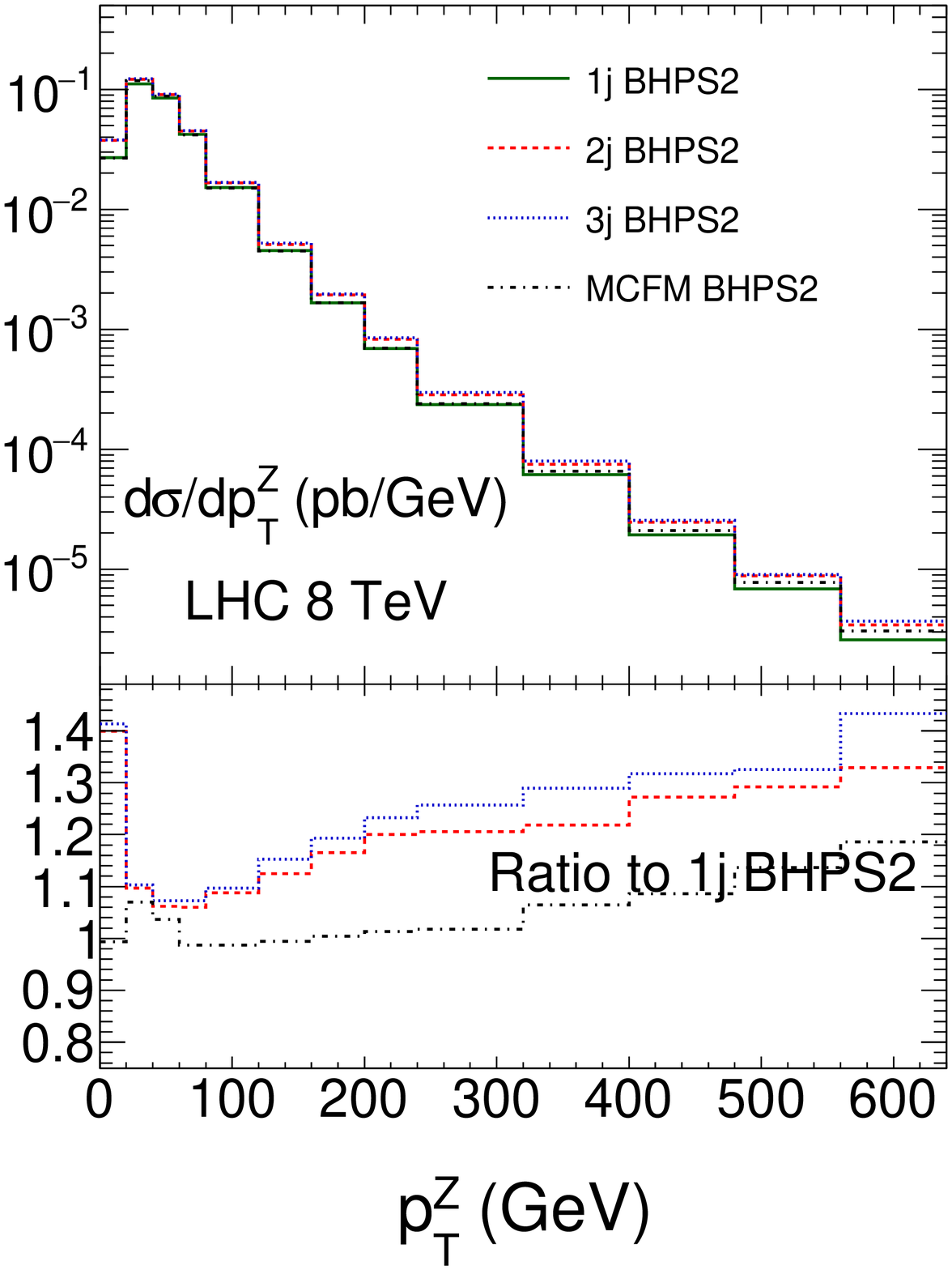}
\vskip-5mm
\caption{\label{fig:zclhc-meps-extra}
  Transverse momentum distribution of $Z$ bosons produced in
  association with at least one charm jet at the LHC for
  $\sqrt{S}=8\tev$. Both panels show \textsc{Sherpa} MEPS@LO
  predictions (obtained by using proper charm tagging) for $Z$+jets
  production with a successively increasing number of multileg matrix
  elements taken into account (i.e.~$n_\textrm{ME}=1,2,3$ where the
  $n_\textrm{ME}=1$ curves serve as the reference). In the left panel,
  Monte Carlo predictions for CT14NNLO are compared with each other
  and the corresponding \textsc{MCFM} result, while in the right
  panel, the same set of curves is shown for using the IC
  parametrization BHPS2.}
\end{figure}

As discussed previously,
we expect the sensitivity to the intrinsic charm component to decrease in a
realistic multijet environment.
The $p_T^Z$ distributions provided by the MEPS@LO method for the
various PDFs are presented in Figures~\ref{fig:zclhc-meps} and
\ref{fig:zclhc-meps-extra},
to be compared with Figures~\ref{fig:zclhc-8} and~\ref{fig:zclhc-13}
depicting the corresponding \textsc{MCFM} results. To support a direct
comparison, the main panels of Figure~\ref{fig:zclhc-meps} also
contain the \textsc{MCFM} prediction for CT14NNLO. While there are no
large deviations between the \textsc{Sherpa} and \textsc{MCFM}
predictions for lower $p_T^Z$ values, the \textsc{Sherpa}
predictions show the expected hardening in the tail of the $p_T^Z$
distributions. In the MEPS@LO simulations, the IC increases
the cross sections in the same way as in the fixed-order case,
although by a smaller factor
(roughly half as much), which is most prominently visible in the associated
ratio plots.

Apart from reconfirming the dilution effect,
Figure~\ref{fig:zclhc-meps-extra} provides us with additional
information. First, the Sudakov (low $p_T^Z$) region is described
in a more sophisticated and therefore robust way (as a result of the
inclusion of resummation effects). Second, regardless of whether CT14NNLO
(Figure~\ref{fig:zclhc-meps-extra}-left) or BHPS2 is used as reference
(Figure~\ref{fig:zclhc-meps-extra}-right), the inclusion of additional
layers of multileg matrix elements leads to relative enhancement and
saturation effects of similar size at larger $p_T^Z$ values.
This is an expression of the fact that although the intrinsic charm
models investigated here do change the initial conditions of the charm
content in the proton, they do not alter the nominal QCD evolution.
The parton shower evolves in the same way as encoded by the DGLAP
theory in the absence of any intrinsic charm.

\begin{table}
\centering
\begin{tabular}{lccccc}\hline\\[-3mm]
  $\left[\,p_{T\mathrm{min}}^Z,p_{T\mathrm{max}}^Z\,\right]$ & $[\gev\,]$ &
  & $[0,30]$ & $[30,60]$ & $[60,200]$\\[2mm]\hline\\[-4mm]
  \multirow{2}{*}{$\Delta\sigma(Zc)/\Delta p_T^Z$} &
  \multirow{2}{*}{$~[\,\mathrm{pb/GeV}\,]~$} & \multirow{2}{*}{CMS} &
  $0.075$            & $0.133$            & $0.017$\\[-1mm]&&&
  $\pm0.011\pm0.012$ & $~~\pm0.013\pm0.018~~$ & $\pm0.002\pm0.002$\\[1mm]
  Rel.\:uncertainty & & CMS & $22\%$ & $17\%$ & $17\%$\\[1mm]\hline\\[-4mm]
  \multirow{4}{*}{\parbox[c]{30mm}{Rel.\:deviation\\wrt.~CT14:\\[2mm]
      $\frac{[d\sigma(Zc)/d p_T^Z]_{\rm IC}}
      {[d\sigma(Zc)/d p_T^Z]_\text{CT14}}-1$}}
  & BHPS2 & \textsc{~MCFM~}      & $4.3\%$ & $4.9\%$ & $9.1\%$\\
  & BHPS2 & \textsc{~Sherpa 3j~} & $3.9\%$ & $4.3\%$ & $6.6\%$\\
  & SEA2  & \textsc{~MCFM~}      & $18\%$  & $19\%$  & $22\%$\\
  & SEA2  & \textsc{~Sherpa 3j~} & $14\%$  & $15\%$  & $16\%$\\[1mm]
  \hline
  \end{tabular}
  \caption{\label{tab:zcpTZ}
    Results of the CMS measurement for the differential $Z$+charm-jet
    cross section as a function of the $Z$ transverse momentum at the
    LHC for $\sqrt{S}=8\tev$~\cite{CMS:2016dyh}. The first uncertainty
    of each data point denotes the statistical error, while the second
    one denotes the systematic error. The relative uncertainties
    associated with the three data points are compared to the size of
    the relative deviations generated by selected IC models with
    respect to the
    CT14NNLO baseline. The theoretical predictions have been obtained
    from \textsc{MCFM} at NLO and \textsc{Sherpa 3j} at MEPS@LO
    accuracy. The details of the calculations are given in the text.}
\end{table}

Similarly to the case of the $Z$+$c$-jet cross section, the CMS data
for the $p_T^Z$ distribution~\cite{CMS:2016dyh} can be used to
estimate the current potential for discriminating possible intrinsic
charm models. The CMS measurement provides cross sections for three
different $p_T^Z/\!\gev$ bins, which are shown in the upper part of
Table~\ref{tab:zcpTZ}, together with their associated relative
uncertainties. These uncertainties are to be compared with the size of
the deviations induced by the intrinsic charm fits with respect to ~the CT14
baseline. According to Figures~\ref{fig:zclhc-8} and
\ref{fig:zclhc-meps}, we can focus on the BHPS2 and SEA2 predictions,
as only those feature differential rates significantly exceeding the
uncertainty range of the CT14 prediction. However, as shown in the
lower part of Table~\ref{tab:zcpTZ}, the deviations generated by both
the BHPS2 as well as the SEA2 model do not exceed the $1\sigma$
variation of the data, in particular if the dilution effect is taken
into account as simulated by MEPS@LO. Thus, none of these changes
reach a magnitude that is distinguishable from the experimental and
theoretical systematic errors at $8\tev$. The discriminating power of
the current CMS data is simply not sufficient to test the IC models,
either in terms of the differential $p_T^Z$ cross section or in terms
of the total $Z$+charm-jet cross section.\footnote{The available
  measurements are still more sensitive to deviations in the total
  cross section. Thus there is a small chance that current data is in
  disfavor of the SEA2 model.}

Owing to the rather low impact of current LHC data, it is important to
better understand the prospects for new measurements of detecting or
excluding a high-$x$ IC component. To this end we extrapolate what we
have learned at $8\tev$ to the case of the $13\tev$ LHC. The CMS
result for $19.7\:\mathrm{fb}^{-1}$ of data at $8\tev$ extends up to a
$Z$ boson transverse momentum range of $200\gev$. The last bin is
fairly wide, from $60\gev$ to $200\gev$, and its associated
differential cross section has been measured as
$\Delta\sigma/\Delta p_T^Z=(0.017\pm0.003)\:\mathrm{pb/GeV}$,
i.e.~it is reasonable to assume that cross sections as low as
$0.01$\,---\,$0.02\:\mathrm{pb/GeV}$ can be measured with
$\sim20\:\mathrm{fb}^{-1}$ of integrated luminosity. In
Figure~\ref{fig:zclhc-8}, a cross section of this size corresponds to
a $p_T^Z$ value of about $120\gev$, which translates into
$x\sim0.03$ on average.
Thus, current measurements probe relatively low values of $x$ compared
to the range ($x\ge0.1$\,---\,$0.2$) where the BHPS models start to
have a significant impact (as shown in Figure~\ref{fig:cratios}). The
cross section for $Z$+charm production of course is larger at
$13\tev$, but for the same low cross section target of
$0.01\:\mathrm{pb/GeV}$, the accessible $p_T^Z$ range would only
be extended by $30\gev$ (according to Figure~\ref{fig:zclhc-13})
pulling the mean $x$ towards $0.02$, which means we would not even
achieve the same sensitivity as for the $8\tev$ case. To reach a
similar $x$ range would require a $Z$ transverse momentum of the order
of $200\gev$ corresponding to a cross section of about
$0.002\:\mathrm{pb/GeV}$. One therefore needs an integrated luminosity
of about $100\:\mathrm{fb}^{-1}$, in order to determine this cross
section with an accuracy comparable to the $8\tev$ case. In other
words, it will take the full Run\:2 cycle to barely get a first
$2\sigma$ sign of deviations at $p_T^Z\sim200\gev$ or probe
transverse momenta of the order of $300\gev$. Needless to say that
definitive confirmation/exclusion will require us to go considerably
beyond the Run\:2 luminosity budget.

The challenging environment for the $Z+c$ analysis forces us
to search for ways to increase the impact of an intrinsic charm
component on the $Z$+$c$-jet cross section. As this cross section is
diluted by the presence of the radiative corrections, for example,
limiting the number of jets in the event could reduce this dilution.
The $Z$+$c$-jet rate could also be measured as a function of the leading
(charm) jet transverse momentum, which in fact has been carried out by CMS in the same publication.
Our studies suggest that this differential cross section is somewhat
more sensitive to the intrinsic charm modeling investigated here, but its sensitivity must be weighed against the size of the relative uncertainties on the measurement of
the charm jet $p_T$, in a similar fashion as shown for $p_T^Z$ in Table III.
In addition, deviations are also found for $Z$ boson rapidities
outside the central phase-space region, such as might be measured at
LHCb \cite{Boettcher:2015sqn}.

\section{Summary and conclusions
\label{sec:Conclusions}}

We have explored the possibility of having a sizable nonperturbative
contribution to charm parton distribution function (PDF), i.e., the
intrinsic charm (IC) quark component, in the proton, using the
CTEQ-TEA (CT) global analysis. In Sec.~\ref{sec:QCDFactorization}, we
reviewed the theoretical framework used in the CT global analysis,
and discussed the conditions under which our formalism, Eq.~(\ref{Fpheno}), can
better approximate the QCD factorization theorem, Eq.~(\ref{F}).

The notion of ``intrinsic charm'' refers to
contributions to charm quark production and scattering that arise
besides twist-2 ``perturbative'' contributions. In DIS,
the twist-4 cross sections for charm production may numerically compete
with ``perturbative'' twist-2 cross sections at a high enough order in
$\alpha_s$. For example, in Fig.~\ref{fig:ICdiagrams}, we show the
relevant squared amplitudes for a DIS structure function
$F(x,Q)$ from the $\gamma^*+gg\rightarrow c+\bar c$ process.
The flavor-creation diagrams $F_{h,gg}^{(n)}$
render most of the twist-4 charm production rate in the HERA kinematical
region ($Q \gtrsim m_c$). But, at very high photon virtualities, $Q^2 \gg
m_c^2$, their dominant part is approximated
in a variable-flavor number scheme by
a twist-2 coefficient function $c_{h,h}^{(k)}$ convoluted with a
universal charm
PDF $c(x,Q)$. A non-zero boundary condition for $c(x,Q)$
at $Q=Q_c\sim m_c$ is obtained
by perturbative matching from light-parton nonperturbative
twist-two and twist-four functions,
such as $f_{g/p}(x,Q_c)$ and $f_{gg/p}(x_1,x_2,Q_c)$.

In the context of the phenomenological PDF analyses, on the other
hand, the ``intrinsic
charm'' PDF is often conflated with a ``fitted charm'' PDF parametrization that
plays a dual role of the approximant for the above
power-suppressed contribution to charm scattering
and of a parametric surrogate for unrelated radiative contributions
that were not explicitly included. At the moment the fitted PDF is
determined solely using the fixed-order convolutions with the twist-2
coefficient functions, without including explicit twist-4 terms.
While the ``fitted charm'' PDF provides a good description of the
cumulative QCD data in the CT fit, care is necessary when making predictions
for new processes based on its parametrization, as it may absorb a
host of process-dependent corrections, notably the contribution of
DIS-specific twist-4 coefficient functions like in
Fig.~\ref{fig:ICdiagrams}. We, as well as the other global
analysis groups, treat the ``fitted charm PDF'' obtained this way
as though it is mostly process-independent, until it is
demonstrated otherwise.

For example, in neutral-current DIS charm
production the twist-4 charm cross section is
 of the same order in the QCD coupling strength as the NNLO twist-2
 one. To estimate the magnitude of the twist-4 IC cross section from
 the DIS data, using its model given by the fitted charm, the twist-2
 DIS contributions in the fit must be evaluated at least to NNLO. Furthermore,  it is
 necessary to study the contributions from the strange (and bottom)
 PDF, dependence on the charm quark mass ($m_c$), and to accurately
 implement suppression of charm production at the mass threshold. In
 the case when low-$Q$ fixed-target data are included, the IC
 component must be further discriminated from the $1/Q^2$
 and nuclear-target effects.

Hence, in this study, we have used both the CT14 NNLO and CT14HERA2 NNLO
analyses, differing mainly in their strange PDFs.
CT14HERA2 has a softer strange quark component
than CT14 at most $x$ values. We have carried out a series of
fits with a varied charm quark pole mass $m_c$ between 1.1 and 1.5
GeV, within the preferred $m_{c}^{pole}$ range of our global fits,
see Fig.~\ref{fig:Mc_dchi2Vxic}.

The NNLO heavy-quark mass effects are implemented in our calculation
using the S-ACOT-$\chi$ factorization
scheme.\footnote{The massive NC DIS perturbative coefficients
  are known in their entirety to NNLO
  in the S-ACOT-$\chi$ \cite{Guzzi:2011ew}, TR'
  \cite{Martin:2010db}, and FONLL-C \cite{Forte:2010ta}
  schemes.  In contrast, some of these NNLO coefficients are still unknown
  in the ``fully massive'' ACOT scheme \cite{Aivazis:1993pi} and its
  FONLL equivalent \cite{Ball:2015tna,Ball:2015dpa} adopted in NNPDF3.1.}
In Sec.~\ref{sec:QCDFactorization}, we have given detailed
arguments showing that it is a self-consistent and sufficient
scheme for predicting massive-quark DIS contributions both in the
twist-2 and twist-4 channels.

The charm content in a hadronic bound state, quantified by an operator
matrix element identified with the charm PDF, can in principle be
predicted by QCD. We examine which ``intrinsic charm'' models predict
the fitted charm PDF compatible with the global QCD data.
Two generic types of the charm models introduced in
Sec.~\ref{sec:ModelsForFittedCharm}, a valence-like BHPS model
and a sea-like SEA model, predict a non-zero
$c(x,Q_0)$ at large $x$ and across all $x$,
respectively. The BHPS model is solved either approximately in the BHPS1
and BHPS2 PDF sets, or exactly in the BHPS3 set.
To better predict the PDF ratios of charm to up and down PDFs,
in the BHPS3 model we also allowed for small intrinsic contributions
to the $\bar{u}$ and $\bar{d}$ (anti-)quarks generated from the
$|uudu\bar{u}\rangle$ and $|uudd\bar{d}\rangle$ Fock states,
included together with the charm intrinsic contribution.
Though we did not present its details, we
have also studied a mixed model of BHPS and SEA and arrived at similar
conclusions.

Figure~\ref{fig:delta_chisqVxic}
shows that, at $Q_0=1.3$ GeV, the charm quark momentum fraction
$\langle x\rangle_{\rm IC}$, as defined in Eq.~\ref{xICdef},
is found to be less than about 2\% and 1.6\%, for the
BHPS IC and SEA IC models, respectively, in the CT14NNLO analysis, at the 90\% C.L.
We note that by its definition, $\langle x\rangle_{\rm IC}$ is evaluated
at the initial scale $Q_{0}$. It is to be distinguished from the full
charm momentum fraction $\langle x\rangle_{c+\bar{c}}$ at $Q > Q_c$,
which rapidly increases with $Q$ because $c(x,Q)+\bar{c}(x,Q)$  also
includes the perturbative contribution. The dependence of the
outcomes on $m_c^{pole}$ was reviewed in Sec.~\ref{sec:McDependence},
and the resulting BHPS and
SEA PDFs and parton luminosities, as well as $Q$ dependence of
$\langle x\rangle_{c+\bar{c}}$, were explored in Section~\ref{sec:PDFs}.

A significant IC component in the proton wave
function could influence observables measured at the
LHC, either directly through enhanced cross sections
via the charm PDF, or indirectly via the momentum sum rule leading
to a change in the momentum fraction carried by the gluons.
Modifications in the light-flavor PDFs are generally mild in the
considered BHPS/SEA models, although the gluon-gluon luminosities can
be suppressed at the highest final-state invariant masses $M_X$, as
observed in Fig.~\ref{ggLumi}. The allowed momentum fraction $\langle
x\rangle_{\rm IC}$ is correlated with the charm pole mass
$m_{c}^{pole}$, especially in the SEA model.
When the charm PDF is purely
perturbative, the inclusive $Z$ cross section increases as $m_c^{pole}$
increases, due to the larger $\bar u$ and $\bar d$ PDFs that compensate
for the smaller perturbative charm PDF contribution. We also observe
reduction in $g(x,Q)$ at large $x$, and consequently some reduction in
cross sections sensitive to large-$x$ gluon scattering. For example, increasing
$m_{c}^{pole}$ from the nominal 1.3 to 1.5 GeV increases the $W/Z$
inclusive total cross sections at 13 TeV, reduces the normalized
high-$p_T$ $Z$ production cross section at the LHC 7 TeV,
and has vanishing effect on the $gg\rightarrow
H^0$ cross sections, see Sec.~\ref{sec:key-obs}. These changes can be
partly offset by introducing the IC, possibly at the expense of some
tension with the non-LHC fitted experiments, and generally within the
regular CT14 PDF uncertainty.

There is much discussion in the literature about the impact of
the EMC measurement \cite{Aubert:1982tt}
of semi-inclusive DIS charm production on
the intrinsic charm PDF. Although our standard analysis does
not include the EMC data, we have examined their impact in several IC models.
Section~\ref{sec:EMCdata} argues that fitting the EMC data is
not expedient, their persistent tension with the other fitted data
sets may reflect the systematic errors that were not documented in the EMC publication.
The level of (dis)agreement with the purely
perturbative charm and the exclusion limits on the intrinsic charm depend
on the assumed magnitude of systematic effects in the EMC measurement.
As shown in Table~\ref{TAB1}, even without the IC contribution,
the $\chi^2/N_{\rm pts}$ of the EMC data
varies from about 3.5 to 2.3 when it is excluded or included with a
large statistical weight in the CT14 fits.
Including the intrinsic charm component
does not significantly change $\chi^2/N_{\rm pts}$ for the EMC.
For the BHPS models, including the EMC data with the nominal errors
reduces the tolerated range of $\langle x\rangle_{\rm IC}$ by about a factor of
two.  The impact of EMC data is small within the SEA model.

Besides the LHC electroweak boson production cross sections, we
examined the implications of the IC for
associate production of $Z$ boson and charm-jet at the LHC,
and summarized our findings in Table~\ref{tab:zclhc}
and Figs.~\ref{fig:zclhc-8}-\ref{fig:zclhc-meps-extra}.
A fixed-order calculation for $Z+c$ production, MCFM at NLO,
was compared to a merged parton showering calculation in Sherpa,
which also generates charm jets in the final state via
gluon splittings. In general, in a fixed-order calculation for $Z+c$,
the various IC models predict enhanced rate
in the transverse momentum distribution of a $Z$ boson ($p_T^Z$)
\cite{Boettcher:2015sqn}. The SEA models tend to
predict a higher differential cross section across all $p_T^Z$,
while the BHPS models suggest the increased rate only at the highest $p_T^Z$.
The predictions based on the NNPDF3IC and NNPDF3IC (no EMC) PDFs
are close to our BHPS3 and BHPS2 predictions, respectively,
they predict a larger rate in the high $p_T^Z$ region.

Inclusion of the final-state parton showering typically dampens the
fixed-order enhancement induced by the IC contribution, as can be observed from the comparison
of Sherpa to MCFM predictions. The dampening is mainly attributed to
the gluon-splitting contributions in the final state which reduce the
relative impact of the IC contribution in the hard $p_T^Z$ tail,
especially for the predictions from the BHPS models.

The analysis of QCD factorization indicates that
the power-suppressed ``intrinsic'' component in semi-inclusive DIS charm
production may be comparable in magnitude to some NNLO and N3LO
leading-power contributions.
Hence, a serious study needs to be carried out at least at
the NNLO, such as in this work.
(It is not possible to draw a definite conclusion from an NLO analysis.)
As of today, the experimental
confirmation of the IC component in the proton is
still missing, and data from far more sensitive measurements are required.
An analysis of very low-$Q$ fixed-target data,
such as the one presented at NLO
in Refs.~\cite{Jimenez-Delgado:2014zga,Jimenez-Delgado:2015tma}, must
meet the challenge of the reliable separation of the IC from the other
relevant factors, including higher-order twist-2
contributions, the $1/Q^2$ terms, $m_c$ dependence,
and nuclear effects. The constraints on the IC from the higher-energy
data are largely compatible between the CT14 IC and NNPDF3.x analyses
\cite{Ball:2016neh,Ball:2017nwa}. Our limits on $\langle x
\rangle_{c+\bar{c}}$ are moderately more conservative than those of NNPDF3.1,
as we do not include the EMC $F_{2c}$ data and acknowledge competing
preferences for $m_{c}^{pole}$ and $\langle x \rangle_{c+\bar{c}}$ among the
various non-LHC and LHC experiments, as outlined in
Secs.~\ref{sec:McDependence}, \ref{sec:DataImpact}, and
\ref{sec:key-obs}. Ultimately,
a combination of high-luminosity measurements at the Large Hadron Collider,
such as $Z+c$ production, and charm SIDIS at the Electron-Ion Collider
\cite{GuzziIC2011} will be desirable to test intrinsic
charm scattering contributions at NNLO and beyond.

\section*{Acknowledgments}
We thank J. Collins and D. Soper for valuable insights on massive
power-suppressed contributions in the QCD factorization framework,
K.-F. Liu for a discussion of massive-quark production in lattice QCD,
and J. Pumplin, S. Alekhin, L. Del Debbio, T. Hobbs, W. Melnitchouk, F. Olness,
J. Rojo, and M. Ubiali for stimulating communications. P.N. thanks the
Kavli Institute for Theoretical Physics at Santa Barbara, CA and
organizers of the ``LHC Run II and Precision Frontier'' research
program during which this work was initiated. This work was supported
in part  by the U.S. Department of Energy under Grant No. \protect{DE-SC0010129};
by the U.S. National Science Foundation under Grant No. PHY-1417326;
by the  National Natural Science Foundation of China under the Grant
No. 11465018; and by the Lancaster-Manchester-Sheffield Consortium for
Fundamental Physics under STFC Grant No. ST/L000520/1. The work of J.G.
is sponsored by Shanghai Pujiang Program.

\end{document}